\pdfoutput=1
\documentclass[12pt]{article}
\usepackage{tikz}
\usepackage{tikz-cd}

\usetikzlibrary{decorations.pathmorphing}

\usepackage{jheppub}

\usepackage{pgfplots}
\usepackage{pgfplotstable}
\usepackage{pgfkeys}
\allowdisplaybreaks
\usepackage{ytableau}
\usepackage{comment}
\usepackage[vcentermath]{youngtab}
\usepackage{bm}
\usepackage{ulem}
\usepackage{xcolor}
\definecolor{nosaka}{rgb}{0.0, 0.7, 0.0}

\newcommand\be{\begin{equation}}
\newcommand\ee{\end{equation}}

\newcommand\Tr{\mathrm{Tr}}

\usepackage{braket}
\usepackage{subfigure}

\title{Abelian dualities and line defect indices for 3d gauge theories}
\abstract{
We find matching pairs of the line defect indices for 3d supersymmetric Abelian gauge theories 
as strong evidence of dualities of the BPS line operators. 
They lead to novel duality maps of the BPS line operators 
for $\mathcal{N}\ge 4$ supersymmetric circular and linear quiver gauge theories which can be realized as brane configurations in Type IIB string theory, 
including SQED, ADHM and ABJM theories. 
}

\author[a]{Hirotaka Hayashi,}
\emailAdd{h.hayashi@tokai.ac.jp}
\affiliation[a]{Department of Physics, School of Science, Tokai University,\\
4-1-1 Kitakaname, Hiratsuka-shi, Kanagawa 259-1292, Japan}

\author[b,c]{Tomoki Nosaka}
\emailAdd{nosaka@simis.cn}
\affiliation[b]{
Fudan Center for Mathematics and Interdisciplinary Study, Fudan University, Shanghai 200433, China
}
\affiliation[c]{
Shanghai Institute for Mathematics and Interdisciplinary Sciences,\\
Block A, International Innovation Plaza, No.~657 Songhu Road, Yangpu District, Shanghai, China
}
\author[d]{and Tadashi Okazaki}
\emailAdd{tokazaki@seu.edu.cn}
\affiliation[d]{
Shing-Tung Yau Center of Southeast University,\\
Yifu Architecture Building, No.2 Sipailou, Xuanwu district, Nanjing, Jiangsu, 210096, China}

\begin{document}
\maketitle

\ytableausetup{boxsize=1.5mm}

\section{Introduction and summary}
Wilson lines and vortex lines are two basic types of line operators in 3d gauge theories. 
They are non-local operators supported on a one-dimensional line. 
The Wilson lines are electric operators labeled by representations of gauge or flavor group. 
They serve as important order parameters \cite{Wilson:1974sk}. 
On the other hand, the vortex lines are magnetic codimension-two disorder operators. 
Such operators were first introduced in the context of Chern-Simons (CS) theories \cite{Moore:1989yh,Witten:1988hf}.
They can be defined by considering the theory with prescribed singular configurations of gauge fields and matter fields as one approaches the line. 

In 3d supersymmetric gauge theories, line operators can preserve part of supersymmetry. 
It is pointed out \cite{Blau:1996bx,Baulieu:1997nj,Maldacena:1998im,Rey:1998ik,Zarembo:2002an,Gaiotto:2007qi,Drukker:2008zx,Chen:2008bp,Rey:2008bh,Assel:2015oxa} 
that the supersymmetric Wilson lines can be constructed in terms of an appropriate modified connection comprising 
an ordinary connection and certain matter fields as it can be annihilated by supercharges. 
On the other hand, the supersymmetric vortex lines can be found by taking singular limits of the solutions to the BPS equations in a plane transverse to the line, 
which can be identified with vortex equations \cite{Nielsen:1973cs}. 
\footnote{See \cite{MR614447} for the details. 
Also see e.g.~\cite{MR573986,MR614447} for Abelian vortex equations and \cite{MR1085139,MR1124279,MR1250254} for non-Abelian vortex equations.} 
The supersymmetric vortex lines were first examined in \cite{Drukker:2008jm} for the ABJM theory \cite{Aharony:2008ug}. 
The prescription for computing the correlation function of the BPS line operators in 3d $\mathcal{N}=2$ supersymmetric gauge theories 
were studied in \cite{Kapustin:2009kz,Kapustin:2012iw,Dimofte:2011py,Drukker:2012sr,Hosomichi:2021gxe} 
with the aid of a supersymmetric localization technique. 
For 3d $\mathcal{N}=4$ supersymmetric gauge theories, the half-BPS line operators can be engineered 
by introducing strings in the brane configurations \cite{Hanany:1996ie,deBoer:1996mp,deBoer:1996ck} of Type IIB string theory as discussed in \cite{Hanany:2003hp,Assel:2015oxa}. 
Accordingly, it was proposed in \cite{Assel:2015oxa} that 
S-duality in Type IIB string theory leads to dualities of the line operators 
under which the Wilson and vortex lines are exchanged for 3d mirror pairs \cite{Intriligator:1996ex}. 
The dualities between the half-BPS Wilson lines and the vortex lines are tested by matching of the expectation values of the line operators on $S^3$ \cite{Assel:2015oxa} 
(See \cite{Dey:2021jbf,Nawata:2021nse} for further extensions). 
The BPS Wilson and vortex lines in supersymmetric gauge theories can be also described 
in terms of 1d supersymmetric quantum mechanics (SQM) on their worldlines \cite{Assel:2015oxa,Hosomichi:2021gxe}. 
While the BPS vortex lines have not yet been fully classified in general gauge theories, 
it was proposed in \cite{Dimofte:2019zzj} that the BPS vortex lines can be characterized by algebraic data 
which can be obtained from a coupling of the 3d $\mathcal{N}=4$ bulk fields to the 1d $\mathcal{N}=4$ SQM. 

In this paper, we examine the \textit{line defect indices} or \textit{line defect correlators} 
in 3d $\mathcal{N}\ge 4$ Abelian circular and linear quiver gauge theories which can be realized as brane configurations in Type IIB string theory. 
While the supersymmetric indices can be defined as a supersymmetric partition function on $S^1\times S^2$ \cite{Kim:2009wb,Imamura:2011su,Kapustin:2011jm}, 
they can be decorated as the line defect indices by introducing the BPS line operators wrapping the $S^1$ and localized on the $S^2$ \cite{Drukker:2012sr}. 
We obtain strong evidence of dualities of line operators in 3d $\mathcal{N}\ge 4$ Abelian circular and linear quiver gauge theories which can be realized in Type IIB string theory 
by finding matching pairs of the line defect indices. 
For mirror pairs of the $\mathcal{N}=4$ SQED and its mirror linear quiver gauge theory 
and for those of the $\mathcal{N}=4$ ADHM theory and its mirror circular quiver gauge theory, 
we find strong evidence of the dualities proposed by Assel and Gomis \cite{Assel:2015oxa}. 
The dualities map the gauge Wilson lines to the flavor vortex lines and vice versa. 
More generally, corresponding to the $STS$ duality transformation in Type IIB string theory, 
we propose dualities of line operators in 
the SQED (resp.~the ADHM theory) and those in the Abelian linear (resp.~circular) quiver CS theories \cite{Jafferis:2008em,Imamura:2008nn,Imamura:2008dt}, including the ABJ(M) theory \cite{Aharony:2008ug,Aharony:2008gk}.
In addition, we find simple examples of $\mathcal{N}=4$ Seiberg-like dualities \cite{Gaiotto:2008ak} of the line operators. 
The matching of the line defect indices generalize the results in \cite{Okazaki:2019ony,Hayashi:2022ldo} for the ordinary indices. 
We have checked all the proposed identities of the full line defect indices by using Mathematica up to certain high orders in $q$. 
In particular, we prove the identities of the Coulomb and Higgs limits. 

\subsection{Structure}
The organization of the paper is as follows. 
In section \ref{sec_lineindex} we begin by reviewing the line operators and the line defect indices 
in $\mathcal{N}\ge 4$ supersymmetric Abelian gauge theories. 
We then summarize the brane construction of the BPS line operators in Type IIB string theory 
and known dualities of these theories. 
New features of our discussion contain the $STS$ transformation of $SL(2,\mathbb{Z})$ duality in Type IIB string theory 
that maps $\mathcal{N}=4$ supersymmetric quiver CS theories. 
In section \ref{sec_SQED} we analyze the line defect indices for the SQED and its mirror. 
Also we investigate the Seiberg-like dualities for the ``ugly'' theories. 
In section \ref{sec_ADHM} the dualities of the line operators for mirror pairs of the $U(1)$ ADHM theory and its mirror are tested. 
In section \ref{sec_linearCS} we find the evidence for 
the dualities of line operators for $\mathcal{N}=4$ linear quiver CS theories and those for the SQED upon the $STS$ transformation. 
In section \ref{sec_ABJM} we obtain exact closed-forms for the line defect correlators for the Abelian ABJM theories. 
We propose that the line operators in the Abelian gauge theories are dual to the flavor vortex lines in discrete gauge theories.
We also consider the generalization to the circular quiver CS theories with more gauge nodes and find the duality relation of the line defect indices to those in the ADHM theory.
In appendix \ref{app_proofHC} we give proofs of the identities of the line defect indices in the Coulomb and Higgs limits. 

\subsection{Future works}

\begin{itemize}

\item The proposed prescription of the vortex line defect indices would be rather general 
as it can be also applied to the 3d Abelian gauge theories with at least $\mathcal{N}=2$ supersymmetry. 
It would be interesting to examine the line defect indices to test further dualities of line operators. 

\item While we focus on the analysis for the Abelian gauge theories in this work, 
the analysis can be generalized to the non-Abelian gauge theories, for which there exist dynamical vortex lines. 
Also there exist non-trivial two-point functions of the line operators, e.g.~pairs of the Wilson lines,  
for which the configuration can map to the straight line in the flat space upon the conformal map. 
It would be intriguing to analyze the large $N$ limits 
\footnote{
The Coulomb and Higgs limits of the Wilson line defect indices of the $U(N)$ ADHM theories were examined in \cite{Hayashi:2024jof}. 
}
and the giant graviton expansions 
\footnote{
See \cite{Imamura:2024lkw,Imamura:2024pgp,Beccaria:2024oif,Hatsuda:2024uwt,Imamura:2024zvw} for the giant graviton expansions for the line defect indices for 4d $\mathcal{N}=4$ SYM theories 
and \cite{Gaiotto:2021xce,Arai:2020uwd,Beccaria:2023sph,Hayashi:2024aaf} for the giant graviton expansions for the M2-brane SCFTs. 
}
of these line defect correlators to derive the spectra of the fluctuation modes on the $AdS_2$ gravity duals (See e.g.~\cite{Drukker:2008jm,Drukker:2008zx,Rey:2008bh,Farquet:2013cwa,Chen:2014gta,Aguilera-Damia:2014bqa,Muck:2016hda,Cookmeyer:2016dln,
Lietti:2017gtc,David:2019lhr,Correa:2019rdk,Giombi:2020mhz,Giombi:2023vzu} for the line defects in the M2-brane SCFTs). 
We hope to report results for the non-Abelian gauge theories in the upcoming work. 

\item The configuration examined in this paper can be generalized by coupling the 3d $\mathcal{N}=4$ gauge theories 
to the 4d $\mathcal{N}=4$ SYM theory with boundary so that the index can be lifted to the ``line defect half-index'' of 4d theories \cite{Gang:2012yr,Gaiotto:2019jvo,Okazaki:2019ony,Hatsuda:2025yzp}. 
In addition to the vortex-Wilson lines in the 3d theories, one can introduce the 't Hooft-Wilson lines in the 4d theories. 
Our results can be used to study line defect half-indices involving the 3d Abelian gauge theories. 

\item Another interesting extension of our setup is to introduce the BPS boundary conditions or interfaces in the 3d $\mathcal{N}=4$ gauge theories. 
The dualities of $\mathcal{N}=(0,4)$ boundary conditions and those of $\mathcal{N}=(2,2)$ boundary conditions in the Abelian gauge theories 
were found in \cite{Okazaki:2019bok} and in \cite{Okazaki:2020lfy} respectively. 
The half-indices of these boundary conditions should be generalized by introducing the boundary line operators. 

\end{itemize}

\section{Line defects in 3d $\mathcal{N}=4$ Abelian gauge theories}
\label{sec_lineindex}

\subsection{Wilson line}
\label{sec_Wline}
There are two basic types of line operators in 3d Abelian gauge theories. 
The first one is a Wilson line operator $W_{\mathsf{p}}$ of charge $\mathsf{p}$ $\in \mathbb{Z}$ of the Abelian gauge or flavor group. 

Let us consider 3d $\mathcal{N}=4$ supersymmetric field theories with eight supercharges $Q_{\alpha}^{A\dot{A}}$, 
where $\alpha=+,-$ is the spinor index for the $Spin(3)$ $\cong$ $SU(2)_J$, 
$A=1,2$, $\dot{A}=\dot{1},\dot{2}$ are the spinor indices for the $SU(2)_H$, $SU(2)_C$ R-symmetry. 
When we choose $U(1)_J\times U(1)_H\times U(1)_C$ subgroup of $Spin(3)_J\times SU(2)_H\times SU(2)_C$ by fixing the complex structures, 
the supercharges transform as
\begin{align}
\begin{array}{c|cccccccc}
&Q_{-}^{1\dot{1}}&Q_{-}^{1\dot{2}}&Q_{-}^{2\dot{1}}&Q_{-}^{2\dot{2}}&Q_{+}^{1\dot{1}}&Q_{+}^{1\dot{2}}&Q_{+}^{2\dot{1}}&Q_{+}^{2\dot{2}} \\ \hline 
U(1)_J&-&-&-&-&+&+&+&+ \\
U(1)_H&+&+&-&-&+&+&-&- \\
U(1)_C&+&-&+&-&+&-&+&- \\
\end{array}
\end{align}
The half-BPS charged Wilson line of charge $\mathsf{p}$ in 3d $\mathcal{N}=4$ supersymmetric Abelian gauge theory can be defined by 
\cite{Assel:2015oxa}
\begin{align}
W_{\mathsf{p}}&= \textrm{exp} \left[i\mathsf{p} \oint ds \left(A_{\mu}\dot{x}^{\mu}-i |\dot{x}|\sigma \right)\right], 
\end{align}
where $A_{\mu}$ is a $U(1)$ connection and $\sigma$ is one of the three scalar fields which are contained in the $\mathcal{N}=4$ $U(1)$ vector multiplet. 
$x^{\mu}(s)$ specifies the path embedded in the three-dimensional space-time. 
By construction the half-BPS Wilson line splits the three scalar fields in the $\mathcal{N}=4$ vector multiplet into a real scalar $\sigma$ and a complex scalar $\varphi$ 
so that it breaks the $SU(2)_C$ down to $U(1)_C$, while it preserves the $SU(2)_H$, 
where $SO(4)_R$ $\cong$ $SU(2)_C$ $\times$ $SU(2)_H$ is the R-symmetry of 3d $\mathcal{N}=4$ supersymmetric field theory. 
The half-BPS Wilson line can be also described by coupling 1d $\mathcal{N}=4$ SQM 
which is obtained from 2d $\mathcal{N}=(0,4)$ Fermi multiplet \cite{Assel:2015oxa}. 
The four supercharges which generate the 1d $\mathcal{N}=4$ SQM are chosen as 
\begin{align}
\begin{array}{c|cccccccc}
&Q_{-}^{1\dot{1}}&Q_{-}^{1\dot{2}}&Q_{-}^{2\dot{1}}&Q_{-}^{2\dot{2}}&Q_{+}^{1\dot{1}}&Q_{+}^{1\dot{2}}&Q_{+}^{2\dot{1}}&Q_{+}^{2\dot{2}} \\ \hline 
\textrm{Wilson}&\circ&&\circ&&&\circ&&\circ \\
\end{array}. 
\end{align}

Similarly, for the CS matter theories, 
supersymmetric Wilson lines can be constructed in terms of the complexified connections  
by combining the connection with bosonic matter fields. 
For the $U(1)_k\times U(1)_{-k}$ ABJM theory \cite{Aharony:2008ug}, 
one can define the supersymmetric Wilson line with a pair $(\mathsf{p}^{(1)},\mathsf{p}^{(2)})$ of electric charges by 
\begin{align}
\label{ABJMW1/6}
W_{\mathsf{p}^{(1)},\mathsf{p}^{(2)}}&=
\textrm{exp} 
\left[ i\mathsf{p}^{(1)} \oint ds \left(A_{\mu}\dot{x}^{\mu}-\frac{2\pi i}{k} |\dot{x}|{M^I}_{J} C_I\overline{C}^J \right)\right] 
\nonumber\\
&\times 
\textrm{exp} 
\left[ i\mathsf{p}^{(2)} \oint ds \left(\hat{A}_{\mu}\dot{x}^{\mu}-\frac{2\pi i}{k} |\dot{x}|{\overline{M}_J}^{I} \overline{C}^J C_I\right)\right]. 
\end{align}
Here $A_{\mu}$ and $\hat{A}_{\mu}$ are connections for the $U(1)_k$ and $U(1)_{-k}$ gauge group respectively. 
$C_I$, $I=1,2,3,4$ are four complex scalar fields and $\overline{C^I}$ $=$ $(C_I)^{\dag}$. 
Again the path is parametrized by $x^{\mu}(s)$. 
${M^I}_J(s)$ and ${\overline{M}_J}^I(s)$ are matrices for the $SU(4)$. 
When the path is a straight line or a circle, 
the Wilson line (\ref{ABJMW1/6}) can preserve $1/6$ of the supersymmmetry 
with $M=\mathrm{diag}(-1,-1,1,1)$ \cite{Drukker:2008zx,Chen:2008bp,Rey:2008bh}. 
It is pointed out that one can construct the Wilson line by incorporating fermionic matter fields in the connections for the CS matter theories. 
Such ``fermionic'' Wilson lines were first proposed in \cite{Drukker:2009hy} (also see \cite{Lee:2010hk}) as the half-BPS Wilson line in the ABJM theory 
and then extended to less supersymmetric ones \cite{Ouyang:2015iza}. 
These ``fermionic'' Wilson lines are shown to be cohomologically equivalent to a specific choice of the ``bosonic'' $1/6$-BPS Wilson lines \cite{Drukker:2009hy}. 
Accordingly, if we know the line defect correlators of the $1/6$-BPS Wilson lines, those of the ``fermionic'' BPS Wilson lines can be obtained straightforwardly.  
The BPS Wilson lines in $\mathcal{N}\ge 4$ CS matter theories 
can be constructed in the same spirit of the ABJM theory 
\cite{Ouyang:2015iza,Ouyang:2015bmy,Ouyang:2015qma,Cooke:2015ila,Mauri:2017whf,Mauri:2018fsf,Drukker:2020opf,Drukker:2020dvr,Drukker:2022ywj,Drukker:2022bff} (see \cite{Drukker:2019bev} for review). 

\subsection{Vortex line}
\label{sec_vline}
The other basic line operator is a vortex line $V_{\mathsf{q}}$ of charge $\mathsf{q}$ for the Abelian gauge or flavor group. 
It is a disorder line operator as it can be defined by a path integral with singular profile of the Abelian gauge field $A$ and matter fields. 
The charge $\mathsf{q}$ is called the \textit{vortex number}. 
It can be expressed as an integral of the curvature $F=dA$ 
\begin{align}
\label{vortexN}
\mathsf{q}&=\frac{1}{2\pi}\int F \in \mathbb{Z}. 
\end{align}
The vortex number can be also defined as the winding number of a gauge transformation. 
If the theory involves matter fields charged under the associated group, 
the vortex line requires them to have zeros or poles whose multiplicities are determined by the vortex number.  
Such singularities can be found as certain solutions to the vortex equations satisfied by the matter fields 
(See e.g. \cite{Nielsen:1973cs,MR614447,MR573986,MR1085139,MR1124279,MR1250254}). 
Alternatively, the vortex line can be defined by adding certain 1d quantum mechanics along the line coupled to 3d bulk fields by gauging flavor symmetries 
in such a way that one finds the singularities after integrating out the 1d fields. 
The full classification of the singularities in the path integral that is sufficient to describe the vortex lines is out of reach. 
Nevertheless, one can reach the descriptions of the singularities by making use of the dualities. 
In pure CS theory, they are equivalent to the Wilson lines \cite{Moore:1989yh}. 

In 3d $\mathcal{N}=4$ Abelian gauge theory, the half-BPS vortex line is either a flavor vortex line 
or it becomes isomorphic to a flavor vortex line upon screening by dynamical vortices (see e.g.~\cite{Dimofte:2019zzj}).
The half-BPS flavor vortex line of charge $\mathsf{q}$ can be obtained by specifying a singular profile of the connection $A$ 
for the associated flavor group to be a fixed flat connection in such a way that it gives rise to the vortex charge $\mathsf{q}$ according to (\ref{vortexN}). 
Besides, the BPS equations state that the chiral scalar fields $X$ and $Y$ in the hypermultiplet are covariantly holomorphic in the plane \cite{Hanany:2003hp,Bullimore:2016hdc}. 
There exists a solution to the BPS equations unique up to gauge equivalence 
with the property that the zeros or poles of the charged scalar fields have order $|\mathsf{q}|$ near the line (see e.g.~\cite{MR614447}). 
Therefore one can obtain the flavor vortex line by exciting hypermultiplet scalar fields so that they have codimension-two singularities near the line. 
It breaks down the $SU(2)_H$ down to $U(1)_H$. 
Alternatively, the flavor vortex line can be described by coupling 1d $\mathcal{N}=4$ SQM 
which is obtained by dimensionally reducing 2d $\mathcal{N}=(2,2)$ gauge theories \cite{Assel:2015oxa}. 
It preserves the $SU(2)_C$ R-symmetry and breaks the $SU(2)_H$ R-symmetry down to $U(1)_H$. 
Besides, there is an attempt to classify the singularities associated to the BPS vortex line in terms of algebraic data \cite{Dimofte:2019zzj}.  
The half-BPS vortex lines can preserve the following four supercharges: 
\begin{align}
\begin{array}{c|cccccccc}
&Q_{-}^{1\dot{1}}&Q_{-}^{1\dot{2}}&Q_{-}^{2\dot{1}}&Q_{-}^{2\dot{2}}&Q_{+}^{1\dot{1}}&Q_{+}^{1\dot{2}}&Q_{+}^{2\dot{1}}&Q_{+}^{2\dot{2}} \\ \hline 
\textrm{vortex}&\circ&\circ&&&&&\circ&\circ \\
\end{array}. 
\end{align}

Analogously, the flavor BPS vortex lines in the Abelian ABJM theory can be constructed 
by allowing the scalar fields to acquire codimension-two singularities along the line, as discussed in \cite{Drukker:2008jm}. 
If either the hypermultiplet scalar fields or twisted hypermultiplet scalar fields are excited, the vortex line can preserve $1/3$ of supersymmetry. 
If both are excited, it can preserve $1/6$ of supersymmetry. 
The amount of supersymmetry can be enhanced for $k$ $=$ $1$ and $2$ as the theory has enhanced $\mathcal{N}=8$ supersymmetry. 

\subsection{Vortex-Wilson line}
There is the $SL(2,\mathbb{Z})$ action on the space of 3d conformal field theories with $U(1)$ flavor symmetry \cite{Witten:2003ya}. 
For 3d $\mathcal{N}=4$ Abelian gauge theories it manifests mirror symmetry \cite{Kapustin:1999ha}. 
It can be shown \cite{Kapustin:2012iw} that
the $SL(2,\mathbb{Z})$ action can lead to the Abelian duality between the gauge Wilson line and the flavor vortex line. 
Here we briefly review the discussion. 
Consider the 3d conformal field theory that has an action $S[\Phi]$ with fields $\Phi$ and an Abelian global symmetry current $J$. 
We couple $J$ to a background gauge field $A$. 
Then the partition function is given by the functional of $A$
\begin{align}
Z_J[A]&=\int [D\Phi]e^{iS[\Phi]+i\int d^3x J^{\mu}A_{\mu}}. 
\end{align}
While the generator $T$ simply adds a background CS term of level $+1$ to the theory 
\begin{align}
T:\qquad S[\Phi]&\rightarrow S[\Phi,A]+\frac{i}{4\pi}\int A \wedge dA, 
\end{align}
the generator $S$ makes the background gauge field $A$ dynamical and then couple it to a background gauge field $B$ for the corresponding topological symmetry
\begin{align}
S:\qquad S[\Phi]+\frac{i}{2\pi}\int A\wedge dB.
\end{align}
The Wilson line of charge $\mathsf{p}=n$ is defined by inserting
\begin{align}
\label{Wline}
e^{in \oint A}
\end{align}
in the path integral. 
Introducing a connection $A_{\omega}$ $=$ $nd\theta$, 
where $(r,\theta,z)$ are local coordinates and the line lies along the $z$-axis, 
the operation inserting the Wilson line (\ref{Wline}) can be defined as the BF coupling to a new background gauge field $A_{\omega}$
\begin{align}
\label{Wop}
W:\qquad 
S[\Phi]&\rightarrow S[\Phi]+\frac{i}{2\pi}\int A_{\omega}\wedge dA. 
\end{align}
Then under the operation $S^{-1}WS$, the action $S[\Phi]$ transforms as
\begin{align}
S[\Phi]&\rightarrow 
S[\Phi]+\frac{i}{2\pi}\int A_1\wedge dA_2+\frac{i}{2\pi}\int A_{\omega}\wedge dA_2-\frac{i}{2\pi}\int A_2\wedge dA, 
\end{align}
where we have used $S^{-1}=-S$. 
Here $A_1$ and $A_2$ are dynamical gauge fields  and $A$ is a background gauge field. 
Performing the path integral over $A_2$ gives a delta function setting $A_1$ $=$ $A-A_{\omega}$. 
Thus we find the relation
\begin{align}
\label{SWSrelation}
S^{-1}WS&=D_{\omega}, 
\end{align}
where we have defined the operation
\begin{align}
D _{\omega}:\qquad Z_J[A]
&\rightarrow Z_J[A-A_{\omega}]. 
\end{align}
Note that $A_{\omega}$ is a fixed $U(1)$ flat connection so that it generates the flavor vortex of charge $\mathsf{q}$ $=$ $n=\frac{1}{2\pi}\int F_{\omega}$. 
It appears through the coupling $J^{\mu}{A_{\omega}}_{\mu}$ in the path integral, 
which requires all fields charged under the current $J$ to acquire a fixed monodromy around the line. 
Notice that the expectation value of the dynamical Wilson line can be obtained by performing the path integral of the $S$ transformed Wilson line $S\cdot W$ as the operation $S$ gauge the global symmetry. 
On the other hand, according to the relation (\ref{SWSrelation}) and the relation $S^2=C$, where $C$ is the charge conjugation, 
it is shown to be equivalent to the operation $D_{-\omega} \cdot S$ up to the phase factor. 
The latter can be identified with the expectation value of the flavor vortex line. 
Therefore one can see that the $S$ transformed Wilson line of charge $n$ for the dynamical gauge field is equivalent to the flavor vortex line of charge $n$.

More generally, the $SL(2,\mathbb{Z})$ action extends the Abelian dualities of the line operators. 
While vortices can be created or annihilated at the location of a monopole operator, 
upon more general operation including both $S$ and $T$, the theory obtains the CS coupling of a dynamical gauge field, 
for which the monopole operator can carry electric charge. 
This fact motivates us to introduce more general mixed vortex-Wilson line operator $L_{\mathsf{p};\mathsf{q}}$. 
It can be defined by a path integral for the vortex line of vortex charge $\mathsf{q}$ with an insertion of a Wilson line of electric charge $\mathsf{p}$. 
While the Wilson line can be defined by introducing the trace of the holonomy matrix in the path integral as discussed in section \ref{sec_Wline}, 
the vortex line is defined as a disorder operator that constrains the gauge and matter fields in the path integral to have singularities along the line. 
In general, the singularities break down the associated gauge group or global symmetry group. 
Again, the classification of such singularities is beyond the scope of the present paper. 
In the following, we consider the vortex-Wilson line operator $L_{\mathsf{p};\mathsf{q}}$ 
involving the flavor vortex line which is compatible with the coexisting gauge Wilson line. 

For 3d $\mathcal{N}=4$ supersymmetric gauge theories, 
Type IIB string theory can realize the mixed BPS vortex-Wilson line and predict dualities of more general line operators, as we will discuss. 
The vortex-Wilson lines can generically preserve two supercharges
\begin{align}
\begin{array}{c|cccccccc}
&Q_{-}^{1\dot{1}}&Q_{-}^{1\dot{2}}&Q_{-}^{2\dot{1}}&Q_{-}^{2\dot{2}}&Q_{+}^{1\dot{1}}&Q_{+}^{1\dot{2}}&Q_{+}^{2\dot{1}}&Q_{+}^{2\dot{2}} \\ \hline 
\textrm{vortex-Wilson}&\circ&&&&&&&\circ \\
\end{array}. 
\end{align}
We note that a potentially relevant description for the mixed BPS vortex-Wilson line was proposed in \cite{Griguolo:2021rke} 
by deforming the 1d $\mathcal{N}=4$ SQM \cite{Assel:2015oxa}. 
At this stage we are not sure whether it is equivalent to our vortex-Wilson line operators $L_{\mathsf{p};\mathsf{q}}$. 
It would be an interesting future work to understand the relation to the description of the vortex-Wilsons in terms of the SQM.
On the other hand, since such vortex-Wilson lines in the Abelian gauge theories can be directly defined 
by fixing the flat connection for the flavor group so as to give the vortex number $\mathsf{q}$ 
and by introducing the Wilson line operator of charge $\mathsf{p}$ in the path integral, 
the line defect index for the mixed vortex-Wilson line can be computed by combining the prescriptions for the vortex and Wilson lines. 
We test the conjectured dualities involving the vortex-Wilson lines by evaluating the indices in the following.  

\subsection{Brane setup}
\label{sec_branesetup}
\begin{figure}[t]
\centering
\scalebox{0.6}{
\begin{tikzpicture}
\draw[color=red] (-3,-2)--(-3,2);
\draw [color=red] (3,-2)--(3,2);
\draw[thick] (-3,0)--(3,0);
\draw[thick] (-6,1.5)--(-3,1.5);
\node[label=below:{$N$ D3}] at (0,0) {};
\node[label=below:{$\ell$ D3}] at (-4.5,1.5) {};
\end{tikzpicture}}\hspace{2cm}
\scalebox{0.6}{
\begin{tikzpicture}
\draw[color=red] (-3,-2)--(-3,2);
\draw[color=red] (3,-2)--(3,2);
\draw[thick] (-3,0)--(3,0);
\draw[color=blue] (-2.5,1)--(-0.5,2);
\node at (0,1.5) {$\cdots$};
\draw[color=blue] (0.5,1)--(2.5,2);
\draw[decorate, decoration={brace}] (-0.6,2.1)--(2.6,2.1);
\node[label=above:{$l$}] at (1,2) {};
\node[label=below:{$N$ D3}] at (0,0) {};
\end{tikzpicture}}
\scalebox{0.6}{
\begin{tikzpicture}
\draw[arrows=->] (5.5,-2)--(6,-2);
\draw[arrows=->] (5.5,-2)--(5.5,-1.5);
\draw[arrows=->] (5.5,-2)--(5.5-1/1.25/2,-2-0.5/1.25/2);
\node[label=right:{$x^3$}] at (6,-2) {};
\node[label=above:{$x^6$}] at (5.5,-1.5) {};
\node[label=left:{$x^9$}] at (5.5-1/1.25/2,-2-0.5/1.25/2) {};
\end{tikzpicture}}\hspace{1cm}
\scalebox{0.6}{
\begin{tikzpicture}
\path 
(0,3) node[minimum size=32, fill=cyan!10, draw](CG1) {$l$}
(0,0) node[circle, minimum size=32, fill=yellow!20, draw](CG2) {$U(N)$};
\draw (CG1) -- (CG2);
\end{tikzpicture}
}
\caption{Two equivalent brane configurations for the $U(N)$ gauge theory with $l$ flavors. A red line is an NS5-brane and a blue line is a D5-brane. We will also use the same convention in the later figures. The two configurations are related to each other by the Hanany-Witten transitions. The rightmost diagram shows the quiver diagram of the the $U(N)$ gauge theory with $l$ flavors.}
\label{fig:SQCDbrane}
\end{figure}
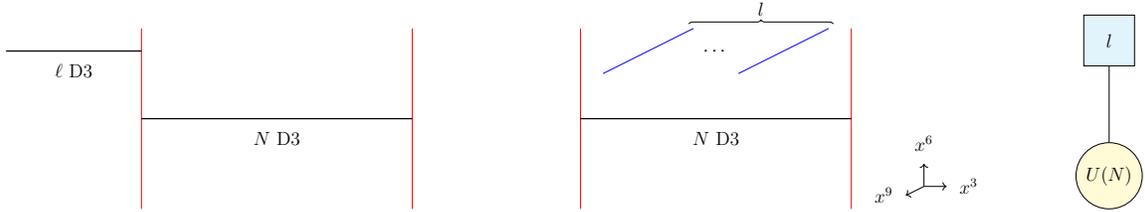

BPS line defect operators in 3d $\mathcal{N}=4$ supersymmetric gauge theories can be constructed by branes in Type IIB string theory \cite{Assel:2015oxa,Hanany:2003hp}. Let us first review the brane construction before introducing line defect operators \cite{Hanany:1996ie, deBoer:1996mp,deBoer:1996ck,Aharony:1997ju, Bergman:1999na, Kitao:1998mf}. The configuration of branes in the ten-dimensional spacetime of Type IIB string theory is summarized in Table \ref{tb:brane} where the complexified string coupling $\tau = \chi + ig_s^{-1}$ is chosen to be $\tau = i$. $\chi$ is the R-R scalar and $g_s$ is the string coupling.
\begin{table}[t]
\centering
\begin{tabular}{c|ccc|c|ccc|ccc}
&0&1&2&3&4&5&6&7&8&9\\
\hline
D3&$\times$&$\times$&$\times$&$\times$&&&&&&\\
\hline
NS5&$\times$&$\times$&$\times$&&$\times$&$\times$&$\times$&&&\\
\hline
D5&$\times$&$\times$&$\times$&&&&&$\times$&$\times$&$\times$\\
\hline
$(1, k)$ 5&$\times$&$\times$&$\times$&&\multicolumn{6}{c}{$(47)_k, (58)_k, (69)_k$}\\
\hline
\end{tabular}
\caption{The directions where branes are extended in the ten-dimensional spacetime of Type IIB string theory. The complexified string coupling is chosen to be $\tau = i$. $(ab)_k$ means a line with the slope $k$ on the $ab$-plane. }
\label{tb:brane}
\end{table}
Open strings ending on $N$ D3-branes suspended between two NS5-branes give an $\mathcal{N}=4$ $U(N)$ vector multiplet. On the other hand, when a stack of $N_1$ D3-branes ends on an NS5-brane and the other stack of $N_2$ D3-branes ends on the NS5-brane from the other side in the $x^3$-direction, open strings ending on the $N_1$ D3-branes and the $N_2$ D3-branes give rise to an $\mathcal{N}=4$ hypermultiplet in the bifundamental representation of $U(N_1) \times U(N_2)$. To illustrate an example, the brane configuration for the $U(N)$ gauge theory with $l$ flavors is depicted in the leftmost figure of Figure \ref{fig:SQCDbrane}. 
One can also terminate each of the $l$ semi-infinite D3-branes on a D5-brane without changing the 3d theory and then move the $l$ D5-branes in the $x^3$-direction until all the D5-branes are located in between the two NS5-branes. All the D3-branes between an NS5-brane and D5-branes are annihilated by the Hanany-Witten effect and one arrives at the configuration in the middle figure of Figure \ref{fig:SQCDbrane}. The field theory content can be conveniently summarized as a quiver diagram as shown in the rightmost diagram in Figure \ref{fig:SQCDbrane}. The notation of quiver diagrams is as follows. A circle filled with yellow color represents an $\mathcal{N}=4$ vector multiplet and the gauge group is written in the circle. In later examples, we will encounter cases where there are $\mathcal{N}=2$ vector multiplets and $\mathcal{N}=4$ twisted vector multiplets. In those cases, an $\mathcal{N}=2$ vector multiplet is denoted by a circle filled with green color while an $\mathcal{N}=4$ twisted vector multiplet is represented by a circle filled with orange color. A line between a box and a circle stands for $\mathcal{N}=4$ hypermultiplets in the fundamental representation of a gauge group written in the circle and the number of the hypermultiplets is given by the number written in the box. Similarly, we will denote  an $\mathcal{N}=4$ hypermultiplet or an $\mathcal{N}=4$ twisted hypermultiplet in the bifundamental representation of $G^{(1)} \times G^{(2)}$ by a line segment which connects a circle with a gauge group $G^{(1)}$ to another circle with a gauge group $G^{(2)}$.

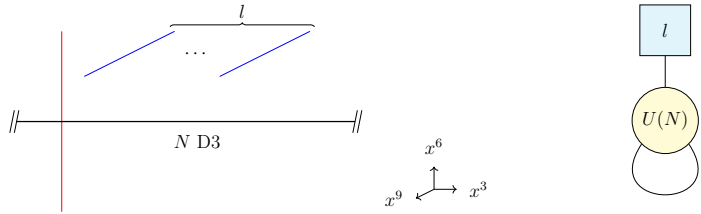
\begin{figure}[t]
\centering
\scalebox{0.6}{
\begin{tikzpicture}
\draw[color=red] (-3,-2)--(-3,2);
\draw[thick] (-4,0)--(3.5,0);
\draw[color=blue] (-2.5,1)--(-0.5,2);
\node at (0,1.5) {$\cdots$};
\draw[color=blue] (0.5,1)--(2.5,2);
\draw[decorate, decoration={brace}] (-0.6,2.1)--(2.6,2.1);
\node[label=above:{$l$}] at (1,2) {};
\node[label=below:{$N$ D3}] at (0,0) {};
\draw[very thin] (3.45,-0.25)--(3.55,0.25);
\draw[very thin] (3.55,-0.25)--(3.65,0.25);
\draw[very thin] (-4.05,-0.25)--(-3.95,0.25);
\draw[very thin] (-4.15,-0.25)--(-4.05,0.25);
\end{tikzpicture}}
\scalebox{0.6}{
\begin{tikzpicture}
\draw[arrows=->] (5.5,-2)--(6,-2);
\draw[arrows=->] (5.5,-2)--(5.5,-1.5);
\draw[arrows=->] (5.5,-2)--(5.5-1/1.25/2,-2-0.5/1.25/2);
\node[label=right:{$x^3$}] at (6,-2) {};
\node[label=above:{$x^6$}] at (5.5,-1.5) {};
\node[label=left:{$x^9$}] at (5.5-1/1.25/2,-2-0.5/1.25/2) {};
\end{tikzpicture}}
\hspace{1cm}
\scalebox{0.6}{
\begin{tikzpicture}
\path 
(0,2) node[minimum size=32, fill=cyan!10, draw](CG1) {$l$}
(0,0) node[circle, minimum size=32, fill=yellow!20, draw](CG2) {$U(N)$};
\draw (CG1) -- (CG2);
\draw (CG2.south west) .. controls ++(-1, -1.5) and ++(1, -1.5) .. node[below]{} (CG2.south east);
\end{tikzpicture}
}
\caption{The brane configuration and the quiver diagram for the $U(N)$ ADHM theory with $l$ flavors. The space in the $x^3$-direction is compactified on a circle and the D3-branes on the left are connected to the D3-branes on the right. }
\label{fig:ADHMbrane}
\end{figure}
A hypermultiplet in the adjoint representation can be introduced by compactifying the space in the $x^3$-direction on a circle. For example, the brane configuration for the $U(N)$ ADHM theory with $l$ flavors is given in the left figure in Figure \ref{fig:ADHMbrane}. The diagram on the right in Figure \ref{fig:ADHMbrane} is the quiver diagram of the $U(N)$ ADHM theory with $l$ flavors. A line segment whose two ends end on a circle represents a hypermultiplet in the adjoint representation of a gauge group associated with the circle.

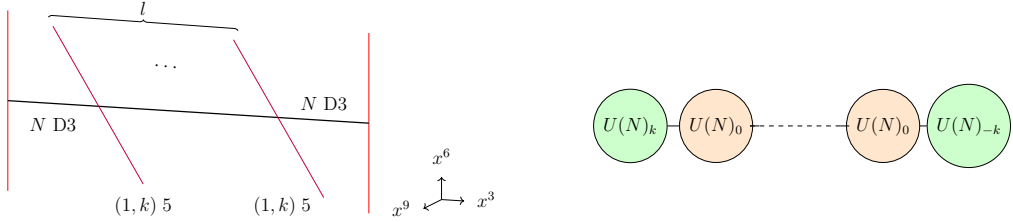
\begin{figure}[t]
\centering
\scalebox{0.6}{
\begin{tikzpicture}
\draw[color=red] (-4,-2)--(-4,2);
\draw[color=red] (4,-2-0.5)--(4,2-0.5);
\draw[thick] (-4,0)--(4,-0.5);
\draw[color=purple] (-3,1.75-0.1)--(-1,-1.75-0.1);
\node[label=below:{$(1, k)\;5$}] at (-1,-1.75-0.1) {};
\draw[color=purple] (1,1.75-0.4)--(3,-1.75-0.4);
\node[label=left:{$(1, k)\;5$}] at (3,-1.75-0.4-0.2) {};
\node at (-0.7,0.75+0.5/8*0.2) {$\cdot$};
\node at (-0.5,0.75) {$\cdot$};
\node at (-0.3,0.75-0.5/8*0.2) {$\cdot$};
\draw[decorate, decoration={brace}] (-3.1,1.75+0.1)--(1.1,1.75-0.2);
\node[label=above:{$l$}] at (-1,1.5+0.1) {};
\node[label=below:{$N$ D3}] at (-3,-0.1) {};
\node[label=above:{$N$ D3}] at (3,-0.5) {};
\end{tikzpicture}}
\scalebox{0.6}{
\begin{tikzpicture}
\draw[arrows=->] (5.5,-2)--(6,-2-0.5*0.5/8);
\draw[arrows=->] (5.5,-2)--(5.5,-1.5);
\draw[arrows=->] (5.5,-2)--(5.5-1/1.25/2,-2-0.5/1.25/2);
\node[label=right:{$x^3$}] at (6,-2) {};
\node[label=above:{$x^6$}] at (5.5,-1.5) {};
\node[label=left:{$x^9$}] at (5.5-1/1.25/2,-2-0.5/1.25/2) {};
\end{tikzpicture}}\hspace{1cm}
\scalebox{0.6}{
\begin{tikzpicture}
\path 
(0,0) node[circle, minimum size=32, fill=green!20, draw](CG1) {$U(N)_k$}
(1.9,0) node[circle, minimum size=32, fill=orange!20, draw](CG2) {$U(N)_0$}
(4.6+1,0) node[circle, minimum size=32, fill=orange!20, draw](CG3) {$U(N)_0$}
(6.5+1,0) node[circle, minimum size=32, fill=green!20, draw](CG4) {$U(N)_{-k}$};
\draw (CG1) -- (CG2);
\draw (CG3) -- (CG4);
\draw (1.9+0.75,0)--(1.9+0.75+0.2,0);
\draw (4.6+1-0.75,0)--(4.6+1-0.75-0.2,0);
\draw[dashed,-]  (1.9+0.75,0) to (4.6+1-0.75-0.2,0);
\node at (0,-2) {};
\end{tikzpicture}
}
\caption{The brane configuration and the quiver diagram for the $U(N)_k \times U(N)^{l-1}_0 \times U(N)_{-k}$ linear quiver CS theory. The purple lines are $(1, k)$ 5-branes and the configuration contains $l$ $(1, k)$ 5-branes and two NS5-branes. }
\label{fig:CSbrane1}
\end{figure}

The inclusion of $(1, k)$ 5-branes in the brane configuration can introduce a CS coupling in the 3d theory realized by the brane setup. A $(p, q)$ 5-brane with coprime $p, q$ represents a 5-brane on which a $(p, q)$-string can end. Here, a $(p, q)$-string is a string which is electrically coupled to $pC_2 + qB_2$ where $C_2$ is the R-R 2-form field and $B_2$ is the NS-NS 2-form field. Open strings ending on $N$ D3-branes suspended between an NS5-brane and a $(1, k)$ 5-brane yield an $\mathcal{N}=2$ $U(N)$ vector multiplet with a level $\pm k$ CS term. The sign of the CS level is related to the order of the positions of the NS5-brane and the $(1, k)$ 5-brane in the $x^3$-direction. On the other hand, open strings ending on $N$ D3-branes suspended between two $(1, k)$ 5-branes give an $\mathcal{N}=4$ $U(N)$ twisted vector multiplet with no CS term. When a stack of $N_1$ D3-branes ends on an $(1, k)$ 5-brane and the other stack of $N_2$ D3-brane ends on the $(1, k)$ 5-brane from the other side in the $x^3$-direction, open strings ending on the $N_1$ D3-branes and the $N_2$ D3-branes give rise to an $\mathcal{N}=4$ twisted hypermultiplet in the bifundamental representation of $U(N_1) \times U(N_2)$ \cite{Imamura:2008nn,Imamura:2008dt}.
The brane configuration for a linear quiver CS theory with the gauge group  $U(N)_k \times U(N)^{l-1}_0 \times U(N)_{-k}$ and $l$ twisted hypermultiplets in the bifundamental representation of $U(N) \times U(N)$ is illustrated in the left figure of Figure \ref{fig:CSbrane1}. The subscript of each gauge group represents the CS level associated with the gauge group. 

\begin{figure}[t]
\centering
\scalebox{0.6}{
\begin{tikzpicture}
\draw[color=red] (-4,-2)--(-4,2);
\draw[thick] (-4-1,0+0.5/8)--(4,-0.5);
\draw[color=purple] (-3,1.75-0.1)--(-1,-1.75-0.1);
\node[label=below:{$(1, k)\;5$}] at (-1,-1.75-0.1) {};
\draw[color=purple] (1,1.75-0.4)--(3,-1.75-0.4);
\node[label=left:{$(1, k)\;5$}] at (3,-1.75-0.4-0.2) {};
\node at (-0.7,0.75+0.5/8*0.2) {$\cdot$};
\node at (-0.5,0.75) {$\cdot$};
\node at (-0.3,0.75-0.5/8*0.2) {$\cdot$};
\draw[decorate, decoration={brace}] (-3.1,1.75+0.1)--(1.1,1.75-0.2);
\node[label=above:{$l$}] at (-1,1.5+0.1) {};
\node[label=below:{$N$ D3}] at (-3,-0.1) {};
\node[label=above:{$N$ D3}] at (3,-0.5) {};
\draw[very thin] (3.95,-0.75)--(4.05,-0.25);
\draw[very thin] (4.05,-0.75-0.5/8*0.1)--(4.15,-0.25-0.5/8*0.1);
\draw[very thin] (-4.05-1,-0.25+0.5/8)--(-3.95-1,0.25+0.5/8);
\draw[very thin] (-4.15-1,-0.25+0.5/8*0.1+0.5/8)--(-4.05-1,0.25+0.5/8*0.1+0.5/8);
\end{tikzpicture}}
\scalebox{0.6}{
\begin{tikzpicture}
\draw[arrows=->] (5.5,-2)--(6,-2-0.5*0.5/8);
\draw[arrows=->] (5.5,-2)--(5.5,-1.5);
\draw[arrows=->] (5.5,-2)--(5.5-1/1.25/2,-2-0.5/1.25/2);
\node[label=right:{$x^3$}] at (6,-2) {};
\node[label=above:{$x^6$}] at (5.5,-1.5) {};
\node[label=left:{$x^9$}] at (5.5-1/1.25/2,-2-0.5/1.25/2) {};
\end{tikzpicture}}\hspace{1cm}
\scalebox{0.6}{
\begin{tikzpicture}
\path 
(1.9,3) node[circle, minimum size=32, fill=green!20, draw](CG1) {$U(N)_k$}
(1.9,0) node[circle, minimum size=32, fill=orange!20, draw](CG2) {$U(N)_0$}
(4.6+1,0) node[circle, minimum size=32, fill=orange!20, draw](CG3) {$U(N)_0$}
(4.6+1,3) node[circle, minimum size=32, fill=green!20, draw](CG4) {$U(N)_{-k}$};
\draw (CG1) -- (CG2);
\draw (CG3) -- (CG4);
\draw (CG1) -- (CG4);
\draw (1.9+0.8,0)--(1.9+0.8+0.2,0);
\draw (4.6+1-0.8,0)--(4.6+1-0.8-0.2,0);
\draw[dashed,-]  (1.9+0.8,0) to (4.6+1-0.8-0.2,0);
\end{tikzpicture}
}
\caption{The brane configuration and the quiver diagram for the 
$U(N)_k \times U(N)^{l-1}_0 \times U(N)_{-k}$ circular quiver CS theory. 
The space in the $x^3$-direction is compactified on a circle and the configuration contains $l$ $(1, k)$ 5-branes and an NS5-brane. }
\label{fig:CSbrane2}
\end{figure}
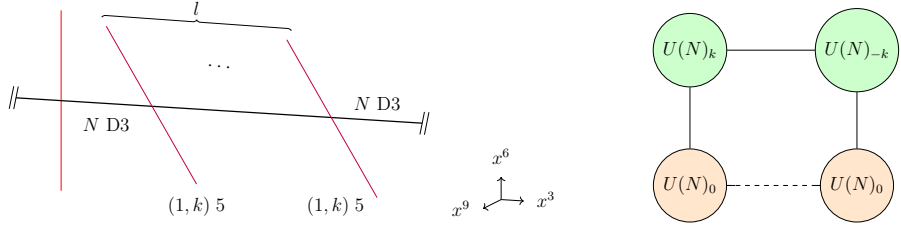

When we consider compactifying the space in the $x^3$-direction on a circle, the brane configuration can yield a circular quiver CS theory. For example, the brane configuration in Figure \ref{fig:CSbrane2} yields a circular quiver CS theory with the gauge group $U(N)_k \times U(N)_0^{l-1} \times U(N)_{-k}$ and $l$ twisted hypermultiplets in the bifundamental representation of $U(N) \times U(N)$ and a hypermultiplet in the bifundamental representation of $U(N) \times U(N)$ \cite{Imamura:2008nn,Imamura:2008dt}.
When $l=1$, the supersymmetry is enhanced to $\mathcal{N}=6$, and the theory becomes the $U(N)_k \times U(N)_{-k}$ ABJM theory \cite{Aharony:2008ug}.

The $SL(2, \mathbb{Z})$ duality\footnote{
The precise form of the duality group is given by the $\mathsf{pin}^+$ version of the double cover of $GL(2, \mathbb{Z})$ \cite{Tachikawa:2018njr}.
}
of Type IIB string theory generates several dual brane configurations. Under the $SL(2, \mathbb{Z})$ transformation, the complexified string coupling $\tau$ and the charge of a $(p, q)$ 5-brane transform as 
\begin{align}
\tau \quad &\to \quad \frac{a\tau + b}{c\tau + d},\\
(p\quad q) \quad &\to \quad (p\quad q)\left(\begin{array}{cc}
d & -b\\
-c & a
\end{array}\right),
\end{align}
with $a, b, c, d \in \mathbb{Z}$ and $ad - bc = 1$. In particular, the S-duality transformation is given by
\be
S\text{: }
\left(\begin{array}{cc}
d & -b\\
-c & a
\end{array}\right)
=
\left(\begin{array}{cc}
0 & -1\\
1 & 0
\end{array}\right)
,
\ee
which exchanges a D5-brane with an NS5-brane. We will focus on Abelian gauge theories from here in this subsection since they are the main examples in the later sections.  

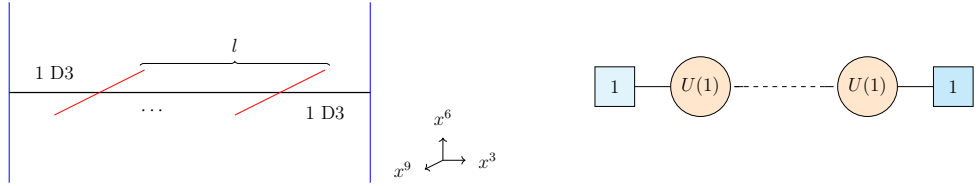
\begin{figure}[t]
\centering
\scalebox{0.6}{
\begin{tikzpicture}
\draw[thick] (-4,0)--(4,0);
\draw[color=blue] (-4,-2)--(-4,2);
\draw[color=red] (-3,-0.5)--(-1,0.5);
\draw[color=red] (1,-0.5)--(3,0.5);
\draw[color=blue] (4,-2)--(4,2);
\node[label=above:{$1$ D3}] at (-3,0) {};
\node[label=below:{$1$ D3}] at (3,0) {};
\draw[decorate, decoration={brace}] (-1-0.1,0.5+0.1)--(3+0.1,0.5+0.1);
\node[label=above:{$l$}] at (1,0.5+0.1) {};
\node at (-0.8,-0.4) {$\cdots$};
\end{tikzpicture}}
\scalebox{0.6}{
\begin{tikzpicture}
\draw[arrows=->] (5.5,-2)--(6,-2);
\draw[arrows=->] (5.5,-2)--(5.5,-1.5);
\draw[arrows=->] (5.5,-2)--(5.5-1/1.25/2,-2-0.5/1.25/2);
\node[label=right:{$x^3$}] at (6,-2) {};
\node[label=above:{$x^6$}] at (5.5,-1.5) {};
\node[label=left:{$x^9$}] at (5.5-1/1.25/2,-2-0.5/1.25/2) {};
\end{tikzpicture}}\hspace{1cm}
\scalebox{0.6}{
\begin{tikzpicture}
\path 
(0,0) node[minimum size=25, fill=cyan!10, draw](CG1) {$1$}
(1.9,0) node[circle, minimum size=32, fill=orange!20, draw](CG2) {$U(1)$}
(4.6+1,0) node[circle, minimum size=32, fill=orange!20, draw](CG3) {$U(1)$}
(6.5+1,0) node[minimum size=25, fill=cyan!20, draw](CG4) {$1$};
\draw (CG1) -- (CG2);
\draw (CG3) -- (CG4);
\draw (1.9+0.75,0)--(1.9+0.75+0.2,0);
\draw (4.6+1-0.75,0)--(4.6+1-0.75-0.2,0);
\draw[dashed,-]  (1.9+0.75,0) to (4.6+1-0.75-0.2,0);
\node at (0,-2) {};
\end{tikzpicture}
}
\caption{The brane configuration and the quiver diagram for the linear quiver theory, $[1] - \widetilde{U(1)^{\otimes (l-1)}} - [1]$, which is S-dual to the $U(1)$ gauge theory with $l$ flavors.}
\label{fig:Sdualbrane1}
\end{figure}

When $N=1$, the S-dual of the brane configuration in the middle figure of Figure \ref{fig:SQCDbrane} becomes the one in Figure \ref{fig:Sdualbrane1}.\footnote{
When an $SL(2,\mathbb{Z})$ transformation results in $(p,q)$ 5-branes with $p=0$, $q<0$ or $p<0$, we identify each of them as $(-p,-q)$ 5-brane, which represent the same object up to the orientation.
}
Before taking the S-duality transformation, we tune the positions of the flavor D5-branes such that the D5-branes intersect with D3-branes. The 3d theory realized on the brane configuration for $l \geq 2$ is a linear quiver theory given by $[1] - U(1) - \cdots - U(1) - [1]$ with $(l-1)$ $U(1)$ gauge nodes. 
Note that NS5-branes are extended in the $x^7, x^8, x^9$-directions and D5-branes are extended in the $x^4, x^5, x^6$-directions in addition to the 3d spacetime directions after the S-duality transformation. Hence, the linear quiver theory consists of $(l-1)$ $\mathcal{N}=4$ $U(1)$ twisted vector multiplets, $(l-2)$ $\mathcal{N}=4$ twisted hypermultiplets in the bifundamental representation of $U(1) \times U(1)$ and two $\mathcal{N}=4$ twisted hypermultiplet in the fundamental representation of $U(1)$. For $l=1$, the S-dual theory is a free $\mathcal{N}=4$ twisted hypermultiplet. We will refer to the linear quiver theory as $[1] - \widetilde{U(1)^{\otimes (l-1)}} - [1]$. 

\begin{figure}[t]
\centering
\scalebox{0.6}{
\begin{tikzpicture}
\draw[thick] (-5,0)--(4,0);
\draw[color=blue] (-4,-2)--(-4,2);
\draw[color=red] (-3,-0.5)--(-1,0.5);
\draw[color=red] (1,-0.5)--(3,0.5);
\node[label=above:{$1$ D3}] at (-3,0) {};
\node[label=below:{$1$ D3}] at (3,0) {};
\draw[decorate, decoration={brace}] (-1-0.1,0.5+0.1)--(3+0.1,0.5+0.1);
\node[label=above:{$l$}] at (1,0.5+0.1) {};
\node at (-0.8,-0.4) {$\cdots$};
\draw[very thin] (3.95,-0.25)--(4.05,0.25);
\draw[very thin] (4.05,-0.25)--(4.15,0.25);
\draw[very thin] (-5.05,-0.25)--(-4.95,0.25);
\draw[very thin] (-5.15,-0.25)--(-5.05,0.25);
\end{tikzpicture}}
\scalebox{0.6}{
\begin{tikzpicture}
\draw[arrows=->] (5.5,-2)--(6,-2);
\draw[arrows=->] (5.5,-2)--(5.5,-1.5);
\draw[arrows=->] (5.5,-2)--(5.5-1/1.25/2,-2-0.5/1.25/2);
\node[label=right:{$x^3$}] at (6,-2) {};
\node[label=above:{$x^6$}] at (5.5,-1.5) {};
\node[label=left:{$x^9$}] at (5.5-1/1.25/2,-2-0.5/1.25/2) {};
\end{tikzpicture}}\hspace{1cm}
\scalebox{0.6}{
\begin{tikzpicture}
\path 
(0,3) node[minimum size=25, fill=cyan!10, draw](CG0) {$1$}
(1.9,3) node[circle, minimum size=32, fill=orange!20, draw](CG1) {$U(1)$}
(1.9,0) node[circle, minimum size=32, fill=orange!20, draw](CG2) {$U(1)$}
(4.6+1,0) node[circle, minimum size=32, fill=orange!20, draw](CG3) {$U(1)$}
(4.6+1,3) node[circle, minimum size=32, fill=orange!20, draw](CG4) {$U(1)$};
\draw (CG1) -- (CG2);
\draw (CG3) -- (CG4);
\draw (CG1) -- (CG4);
\draw (CG0) -- (CG1);
\draw (1.9+0.8,0)--(1.9+0.8+0.2,0);
\draw (4.6+1-0.8,0)--(4.6+1-0.8-0.2,0);
\draw[dashed,-]  (1.9+0.8,0) to (4.6+1-0.8-0.2,0);
\end{tikzpicture}
}
\caption{The brane configuration and the quiver diagram for the circular quiver theory, $\widetilde{U(1)^{\otimes l} \textrm{mADHM}-[1]}$, which is S-dual to the $U(1)$ ADHM theory with $l$ flavors. The space in the $x^3$-direction is compactified on $S^1$.}
\label{fig:Sdualbrane2}
\end{figure}
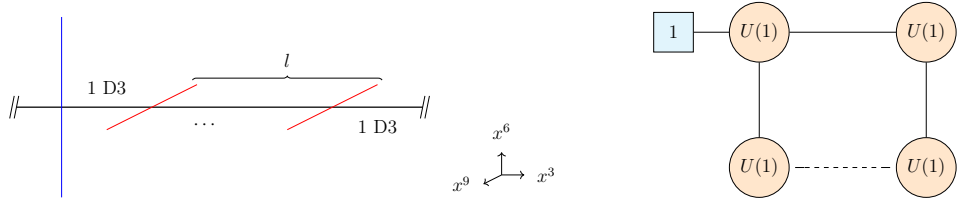

Similarly, the S-dual of the brane configuration in Figure \ref{fig:ADHMbrane} for $N=1$ is given in Figure \ref{fig:Sdualbrane2}. The 3d theory realized on the brane configuration for $l \geq 1$ is a circular quiver theory with $l$ $U(1)$ gauge nodes and a flavor attached to one of the $U(1)$ gauge nodes. The circular quiver theory consists of $l$ $\mathcal{N}=4$ $U(1)$ twisted vector multiplets, $l$ $\mathcal{N}=4$ twisted hypermultiplets in the bifundamental representation of $U(1) \times U(1)$ and an $\mathcal{N}=4$ twisted hypermultiplet in the fundamental representation of $U(1)$. We will refer to the circular quiver theory as $\widetilde{U(1)^{\otimes l} \textrm{mADHM}-[1]}$. 

\begin{figure}[t]
\centering
\scalebox{0.6}{
\begin{tikzpicture}
\draw[thick] (-4,0)--(4,0);
\draw[color=red] (-4,-2)--(-4,2);
\draw[color=purple] (-3,-0.5)--(-1,0.5);
\draw[color=purple] (1,-0.5)--(3,0.5);
\draw[color=red] (4,-2)--(4,2);
\node[label=below:{$(1, 1)\;5$}] at (-3+0.25,-0.5) {};
\node[label=below:{$(1, 1)\;5$}] at (1,-0.5) {};
\node[label=above:{$1$ D3}] at (-3,0) {};
\node[label=below:{$1$ D3}] at (3,0) {};
\draw[decorate, decoration={brace}] (-1-0.1,0.5+0.1)--(3+0.1,0.5+0.1);
\node[label=above:{$l$}] at (1,0.5+0.1) {};
\node at (-0.8,-0.4) {$\cdots$};
\end{tikzpicture}}\hspace{1.5cm}
\scalebox{0.6}{
\begin{tikzpicture}
\draw[thick] (-4-1,0)--(4,0);
\draw[color=red] (-4,-2)--(-4,2);
\draw[color=purple] (-3,-0.5)--(-1,0.5);
\draw[color=purple] (1,-0.5)--(3,0.5);
\node[label=below:{$(1, 1)\;5$}] at (-3+0.25,-0.5) {};
\node[label=below:{$(1, 1)\;5$}] at (1,-0.5) {};
\node[label=above:{$1$ D3}] at (-3,0) {};
\node[label=below:{$1$ D3}] at (3,0) {};
\draw[decorate, decoration={brace}] (-1-0.1,0.5+0.1)--(3+0.1,0.5+0.1);
\node[label=above:{$l$}] at (1,0.5+0.1) {};
\node at (-0.8,-0.4) {$\cdots$};
\draw[very thin] (3.95,-0.25)--(4.05,0.25);
\draw[very thin] (4.05,-0.25)--(4.15,0.25);
\draw[very thin] (-5.05,-0.25)--(-4.95,0.25);
\draw[very thin] (-5.15,-0.25)--(-5.05,0.25);
\end{tikzpicture}}
\scalebox{0.6}{
\begin{tikzpicture}
\draw[arrows=->] (5.5,-2)--(6,-2);
\draw[arrows=->] (5.5,-2)--(5.5,-1.5);
\draw[arrows=->] (5.5,-2)--(5.5-1/1.25/2,-2-0.5/1.25/2);
\node[label=right:{$x^3$}] at (6,-2) {};
\node[label=above:{$x^6$}] at (5.5,-1.5) {};
\node[label=left:{$x^9$}] at (5.5-1/1.25/2,-2-0.5/1.25/2) {};
\end{tikzpicture}}
\caption{Left: The brane configuration after applying the $STS$ transformation to the configuration in the middle figure of Figure \ref{fig:SQCDbrane} with $N=1$. Right: The brane configuration after applying the $STS$ transformation to the configuration in Figure \ref{fig:ADHMbrane} with $N=1$. The space in the $x^3$-direction is compactified on $S^1$. }
\label{fig:STSdualbrane}
\end{figure}
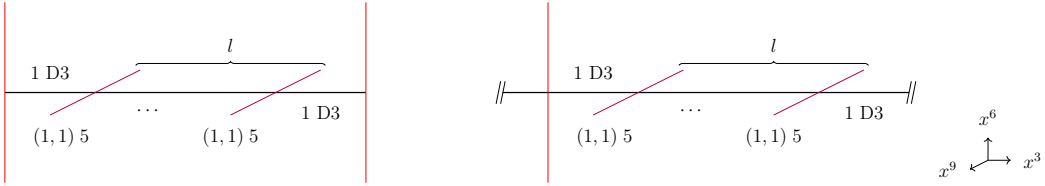

It is also possible to obtain other dual descriptions by using other elements in $SL(2, \mathbb{Z})$. For example, the $STS$ transformation with 
\be
T\text{: }
\left(\begin{array}{cc}
d & -b \\
-c & a
\end{array}\right)
= \left(\begin{array}{cc}
1 & -1 \\
0 & 1
\end{array}\right),
\ee
changes a D5-brane into a $(1, 1)$ 5-brane up to an overall sign while an NS5-brane is unchanged. When $N=1$, the brane configuration in the middle figure of Figure \ref{fig:SQCDbrane} becomes the one in the left figure of Figure \ref{fig:STSdualbrane} by the $STS$ transformation. The configuration is equivalent to the one in Figure \ref{fig:CSbrane1} with $N=1$ and $k=1$ except for the directions in which $(1, 1)$ 5-branes are extended. The difference in the extended directions comes from the fact that the complexified string coupling $\tau$ becomes $\tau = \frac{1}{2} + \frac{1}{2}i$ in Figure \ref{fig:STSdualbrane} due to the $STS$ transformation while $\tau = i$ in Figure \ref{fig:CSbrane1}. It is then natural to expect that the brane configuration in the left figure of Figure \ref{fig:STSdualbrane} and that in Figure \ref{fig:CSbrane1} with $N=1$ and $k=1$ realize the same $U(1)_1 \times U(1)^{l-1}_0 \times U(1)_{-1}$ linear quiver CS theory. Hence the brane construction implies a duality between the $U(1)$ gauge theory with $l$ flavors (SQED$_l$) and the $U(1)_1 \times U(1)^{l-1}_0 \times U(1)_{-1}$ linear quiver CS theory \cite{Jafferis:2008em}.

The application of the $STS$ transformation to the configuration in Figure \ref{fig:ADHMbrane} with $N=1$ yields the brane configuration in the right figure of Figure \ref{fig:STSdualbrane}. The configuration is equivalent to the one in Figure \ref{fig:CSbrane2} with $N=1$ and $k=1$ except for the directions in which $(1, 1)$ 5-branes are extended. Then, the brane configuration in the right figure of Figure \ref{fig:STSdualbrane} and that in Figure \ref{fig:CSbrane2} are expected to realize the same $U(1)_1 \times U(1)^{l-1}_0 \times U(1)_{-1}$ circular quiver CS theory. Namely, the duality between the brane configurations implies a duality between the $U(1)$ ADHM theory with $l$ flavors and the $U(1)_1 \times U(1)^{l-1}_0 \times U(1)_{-1}$ circular quiver CS theory.

\begin{figure}[t]
\centering
\scalebox{0.6}{
\begin{tikzpicture}
\draw[color=red] (-5,-2)--(-5,2);
\draw[color=red] (5,-2)--(5,2);
\draw (-5,0)--(5,0);
\draw[color=blue] (-4.5,1)--(-2.5,2);
\draw[color=blue] (-2.5,1)--(-0.5,2);
\node at (-2.5,1.5) {$\cdots$};
\draw[color=blue] (0.5,1)--(2.5,2);
\draw[color=blue] (2.5,1)--(4.5,2);
\node at (2.5,1.5) {$\cdots$};
\draw[color=blue] (-1,4)--(1,5);
\draw[thick] (0,0)--(0,4.5);
\draw[decorate, decoration={brace}] (-2.5-0.1,2+0.1)--(-0.5+0.1,2+0.1);
\node[label=above:{$l_1$}] at (-1.5,2.1) {};
\draw[decorate, decoration={brace}] (2.5-0.1,2+0.1)--(4.5+0.1,2+0.1);
\node[label=above:{$l-l_1$}] at (3.4,2.1) {};
\node[label=below:{$1$ D3}] at (-1,0) {};
\node[label=right:{$n$ F1}] at (0,3.5) {};
\end{tikzpicture}}
\scalebox{0.6}{
\begin{tikzpicture}
\draw[arrows=->] (5.5,-2)--(6,-2);
\draw[arrows=->] (5.5,-2)--(5.5,-1.5);
\draw[arrows=->] (5.5,-2)--(5.5-1/1.25/2,-2-0.5/1.25/2);
\node[label=right:{$x^3$}] at (6,-2) {};
\node[label=above:{$x^6$}] at (5.5,-1.5) {};
\node[label=left:{$x^9$}] at (5.5-1/1.25/2,-2-0.5/1.25/2) {};
\end{tikzpicture}}
\caption{The brane configuration for a Wilson line with charge $n$ in the $U(1)$ gauge theory with $l$ flavors.}
\label{fig:WilsonSQED}
\end{figure}
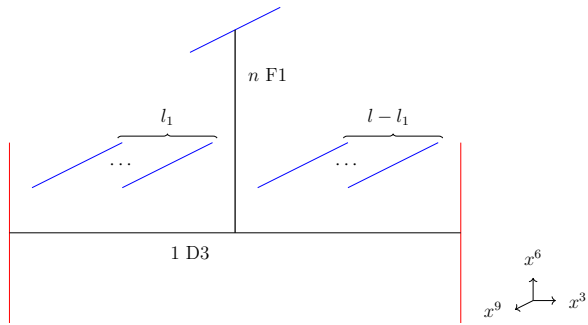

\begin{table}[t]
\centering
\begin{tabular}{c|ccc|c|ccc|ccc}
&0&1&2&3&4&5&6&7&8&9\\
\hline
D3&$\times$&$\times$&$\times$&$\times$&&&&&&\\
\hline
NS5&$\times$&$\times$&$\times$&&$\times$&$\times$&$\times$&&&\\
\hline
D5&$\times$&$\times$&$\times$&&&&&$\times$&$\times$&$\times$\\
\hline
$(1, k)$ 5&$\times$&$\times$&$\times$&&\multicolumn{6}{c}{$(47)_k, (58)_k, (69)_k$}\\
\hline
F1 & $\times$&&&&&&$\times$&&&\\
\hline
\end{tabular}
\caption{The directions where branes and fundamental strings (F1) are extended in the ten-dimensional spacetime of Type IIB string theory. The complexified string coupling is chosen to be $\tau = i$.}
\label{tb:F1string}
\end{table}
BPS Wilson line defect operators can be also introduced by adding semi-infinite fundamental strings in the brane construction \cite{Assel:2015oxa}. A BPS Wilson line defect can be inserted by introducing semi-infinite F1-strings ending on D3-branes in the brane construction as in Table \ref{tb:F1string}.  
More specifically, the insertion of a BPS Wilson line with charge $n$ under a $U(1)$ symmetry is realized by adding semi-infinite $n$ F1-strings ending on a D3-brane which supports the $U(1)$ symmetry. The other ends of the $n$ F1-strings may end on a D5-brane.\footnote{Here, the defect D5-brane is extended in the same directions as those of a flavor D5-brane. In the case of a $U(N)$ gauge theory, F1-strings ending on D3-branes which support the $U(N)$ gauge group and such a defect D5-brane may give a Wilson line in a symmetric representation of $U(N)$. One can also consider F1-strings which end on the D3-branes and another type of a D5-brane which is extended in the $x^0, x^4, x^5, x^7, x^8, x^9$-directions \cite{Assel:2015oxa}. Such F1-strings may yield a Wilson line in an antisymmetric representation of $U(N)$ due to the s-rule. Since we here focus on Abelian gauge theories, it is enough to consider the former type. } The brane configuration for a Wilson line with charge $n$ in the $U(1)$ gauge theory with $l$ flavors is illustrated in Figure \ref{fig:WilsonSQED}.
The flavor D5-branes seem to be split into two sets by the presence of the F1-strings but the splitting does not affect the low energy effective field theory since a flavor D5-brane on one side can move to the other side without colliding with the F1-strings nor the defect D5-brane.

\begin{figure}[t]
\centering
\scalebox{0.6}{
\begin{tikzpicture}
\draw[color=red] (-5,-2)--(-5,2);
\draw[color=red] (5,-2)--(5,2);
\draw (-5,0)--(5,0);
\draw[color=blue] (-3.5,-0.5)--(-1.5,0.5);
\draw[color=blue] (1.5,-0.5)--(3.5,0.5);
\node at (-0.5,-0.4) {$\cdots$};
\draw[color=blue] (3,3)--(5,4);
\draw[thick] (4,0)--(4,3.5);
\draw[decorate, decoration={brace}] (-1.5-0.1,0.5+0.1)--(3.5+0.1,0.5+0.1);
\node[label=above:{$l$}] at (1,0.6) {};
\node[label=above:{$1$ D3}] at (-4,0) {};
\node[label=left:{$n$ F1}] at (4,2) {};
\end{tikzpicture}}\hspace{1.2cm}
\scalebox{0.6}{
\begin{tikzpicture}
\draw[color=blue] (-5,-2)--(-5,2);
\draw[color=blue] (5,-2)--(5,2);
\draw (-5,0)--(5,0);
\draw[color=red] (-3.5,-0.5)--(-1.5,0.5);
\draw[color=red] (1.5,-0.5)--(3.5,0.5);
\node at (-0.5,-0.4) {$\cdots$};
\draw[color=red] (3,3)--(5,4);
\draw[thick] (4,0)--(4,3.5);
\draw[decorate, decoration={brace}] (-1.5-0.1,0.5+0.1)--(3.5+0.1,0.5+0.1);
\node[label=above:{$l$}] at (1,0.6) {};
\node[label=above:{$1$ D3}] at (-4,0) {};
\node[label=left:{$n$ D1}] at (4,2) {};
\end{tikzpicture}}
\scalebox{0.6}{
\begin{tikzpicture}
\draw[arrows=->] (5.5,-2)--(6,-2);
\draw[arrows=->] (5.5,-2)--(5.5,-1.5);
\draw[arrows=->] (5.5,-2)--(5.5-1/1.25/2,-2-0.5/1.25/2);
\node[label=right:{$x^3$}] at (6,-2) {};
\node[label=above:{$x^6$}] at (5.5,-1.5) {};
\node[label=left:{$x^9$}] at (5.5-1/1.25/2,-2-0.5/1.25/2) {};
\end{tikzpicture}}
\caption{Left: The brane configuration for a Wilson line with charge $n$ in the $U(1)$ gauge theory with $l$ flavors. Right: The S-dual of the brane configuration in the left figure.  The D1-branes yield a vortex line in $[1] - \widetilde{U(1)^{\otimes (l-1)}} - [1]$. }
\label{fig:SdualWilsonSQED}
\end{figure}
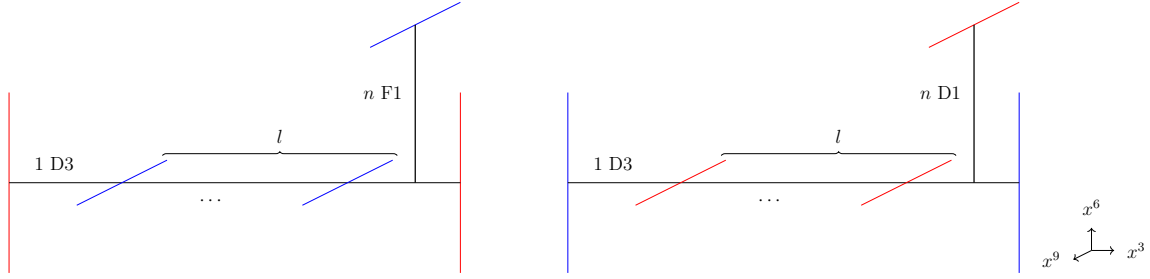

Let us consider configurations which are dual to brane configurations with a Wilson line operator. When we apply the S-duality transformation to the brane configuration in the left figure of Figure \ref{fig:SdualWilsonSQED}, $n$ F1-strings are changed into $n$ D1-branes and the brane configuration becomes the one in the right figure of Figure \ref{fig:SdualWilsonSQED}. 
The configuration without the D1-branes is the same as the one in Figure \ref{fig:Sdualbrane1} and it realizes the linear quiver theory, $[1] - \widetilde{U(1)^{\otimes (l-1)}} - [1]$. The D1-branes ending on a D3-brane may be interpreted as a vortex line operator \cite{Assel:2015oxa}. Therefore, the duality between the two configurations implies that the S-dual of the charge $n$ Wilson line in the SQED$_l$ is certain vortex line operator in the mirror theory, $[1] - \widetilde{U(1)^{\otimes (l-1)}} - [1]$.

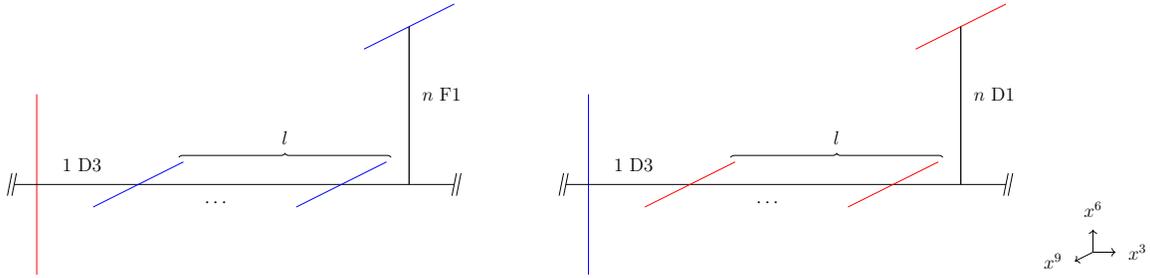
\begin{figure}[t]
\centering
\scalebox{0.6}{
\begin{tikzpicture}
\draw[color=red] (-5+0.25,-2)--(-5+0.25,2);
\draw (-5.25,0)--(5-0.5,0);
\draw[color=blue] (-3.5,-0.5)--(-1.5,0.5);
\draw[color=blue] (1.5-0.5,-0.5)--(3.5-0.5,0.5);
\node at (-0.5-0.25,-0.4) {$\cdots$};
\draw[color=blue] (3-0.5,3)--(5-0.5,4);
\draw[thick] (4-0.5,0)--(4-0.5,3.5);
\draw[decorate, decoration={brace}] (-1.5-0.1,0.5+0.1)--(3.5+0.1-0.5,0.5+0.1);
\node[label=above:{$l$}] at (0.75,0.6) {};
\node[label=above:{$1$ D3}] at (-3.75,0) {};
\node[label=right:{$n$ F1}] at (4-0.5,2) {};
\draw[very thin] (4.95-0.5,-0.25)--(5.05-0.5,0.25);
\draw[very thin] (5.05-0.5,-0.25)--(5.15-0.5,0.25);
\draw[very thin] (-5.05-0.25,-0.25)--(-4.95-0.25,0.25);
\draw[very thin] (-5.15-0.25,-0.25)--(-5.05-0.25,0.25);
\end{tikzpicture}}\hspace{1cm}
\scalebox{0.6}{
\begin{tikzpicture}
\draw[color=blue] (-5+0.25,-2)--(-5+0.25,2);
\draw (-5.25,0)--(5-0.5,0);
\draw[color=red] (-3.5,-0.5)--(-1.5,0.5);
\draw[color=red] (1.5-0.5,-0.5)--(3.5-0.5,0.5);
\node at (-0.5-0.25,-0.4) {$\cdots$};
\draw[color=red] (3-0.5,3)--(5-0.5,4);
\draw[thick] (4-0.5,0)--(4-0.5,3.5);
\draw[decorate, decoration={brace}] (-1.5-0.1,0.5+0.1)--(3.5+0.1-0.5,0.5+0.1);
\node[label=above:{$l$}] at (0.75,0.6) {};
\node[label=above:{$1$ D3}] at (-3.75,0) {};
\node[label=right:{$n$ D1}] at (4-0.5,2) {};
\draw[very thin] (4.95-0.5,-0.25)--(5.05-0.5,0.25);
\draw[very thin] (5.05-0.5,-0.25)--(5.15-0.5,0.25);
\draw[very thin] (-5.05-0.25,-0.25)--(-4.95-0.25,0.25);
\draw[very thin] (-5.15-0.25,-0.25)--(-5.05-0.25,0.25);
\end{tikzpicture}}
\scalebox{0.6}{
\begin{tikzpicture}
\draw[arrows=->] (5.5,-2)--(6,-2);
\draw[arrows=->] (5.5,-2)--(5.5,-1.5);
\draw[arrows=->] (5.5,-2)--(5.5-1/1.25/2,-2-0.5/1.25/2);
\node[label=right:{$x^3$}] at (6,-2) {};
\node[label=above:{$x^6$}] at (5.5,-1.5) {};
\node[label=left:{$x^9$}] at (5.5-1/1.25/2,-2-0.5/1.25/2) {};
\end{tikzpicture}}
\caption{Left: The brane configuration for a Wilson line in the $U(1)$ ADHM theory with $l$ flavors. The $n$ F1-strings give rise to the Wilson line with charge $n$ under the $U(1)$ gauge group. Right: The S-dual of the brane configuration in the left figure. The D1-branes yield a vortex line in $\widetilde{U(1)^{\otimes l} \textrm{mADHM}-[1]}$. }
\label{fig:SdualWilsonADHM}
\end{figure}

Similarly, we can consider a Wilson line in the $U(1)$ ADHM theory with $l$ flavors. The brane configuration for a Wilson line with charge $n$ under the $U(1)$ gauge group of the $U(1)$ ADHM theory with $l$ flavors is depicted in the left figure of Figure \ref{fig:SdualWilsonADHM}. 
The S-dual of the configuration gives rise to the one in the right figure of Figure \ref{fig:SdualWilsonADHM}. The presence of D1-branes again implies that the S-dual of the Wilson line in the $U(1)$ ADHM theory with $l$ flavors yields a vortex line in the mirror theory, $\widetilde{U(1)^{\otimes l} \textrm{mADHM}-[1]}$.

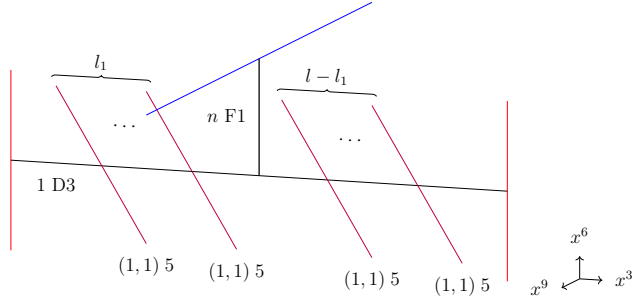
\begin{figure}[t]
\centering
\scalebox{0.6}{
\begin{tikzpicture}
\draw[color=red] (-4,-2)--(-4,2);
\draw[color=red] (4+3,-2-0.5-0.5/8*3)--(4+3,2-0.5-0.5/8*3);
\draw (-4,0)--(4+3,-0.5-0.5/8*3);
\draw[color=purple] (-3,1.75-0.1)--(-1,-1.75-0.1);
\draw[color=purple] (-1,1.75-0.1-0.5/8*2)--(1,-1.75-0.1-0.5/8*2);
\draw[color=blue] (-1,3.25+0.5-5/2-0.25)--(4,3.25+0.5-0.25);
\draw[thick] (1.5,3.25+0.5-0.25-2.5/2)--(1.5,-0.5/8*5.5);
\draw[color=purple] (2,1.75-0.1-0.5/8*5)--(4,-1.75-0.1-0.5/8*5);
\draw[color=purple] (4,1.75-0.1-0.5/8*7)--(6,-1.75-0.1-0.5/8*7);
\node[label=below:{$(1, 1)\;5$}] at (-1,-1.75-0.1) {};
\node[label=below:{$(1, 1)\;5$}] at (1,-1.75-0.1-0.5/8*2) {};
\node[label=below:{$(1, 1)\;5$}] at (4,-1.75-0.1-0.5/8*5) {};
\node[label=below:{$(1, 1)\;5$}] at (6,-1.75-0.1-0.5/8*7) {};
\node at (-2+2/3.5*0.9-0.2,0.75+0.5/8*0.2) {$\cdot$};
\node at (-2+2/3.5*0.9,0.75) {$\cdot$};
\node at (-2+2/3.5*0.9+0.2,0.75-0.5/8*0.2) {$\cdot$};
\node at (3+2/3.5*0.9-0.2,0.75-0.5/8*5+0.5/8*0.2) {$\cdot$};
\node at (3+2/3.5*0.9,0.75-0.5/8*5) {$\cdot$};
\node at (3+2/3.5*0.9+0.2,0.75-0.5/8*5-0.5/8*0.2) {$\cdot$};
\draw[decorate, decoration={brace}] (-3.1,1.75+0.1)--(-0.9,1.75+0.1-0.5/8*2);
\draw[decorate, decoration={brace}] (2-0.1,1.75+0.1-0.5/8*5)--(4.1,1.75+0.1-0.5/8*7);
\node[label=above:{$l_1$}] at (-2,1.5+0.2) {};
\node[label=above:{$l-l_1$}] at (3,1.5+0.2-0.5/8*5) {};
\node[label=below:{$1$ D3}] at (-3,-0.1) {};
\node[label=left:{$n$ F1}] at (1.5,1) {};
\end{tikzpicture}}
\scalebox{0.6}{
\begin{tikzpicture}
\draw[arrows=->] (5.5,-2)--(6,-2-0.5*0.5/8);
\draw[arrows=->] (5.5,-2)--(5.5,-1.5);
\draw[arrows=->] (5.5,-2)--(5.5-1/1.25/2,-2-0.5/1.25/2);
\node[label=right:{$x^3$}] at (6,-2) {};
\node[label=above:{$x^6$}] at (5.5,-1.5) {};
\node[label=left:{$x^9$}] at (5.5-1/1.25/2,-2-0.5/1.25/2) {};
\end{tikzpicture}}
\caption{The brane configuration for a Wilson line in the $U(1)_1 \times U(1)_0^{l-1} \times U(1)_{-1}$ linear quiver CS theory. The Wilson line has charge $n$ under the $(l_1+1)$th $U(1)$ gauge group.}
\label{fig:WilsonlinearCS}
\end{figure}

\begin{figure}[t]
\centering
\scalebox{0.6}{
\begin{tikzpicture}
\draw[color=red] (-4,-2)--(-4,2);
\draw[color=red] (4+3,-2-0.5-0.5/8*3)--(4+3,2-0.5-0.5/8*3);
\draw (-4,0)--(4+3,-0.5-0.5/8*3);
\draw[color=blue] (-3,1.75-0.1)--(-1,-1.75-0.1);
\draw[color=blue] (-1,1.75-0.1-0.5/8*2)--(1,-1.75-0.1-0.5/8*2);
\draw[color=purple] (-1,3.25+0.5-5/2-0.25)--(4,3.25+0.5-0.25);
\draw[thick] (1.5,3.25+0.5-0.25-2.5/2)--(1.5,-0.5/8*5.5);
\draw[color=blue] (2,1.75-0.1-0.5/8*5)--(4,-1.75-0.1-0.5/8*5);
\draw[color=blue] (4,1.75-0.1-0.5/8*7)--(6,-1.75-0.1-0.5/8*7);
\node[label=below:{D5}] at (-1,-1.75-0.1) {};
\node[label=below:{D5}] at (1,-1.75-0.1-0.5/8*2) {};
\node[label=below:{D5}] at (4,-1.75-0.1-0.5/8*5) {};
\node[label=below:{D5}] at (6,-1.75-0.1-0.5/8*7) {};
\node at (-2+2/3.5*0.9-0.2,0.75+0.5/8*0.2) {$\cdot$};
\node at (-2+2/3.5*0.9,0.75) {$\cdot$};
\node at (-2+2/3.5*0.9+0.2,0.75-0.5/8*0.2) {$\cdot$};
\node at (3+2/3.5*0.9-0.2,0.75-0.5/8*5+0.5/8*0.2) {$\cdot$};
\node at (3+2/3.5*0.9,0.75-0.5/8*5) {$\cdot$};
\node at (3+2/3.5*0.9+0.2,0.75-0.5/8*5-0.5/8*0.2) {$\cdot$};
\draw[decorate, decoration={brace}] (-3.1,1.75+0.1)--(-0.9,1.75+0.1-0.5/8*2);
\draw[decorate, decoration={brace}] (2-0.1,1.75+0.1-0.5/8*5)--(4.1,1.75+0.1-0.5/8*7);
\node[label=above:{$l_1$}] at (-2,1.5+0.2) {};
\node[label=above:{$l-l_1$}] at (3,1.5+0.2-0.5/8*5) {};
\node[label=below:{$1$ D3}] at (-3,-0.1) {};
\node[label=left:{$n\;(1,-1)$}] at (1.5,1) {};
\node[label=right:{$(1,-1)\;5$}] at (4,3.25+0.5-0.25) {};
\end{tikzpicture}}
\scalebox{0.6}{
\begin{tikzpicture}
\draw[arrows=->] (5.5,-2)--(6,-2-0.5*0.5/8);
\draw[arrows=->] (5.5,-2)--(5.5,-1.5);
\draw[arrows=->] (5.5,-2)--(5.5-1/1.25/2,-2-0.5/1.25/2);
\node[label=right:{$x^3$}] at (6,-2) {};
\node[label=above:{$x^6$}] at (5.5,-1.5) {};
\node[label=left:{$x^9$}] at (5.5-1/1.25/2,-2-0.5/1.25/2) {};
\end{tikzpicture}}
\caption{The brane configuration after applying the inverse of the $STS$ transformation to the configuration in Figure \ref{fig:WilsonlinearCS}. $(1,-1)$ stands for a $(1,-1)$-string and the purple line is a $(1, -1)$ 5-brane. }
\label{fig:STSWilsonlinearCS}
\end{figure}
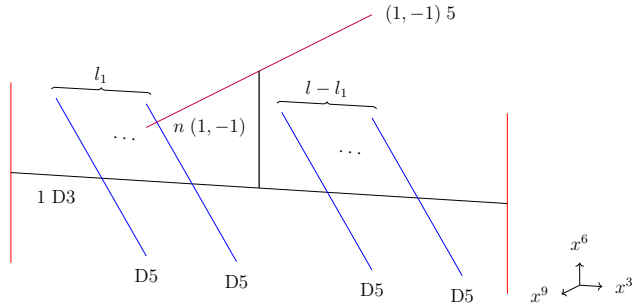

It is also possible to consider a brane setup for BPS Wilson lines in quiver CS theories. For example, the brane configuration in Figure \ref{fig:WilsonlinearCS} gives a Wilson line with charge $n$ under the $(l_1+1)$th $U(1)$ gauge group in the $U(1)_1 \times U(1)_0^{l-1} \times U(1)_{-1}$ linear quiver CS theory. 
The application of an $SL(2,\mathbb{Z})$ transformation to the brane configuration will give a line operator in a dual picture. Applying the inverse of the $STS$ transformation to the configuration in Figure \ref{fig:WilsonlinearCS} yields the brane configuration in Figure \ref{fig:STSWilsonlinearCS}.
Since the $STS$ transformation of SQED$_l$ becomes the $U(1)_1\times U(1)_0^{l-1}\times U(1)_{-1}$ linear quiver CS theory, the inverse of the $STS$ transformation of the $U(1)_1\times U(1)_0^{l-1}\times U(1)_{-1}$ linear quiver CS theory yields SQED$_l$. Hence the brane configuration in Figure \ref{fig:STSWilsonlinearCS} will give rise to a line operator in SQED$_l$. Since the $(1,-1)$-string has the D1-brane charge and also the fundamental string charge, it is natural to conjecture that the line operator is a vortex-Wilson line operator. Therefore, we argue that a Wilson line in the $U(1)_1\times U(1)_0^{l-1}\times U(1)_{-1}$ linear quiver CS theory is dual to a vortex-Wilson line in SQED$_l$. 

\begin{figure}[t]
\centering
\scalebox{0.6}{
\begin{tikzpicture}
\draw[color=blue] (-4,-2)--(-4,2);
\draw[color=blue] (4+3,-2-0.5-0.5/8*3)--(4+3,2-0.5-0.5/8*3);
\draw (-4,0)--(4+3,-0.5-0.5/8*3);
\draw[color=red] (-3,1.75-0.1)--(-1,-1.75-0.1);
\draw[color=red] (-1,1.75-0.1-0.5/8*2)--(1,-1.75-0.1-0.5/8*2);
\draw[color=purple] (-1,3.25+0.5-5/2-0.25)--(4,3.25+0.5-0.25);
\draw[thick] (1.5,3.25+0.5-0.25-2.5/2)--(1.5,-0.5/8*5.5);
\draw[color=red] (2,1.75-0.1-0.5/8*5)--(4,-1.75-0.1-0.5/8*5);
\draw[color=red] (4,1.75-0.1-0.5/8*7)--(6,-1.75-0.1-0.5/8*7);
\node[label=below:{NS5}] at (-1,-1.75-0.1) {};
\node[label=below:{NS5}] at (1,-1.75-0.1-0.5/8*2) {};
\node[label=below:{NS5}] at (4,-1.75-0.1-0.5/8*5) {};
\node[label=below:{NS5}] at (6,-1.75-0.1-0.5/8*7) {};
\node at (-2+2/3.5*0.9-0.2,0.75+0.5/8*0.2) {$\cdot$};
\node at (-2+2/3.5*0.9,0.75) {$\cdot$};
\node at (-2+2/3.5*0.9+0.2,0.75-0.5/8*0.2) {$\cdot$};
\node at (3+2/3.5*0.9-0.2,0.75-0.5/8*5+0.5/8*0.2) {$\cdot$};
\node at (3+2/3.5*0.9,0.75-0.5/8*5) {$\cdot$};
\node at (3+2/3.5*0.9+0.2,0.75-0.5/8*5-0.5/8*0.2) {$\cdot$};
\draw[decorate, decoration={brace}] (-3.1,1.75+0.1)--(-0.9,1.75+0.1-0.5/8*2);
\draw[decorate, decoration={brace}] (2-0.1,1.75+0.1-0.5/8*5)--(4.1,1.75+0.1-0.5/8*7);
\node[label=above:{$l_1$}] at (-2,1.5+0.2) {};
\node[label=above:{$l-l_1$}] at (3,1.5+0.2-0.5/8*5) {};
\node[label=below:{$1$ D3}] at (-3,-0.1) {};
\node[label=left:{$n\;(1,1)$}] at (1.5,1) {};
\node[label=right:{$(1,1)\;5$}] at (4,3.25+0.5-0.25) {};
\end{tikzpicture}}
\scalebox{0.6}{
\begin{tikzpicture}
\draw[arrows=->] (5.5,-2)--(6,-2-0.5*0.5/8);
\draw[arrows=->] (5.5,-2)--(5.5,-1.5);
\draw[arrows=->] (5.5,-2)--(5.5-1/1.25/2,-2-0.5/1.25/2);
\node[label=right:{$x^3$}] at (6,-2) {};
\node[label=above:{$x^6$}] at (5.5,-1.5) {};
\node[label=left:{$x^9$}] at (5.5-1/1.25/2,-2-0.5/1.25/2) {};
\end{tikzpicture}}
\caption{The brane configuration after applying the $S$ transformation to the configuration in Figure \ref{fig:STSWilsonlinearCS}. $(1,1)$ stands for a $(1,1)$-string and the purple line is a $(1, 1)$ 5-brane. }
\label{fig:SSTSWilsonlinearCS}
\end{figure}
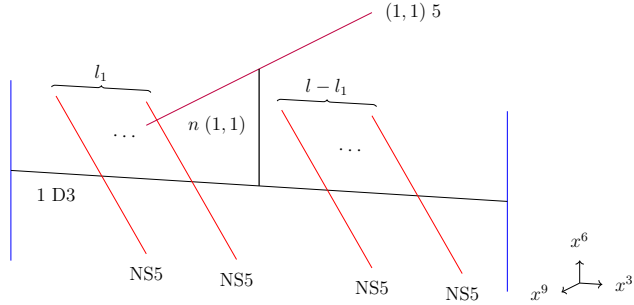

Furthermore, one can apply the $S$ transformation to the configuration in Figure \ref{fig:STSWilsonlinearCS}. The resulting brane configuration is depicted in Figure \ref{fig:SSTSWilsonlinearCS}. 
The $(1,-1)$-strings change into $(1, 1)$-strings up to an overall sign. Since the S-dual of SQED$_l$ is the linear quiver theory, $[1] - \widetilde{U(1)^{\otimes (l-1)}} - [1]$, the configuration in Figure \ref{fig:SSTSWilsonlinearCS} is expected to yield a line operator in $[1] - \widetilde{U(1)^{\otimes (l-1)}} - [1]$. We conjecture that the line operator is a vortex-Wilson line operator which has some electric charge under the $l_1$th $U(1)$ gauge group of $U(1)^{\otimes (l-1)}$ since the $(1, 1)$-strings end on the $(l_1+1)$th segment of the D3-brane from the left and the $(1, 1)$-strings have the D1-brane charge and also the fundamental string charge. Namely, a Wilson line in the linear quiver CS theory, $U(1)_1\times U(1)_0^{l-1}\times U(1)_{-1}$, has another dual description given by a vortex-Wilson line in the linear quiver theory, $[1] - \widetilde{U(1)^{\otimes (l-1)}} - [1]$. 

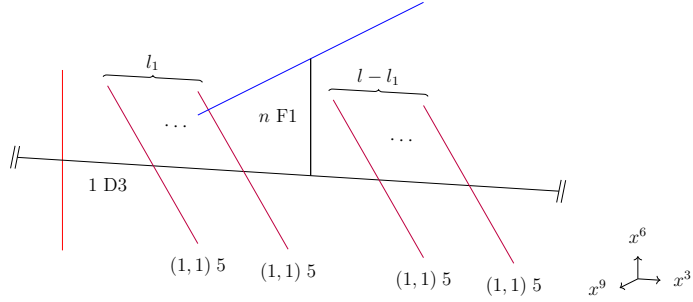
\begin{figure}[t]
\centering
\scalebox{0.6}{
\begin{tikzpicture}
\draw[color=red] (-4,-2)--(-4,2);
\draw (-4-1,0+0.5/8)--(4+3,-0.5-0.5/8*3);
\draw[color=purple] (-3,1.75-0.1)--(-1,-1.75-0.1);
\draw[color=purple] (-1,1.75-0.1-0.5/8*2)--(1,-1.75-0.1-0.5/8*2);
\draw[color=blue] (-1,3.25+0.5-5/2-0.25)--(4,3.25+0.5-0.25);
\draw[thick] (1.5,3.25+0.5-0.25-2.5/2)--(1.5,-0.5/8*5.5);
\draw[color=purple] (2,1.75-0.1-0.5/8*5)--(4,-1.75-0.1-0.5/8*5);
\draw[color=purple] (4,1.75-0.1-0.5/8*7)--(6,-1.75-0.1-0.5/8*7);
\node[label=below:{$(1, 1)\;5$}] at (-1,-1.75-0.1) {};
\node[label=below:{$(1, 1)\;5$}] at (1,-1.75-0.1-0.5/8*2) {};
\node[label=below:{$(1, 1)\;5$}] at (4,-1.75-0.1-0.5/8*5) {};
\node[label=below:{$(1, 1)\;5$}] at (6,-1.75-0.1-0.5/8*7) {};
\node at (-2+2/3.5*0.9-0.2,0.75+0.5/8*0.2) {$\cdot$};
\node at (-2+2/3.5*0.9,0.75) {$\cdot$};
\node at (-2+2/3.5*0.9+0.2,0.75-0.5/8*0.2) {$\cdot$};
\node at (3+2/3.5*0.9-0.2,0.75-0.5/8*5+0.5/8*0.2) {$\cdot$};
\node at (3+2/3.5*0.9,0.75-0.5/8*5) {$\cdot$};
\node at (3+2/3.5*0.9+0.2,0.75-0.5/8*5-0.5/8*0.2) {$\cdot$};
\draw[decorate, decoration={brace}] (-3.1,1.75+0.1)--(-0.9,1.75+0.1-0.5/8*2);
\draw[decorate, decoration={brace}] (2-0.1,1.75+0.1-0.5/8*5)--(4.1,1.75+0.1-0.5/8*7);
\node[label=above:{$l_1$}] at (-2,1.5+0.2) {};
\node[label=above:{$l-l_1$}] at (3,1.5+0.2-0.5/8*5) {};
\node[label=below:{$1$ D3}] at (-3,-0.1) {};
\node[label=left:{$n$ F1}] at (1.5,1) {};
\draw[very thin] (3.95+3,-0.75-0.5/8*3)--(4.05+3,-0.25-0.5/8*3);
\draw[very thin] (4.05+3,-0.75-0.5/8*0.1-0.5/8*3)--(4.15+3,-0.25-0.5/8*0.1-0.5/8*3);
\draw[very thin] (-4.05-1,-0.25+0.5/8)--(-3.95-1,0.25+0.5/8);
\draw[very thin] (-4.15-1,-0.25+0.5/8*0.1+0.5/8)--(-4.05-1,0.25+0.5/8*0.1+0.5/8);
\end{tikzpicture}}
\scalebox{0.6}{
\begin{tikzpicture}
\draw[arrows=->] (5.5,-2)--(6,-2-0.5*0.5/8);
\draw[arrows=->] (5.5,-2)--(5.5,-1.5);
\draw[arrows=->] (5.5,-2)--(5.5-1/1.25/2,-2-0.5/1.25/2);
\node[label=right:{$x^3$}] at (6,-2) {};
\node[label=above:{$x^6$}] at (5.5,-1.5) {};
\node[label=left:{$x^9$}] at (5.5-1/1.25/2,-2-0.5/1.25/2) {};
\end{tikzpicture}}
\caption{The brane configuration for a Wilson line in the $U(1)_1 \times U(1)_0^{l-1} \times U(1)_{-1}$ circular quiver CS theory. The space in the $x^3$-direction is compactified on a circle. The Wilson line has charge $n$ under the gauge group associated to the D3-brane on which the F1-strings end.}
\label{fig:WilsoncircularCS}
\end{figure}

We can also consider a Wilson line with charge $n$ under a $U(1)$ gauge group in the $U(1)_1 \times U(1)_0^{l-1} \times U(1)_{-1}$ circular quiver CS theory as in Figure \ref{fig:WilsoncircularCS}.
Using the similar reasoning, the application of $SL(2,\mathbb{Z})$ transformations to the brane configuration in Figure \ref{fig:WilsoncircularCS} implies a triality among a Wilson line in the $U(1)_1 \times U(1)_0^{l-1} \times U(1)_{-1}$ circular quiver CS theory, a vortex-Wilson line in the $U(1)$ ADHM theory with $l$ flavors and a vortex-Wilson line in $\widetilde{U(1)^{\otimes l} \textrm{mADHM}-[1]}$.

\subsection{Line defect indices}
The supersymmetric index of 3d $\mathcal{N}=4$ supersymmetric gauge theory can be defined by 
\footnote{We follow the convention and definition of the index in \cite{Okazaki:2019ony,Hayashi:2022ldo}.}
\begin{align}
\label{DEF_ind}
\mathcal{I}(t,x,z;q)&:=\Tr (-1)^F q^{J+\frac{H+C}{4}} t^{H-C} x^{f_H} z^{f_C}, 
\end{align}
where the trace is taken over the cohomology with respect to the chosen supercharges $Q$. 
In particular, for the superconformal field theories it is taken over the Hilbert space of the radially quantized theory on $S^2$ 
and it counts the BPS local operators or states in the theory. 
Here $F$ is the Fermion number and $J$ is the spin. 
$H$ and $C$ are the Cartan generators of the $SU(2)_H$ and $SU(2)_C$ R-symmetry respectively. 
$f_H$ and $f_C$ are the Cartan generators of the Higgs and Coulomb branch symmetries. 

The index can be also defined from the UV description as a supersymmetric partition function on $S^1\times S^2$. 
According to the localization technique, it can be evaluated as a matrix integral and sum \cite{Kim:2009wb,Imamura:2011su,Kapustin:2011jm}. 
For the $\mathcal{N}=4$ supersymmetric gauge theory of gauge group $G$ with hypermultiplets transforming in the representation $\bf{R}$, 
it takes the following form: 
\begin{align}
\label{3d_index}
&
I^{\textrm{3d $G$}}(t,x,z;q)
\nonumber\\
&=
\frac{1}{| \textrm{Weyl} (G)|}
\frac{
(q^{\frac12}t^2;q)_{\infty}^{\mathrm{rank}(G)}
}
{
(q^{\frac12}t^{-2};q)_{\infty}^{\mathrm{rank}(G)}
}
\sum_{m\in \mathrm{cochar}(G)}
\oint \prod_{\alpha\in \mathrm{roots} (G) }
\frac{ds}{2\pi is}\nonumber \\
&\times 
\frac{
\left(1-q^{\frac{|m\cdot \alpha|}{2}}s^{\alpha} \right)
\left(q^{\frac{1+|m\cdot \alpha|}{2}} t^2 s^{\alpha};q\right)_{\infty}
}
{
\left(q^{\frac{1+|m\cdot \alpha|}{2}} t^{-2} s^{\alpha};q\right)_{\infty}
}
\prod_{\lambda \in {\bf R}}
\frac{
\left( q^{\frac34+\frac{|m\cdot \lambda|}{2}} t^{-1} s^{\lambda} x;q \right)_{\infty}
}
{
\left( q^{\frac14+\frac{|m\cdot \lambda|}{2}} t s^{\lambda} x;q \right)_{\infty}
}
q^{\frac{\Delta(m)}{2}}
\cdot 
t^{-2\Delta(m)}
\cdot 
z^{m}. 
\end{align}
Here the scaling dimension $\Delta(m)$ of the BPS monopole operator, where $m$ $\in$ $\textrm{cochar}(G)$ is a magnetic flux associated with the monopole, 
is given by \cite{Gaiotto:2008ak,Bashkirov:2010kz,Benna:2009xd}
\begin{align}
\Delta(m)&=\frac12 \sum_{\lambda \in \bf{R}}|\lambda(m)|-\sum_{\alpha\in \Delta_+}|\alpha(m)|, 
\end{align}
where $\lambda$ are the weights of $\bf{R}$ and $\alpha$ $\in$ $\Delta_+$ are positive roots for the Lie algebra of gauge group $G$. 
The first set of factors in the second and third lines of (\ref{3d_index}) is contributed from the fluctuations of the $\mathcal{N}=4$ vector multiplet. 
The other factor labeled by the weight $\lambda$ is the contribution from those of the $\mathcal{N}=4$ hypermultiplets (see e.g. \cite{Okazaki:2019ony} for the details).

One can decorate the index by further introducing the half-BPS line defect operators 
$L_{\mathsf{p}_1;\mathsf{q}_1}$, $\cdots$, $L_{\mathsf{p}_k;\mathsf{q}_k}$ wrapping the $S^1$ \cite{Drukker:2012sr}. 
We refer to the resulting object as the \textit{line defect index} or \textit{line defect correlation function}. 
The line defect operator modifies the Hilbert space on the $S^2$ 
so that the line defect correlation function can enumerate the BPS local operators living at a junction of the line operators. 

In the presence of the charged BPS Wilson line operator $W_{\mathsf{p}}$ in the Abelian gauge theory, 
one can compute the line defect index by inserting the factor $s^{\mathsf{p}}$ in the integrand of the matrix integral (\ref{3d_index}), 
where $s$ is the fugacity for the gauge or flavor group. 
Physically, at the junction of the Wilson lines the electrically charged BPS local operators can exist 
in such a way that the total charge of the Wilson lines and of the local operators vanishes. 
Accordingly, the index decorated by the charged Wilson lines allows us to detect the charged BPS local operators in the theory. 

As discussed in section \ref{sec_vline}, the flavor vortex line $V_{\mathsf{q}}$ in the Abelian gauge theory admits zeros or poles of order $\mathsf{q}$ for the matter fields. 
Accordingly, it shifts the spin $J$ of the BPS local operators carrying the charges for the corresponding gauge or flavor group. 
In the absence of the vortex line, the matter index in (\ref{3d_index}) can be evaluated from the localization technique 
by taking into account the Gaussian integral with respect to the fluctuations of fields. 
The covariant derivative on $S^2$ acting on the fields of spin $s$ is 
\begin{align}
D_i&=\partial_i-is_{\textrm{eff}}\omega_i, 
\end{align}
where 
\begin{align}
s_{\textrm{eff}}&=s+\frac{1}{2}\lambda(m), 
\end{align}
can be viewed as the effective spin 
and $\omega_i$ is the spin connection on $S^2$. 
Then we can use the spherical harmonics $Y_{j,j_3}^{s_{\textrm{eff}}}$ of spin $s_{\textrm{eff}}$ to get the matter index (see \cite{Imamura:2011su,Kapustin:2011jm} for the details). 
Inserting the flavor Abelian vortex line of charge $\mathsf{q}$ which is localized on a point on the $S^2$, 
the shifted spin is modified as
\begin{align}
s_{\textrm{eff}}&=s+\frac{1}{2}\lambda(m)+\frac{1}{2}\mathsf{q}. 
\end{align}
By noting that the actual dependence on the shifted spin appearing in the matter index is simply controlled by the the absolute values of the weights 
after summing over the contributions to the single particle index 
\cite{Imamura:2011su,Kapustin:2011jm}, 
we reach the conclusion that the insertion of the Abelian flavor vortex line acts on the index by shifting the associated flavor fugacity $x$ in the indices of the matter fields 
\footnote{Our convention is different from that in \cite{Drukker:2012sr} 
as we describe the contributions involving the effective spins due to the background flux by using the absolute values of the weights. }
\begin{align}
\label{vortex_shift}
q^{\frac{|m|}{2}}x^{\pm}&\rightarrow q^{\frac{|m+\mathsf{q}|}{2}}x^{\pm}. 
\end{align}
Here the zeros or poles for the matter fields are also generated by a dynamical Abelian monopole operator of magnetic charge $m$ corresponding to a singular gauge transformation. 
Hence the total shift of the spin is given by the magnitude of a net sum of the vortex number and the magnetic charge in the evaluation of the vortex line defect index. 
Furthermore, the junction of the vortex lines can support the BPS flavor monopole operators for the background gauge fields. 
The contributions from the flavor monopole operators to the vortex line defect index may appear as an extra overall factor with the same form as a dynamical monopole operator but with a fixed magnetic charge. 

In the following, we test the dualities of line operators in 3d $\mathcal{N}\ge 4$ Abelian gauge theories 
by computing the line defect indices according to the above methods. 
They are perfectly consistent with the conjectured dualities \cite{Assel:2015oxa,Dimofte:2019zzj} under $\mathcal{N}=4$ mirror symmetry. 
Applying the both prescriptions for the supersymmetric indices decorated by the vortex and Wilson lines, 
we also evaluate the line defect indices for the mixed vortex-Wilson line. 
Consequently, we propose more general dualities of line operators for $\mathcal{N}\ge 4$ supersymmetric Abelian gauge theories 
as matching pairs of the line defect indices.

The line defect index for $\mathcal{N}=4$ supersymmetric gauge theory admits the Coulomb and Higgs limits \cite{Razamat:2014pta}, 
for which the index can enumerate the half-BPS Coulomb and Higgs branch operators in the theory. 
Similarly, we can consider the Coulomb and Higgs limits of the line defect indices
\begin{align}
\langle L_{\mathsf{p}_1;\mathsf{q}_1} \cdots L_{\mathsf{p}_k;\mathsf{q}_k}\rangle_{C}(z;\mathfrak{t})
&:=
\lim_{\begin{smallmatrix}
\textrm{$\mathfrak{t}=q^{\frac14}t^{-1}$: fixed}\\
q\rightarrow \infty\\
\end{smallmatrix}}
\langle L_{\mathsf{p}_1;\mathsf{q}_1} \cdots L_{\mathsf{p}_k;\mathsf{q}_k}\rangle(t,x,z;q), \\
\langle L_{\mathsf{p}_1;\mathsf{q}_1} \cdots L_{\mathsf{p}_k;\mathsf{q}_k}\rangle_{H}(x;\mathfrak{t})
&:=
\lim_{\begin{smallmatrix}
\textrm{$\mathfrak{t}=q^{\frac14}t$: fixed}\\
q\rightarrow \infty\\
\end{smallmatrix}}
\langle L_{\mathsf{p}_1;\mathsf{q}_1} \cdots L_{\mathsf{p}_k;\mathsf{q}_k}\rangle(t,x,z;q). 
\label{CoulombHiggslimit}
\end{align}
Typically, the BPS local operators in the Higgs (resp. Coulomb) branch chiral ring are bound on the half-BPS Wilson (resp. vortex) lines. 
Thus one finds non-trivial Higgs (Coulomb) limits of the line defect correlators of the Wilson (resp. vortex) lines. 
They can be regarded as the Hilbert series for the line operator algebra (See e.g. \cite{Benvenuti:2006qr,Benvenuti:2010pq,Hanany:2011db,Cremonesi:2013lqa,Mekareeya:2015bla,Hayashi:2022ldo} 
for the Hilbert series for the bulk operator algebra). 
The closed-form expressions for the Coulomb and Higgs limits of the Wilson line defect correlators for the $U(N)$ ADHM theory with $l$ flavors were presented in \cite{Hayashi:2024jof}. 
Also some computations for the line defect Hilbert series can be found in e.g. \cite{Dimofte:2019zzj,Crew:2020psc,Nawata:2021nse,Crew:2024fom}. 

\section{Linear quiver gauge theories}
\label{sec_SQED}

\subsection{SQED$_1$}
\label{sec_SQED1}
We begin with the $\mathcal{N}=4$ SQED$_1$, the 3d $\mathcal{N}=4$ $U(1)$ gauge theory with a hypermultiplet of charge $+1$. 
The theory can be constructed as the theory on a single D3-brane suspended between NS5-branes and intersecting with a single D5-brane in Type IIB string theory, as reviewed in section \ref{sec_branesetup}.
The one-point function of the charged Wilson line $W_\mathsf{p}$ can be evaluated as
\begin{align}
\label{SQED1Wn}
\langle W_{\mathsf{p}} \rangle^{U(1)-[1]}(t,z;q)
&=
\frac{(q^{\frac12}t^{2};q)_{\infty}}{(q^{\frac12}t^{-2};q)_{\infty}}
\sum_{m\in \mathbb{Z}}\oint \frac{ds}{2\pi is}
\frac{(q^{\frac34+\frac{|m|}{2}}t^{-1}s^{\mp};q)_{\infty}}{(q^{\frac14+\frac{|m|}{2}}ts^{\pm};q)_{\infty}}
s^{\mathsf{p}} 
q^{\frac{|m|}{4}}t^{-|m|}z^m. 
\end{align}
The integrand has a tower of simple poles in $|s|<1$ at $s=q^{\frac14+\frac{|m|}{2}+l}t$ with $l=0,1,\cdots$. 
Picking up residues at the poles, we obtain
\begin{align}
&\langle W_\mathsf{p} \rangle^{U(1)-[1]}(t,z;q)\nonumber \\
&=\frac{(q^{\frac12}t^2;q)_{\infty}^2}{(q)_{\infty}^2}
\sum_{m\in \mathbb{Z}}
q^{\frac{|m|}{4}}t^{-|m|}z^m
\sum_{l=0}^{\infty}
\frac{(q^{1+l};q)_{\infty}(q^{1+|m|+l};q)_{\infty}}
{(q^{\frac12+l}t^2;q)_{\infty}(q^{\frac12+|m|+l}t^2;q)_{\infty}}
q^{\frac{\mathsf{p}}{4}+\frac{l}{2}+\frac{n|m|}{2}+l\mathsf{p}}
t^{\mathsf{p}-2l}. 
\end{align}

The theory is mirror to the theory of a free twisted hypermultiplet. 
The theory admits non-trivial half-BPS vortex lines. 
The twisted hypermultiplet scalar fields $(\tilde{I},\tilde{J})$ obeying the BPS equations 
are required to be constant in time and holomorphic in the two-dimensional plane. 
As observed in \cite{Okazaki:2019ony}, the one-point function (\ref{SQED1Wn}) coincides with 
the line defect index of the flavor vortex line $V_\mathsf{q}$ in the theory of a free twisted hyper
\begin{align}
\label{thypVn}
\langle V_\mathsf{q} \rangle^{\textrm{thyper}}(t,z;q)
=q^{\frac{|\mathsf{q}|}{4}}t^{|\mathsf{q}|}
\frac{(q^{\frac34+\frac{|\mathsf{q}|}{2}}tz^{\pm};q)_{\infty}}{(q^{\frac14+\frac{|\mathsf{q}|}{2}}t^{-1}z^{\pm};q)_{\infty}}
\end{align}
when $\mathsf{p}=\mathsf{q}$. 
The overall factor $q^{\frac{|\mathsf{q}|}{4}}t^{|\mathsf{q}|}$ will correspond to an insertion of a flavor monopole operator of magnetic charge $\mathsf{q}$ acting on the vortex state. 
While the line defect index (\ref{SQED1Wn}) vanishes for $\mathsf{p}\neq 0$ in the Coulomb limit, it reduces to $\mathfrak{t}^\mathsf{p}$ in the Higgs limit. 
This factor is simply interpreted as the contribution from $\mathsf{p}$ fundamental hypermultiplet scalar fields with total gauge charge $-\mathsf{p}$.

\subsection{$T[SU(2)]$}
\label{sec_TSU2}
The 3d $\mathcal{N}=4$ $U(1)$ gauge theory with two fundamental hypers $(I_1,J_1)$ and $(I_2,J_2)$ is referred to as the $T[SU(2)]$ \cite{Gaiotto:2008ak}. 
It is self-mirror theory in that its Coulomb branch and Higgs branch are equivalent. 
The one-point function of the charged Wilson line in $T[SU(2)]$ is evaluated as
\begin{align}
\label{SQED2Wn}
&
\langle W_\mathsf{p} \rangle^{U(1)-[2]}(t,x,z;q)
\nonumber\\
&=
\frac{(q^{\frac12}t^{2};q)_{\infty}}{(q^{\frac12}t^{-2};q)_{\infty}}
\sum_{m\in \mathbb{Z}}\oint \frac{ds}{2\pi is}
\frac{(q^{\frac34+\frac{|m|}{2}}t^{-1}s^{\mp}x^{\mp};q)_{\infty} (q^{\frac34+\frac{|m|}{2}}t^{-1}s^{\mp}x^{\pm};q)_{\infty}}
{(q^{\frac14+\frac{|m|}{2}}ts^{\pm}x^{\pm};q)_{\infty} (q^{\frac14+\frac{|m|}{2}}ts^{\pm}x^{\mp};q)_{\infty}}
s^\mathsf{p} 
q^{\frac{|m|}{2}}t^{-2|m|}z^{2m}. 
\end{align}
The integral contains two towers of simple poles in $|s|<1$ at 
$s$ $=$ $q^{\frac14+\frac{|m|}{2}+l}tx^{-1}$ and $q^{\frac14+\frac{|m|}{2}+l}tx$ 
with $l=0,1,\cdots$. 
Summing over the residues at these poles, the expression (\ref{SQED2Wn}) becomes 
\begin{align}
&
\langle W_\mathsf{p} \rangle^{U(1)-[2]}(t,x,z;q)
=\frac{(q^{\frac12}t^2;q)_{\infty}^2}{(q)_{\infty}^2}
\sum_{m\in \mathbb{Z}}q^{\frac{|m|}{2}}t^{-2|m|}z^{2m}
\Biggl[
\frac{(q^{\frac12}t^{-2}x^2;q)_{\infty}(q^{\frac12}t^2x^{-2};q)_{\infty}}
{(x^2;q)_{\infty}(qx^{-2};q)_{\infty}}
\nonumber\\
&\times 
\sum_{l=0}^{\infty}
\frac{(q^{1+l};q)_{\infty}(q^{1+|m|+l};q)_{\infty}(q^{1+l}x^{-2};q)_{\infty}(q^{1+|m|+l}x^{-2};q)_{\infty}}
{(q^{\frac12+l}t^2;q)_{\infty}(q^{\frac12+|m|+l}t^2;q)_{\infty}(q^{\frac12+l}t^2x^{-2};q)_{\infty}(q^{\frac12+|m|+l}t^2x^{-2};q)_{\infty}}
\nonumber\\
&\times q^{\frac{\mathsf{p}}{4}+l+\frac{\mathsf{p}|m|}{2}+l\mathsf{p}} t^{\mathsf{p}-4l} x^{-\mathsf{p}}+(x\leftrightarrow x^{-1})
\Biggr]. 
\end{align}
The first term in the expansion is 
\begin{align}
\left(\sum_{i=0}^{\mathsf{p}}x^{\mathsf{p}-2i}\right)q^{\frac{\mathsf{p}}{4}}t^{\mathsf{p}}
\end{align}
This simply enumerates the following $(\mathsf{p}+1)$ Higgs branch operator with charge $-\mathsf{p}$ attached with the charged Wilson line: 
\begin{align}
J_1^\mathsf{p}, \quad J_1^{\mathsf{p}-1}J_2, \quad J_1^{\mathsf{p}-2}J_2^2, \quad \cdots, \quad J_1J_2^{\mathsf{p}-1}, \quad J_2^{\mathsf{p}},
\end{align}
where the hypermultiplet complex scalar fields $J_1$ and $J_2$ carry gauge charge $-1$. 
For $\mathsf{p}=1$ the second term in the expansion is 
\begin{align}
(x^3+x+x^{-1}+x^{-3})q^{\frac34}t^3. 
\end{align}
This again corresponds to the Higgs branch operators
\begin{align}
J_1^2 I_1, \quad, J_1^2 I_2, \quad J_2^2 I_1, \quad J_2^2 I_2, 
\end{align}
where $I_1$ and $I_2$ are the hypermultiplet scalar fields with gauge charge $+1$. 
When $\mathsf{p}>1$, the second term with $q^{\frac{\mathsf{p}}{4}+\frac{1}{2}}$ in the expansion takes the form 
\begin{align}
\left[
-\left(\sum_{i=0}^{\mathsf{p}-2} x^{\mathsf{p}-2i-2}\right)t^{\mathsf{p}-2}
+\left(\sum_{i=0}^{\mathsf{p}+2} x^{\mathsf{p}-2i+2}\right)t^{\mathsf{p}+2}
\right]q^{\frac{\mathsf{p}}{4}+\frac12}, 
\end{align}
which contains further fermionic contributions to the index. 
For example, for $\mathsf{p}=2$ we have the term $-q$ corresponding to the following two fermionic and one bosonic operators carrying charge $-2$: 
\begin{align}
J_1\psi_{I_1},\quad J_2\psi_{I_2},\quad \varphi J_1 J_2, 
\end{align}
where $\psi_{I_i}$ is the fermionic superpartners of $I_i$ and $\varphi$ is the vector multiplet scalar field. 

Next consider the line defect indices for the vortex lines. 
The vortex lines in the $T[SU(2)]$ are characterized by the flavor vortex number $\mathsf{q}$. 
\footnote{In \cite{Assel:2015oxa} the index $i=0,1,2$ is introduced to describe three possible splitting of the two hypers, 
i.e. the number of the hypermultiplets with singular configurations $2=i+(2-i)$. 
It can be determined once we fix all the vortex charges as it simply encodes the splitting of the fundamental hypers. }
The one-point function of the vortex line with vortex number $\mathsf{q}$ is given by 
\begin{align}
\label{SQED2V1_n}
&
\langle V_{\mathsf{q}}\rangle^{U(1)-[2]}(t,x,z;q)
\nonumber\\
&=\frac{(q^{\frac12}t^{2};q)_{\infty}}{(q^{\frac12}t^{-2};q)_{\infty}}
\sum_{m\in \mathbb{Z}}
\oint \frac{ds}{2\pi is}
\frac{(q^{\frac34+\frac{|m+\mathsf{q}|}{2}}t^{-1}s^{\mp}x^{\mp};q)_{\infty}}
{(q^{\frac14+\frac{|m+\mathsf{q}|}{2}}ts^{\pm}x^{\pm};q)_{\infty}}
\frac{(q^{\frac34+\frac{|m|}{2}}t^{-1}s^{\mp}x^{\pm};q)_{\infty}}
{(q^{\frac14+\frac{|m|}{2}}ts^{\pm}x^{\mp};q)_{\infty}}
\nonumber\\
&\times 
q^{\frac{|m|+|m+\mathsf{q}|}{4}}
t^{-|m|-|m+\mathsf{q}|}
z^{2m+\mathsf{q}}. 
\end{align}
Under the mirror transformation
\begin{align}
\label{mirror_trans}
t&\rightarrow t^{-1}, \qquad 
x\rightarrow z, \qquad
z\rightarrow x, 
\end{align}
the Wilson line defect index (\ref{SQED2Wn}) for $\mathsf{p}=n$
agrees with the vortex line defect index (\ref{SQED2V1_n}) for $\mathsf{q}=n$ 
\begin{align}
\label{SQED2_WVid}
\langle W_{\mathsf{p}=n} \rangle^{U(1)-[2]}(t,x,z;q)
&=\langle V_{\mathsf{q}=n}\rangle^{U(1)-[2]}(t^{-1},z,x;q). 
\end{align}
This provides us with strong evidence for the conjecture \cite{Assel:2015oxa} that 
the charged Wilson line $W_{\mathsf{p}=n}$ is dual to the vortex line $V_{\mathsf{q}=n}$!

We can give the proof for the Higgs and Coulomb limits of the proposed identity (\ref{SQED2_WVid}). 
Though the one-point function (\ref{SQED2Wn}) of the charged Wilson line vanishes in the Coulomb limit, 
it is non-trivial in the Higgs limit
\begin{align}
\langle W_\mathsf{p} \rangle^{U(1)-[2]}_H(x;\mathfrak{t})
&=(1-\mathfrak{t}^2)\oint \frac{ds}{2\pi is}
\frac{1}{(1-\mathfrak{t}sx^{\pm})(1-\mathfrak{t}s^{-1}x^{\pm})}s^\mathsf{p}. 
\end{align}
By picking up the residues at two poles $s=\mathfrak{t}x$ and $\mathfrak{t}x^{-1}$, we get
\begin{align}
\label{SQED2Wn_H}
\langle W_\mathsf{p} \rangle^{U(1)-[2]}_H(x;\mathfrak{t})
&=\frac{\mathfrak{t}^\mathsf{p} x^\mathsf{p}}{(1-\mathfrak{t}^2x^2)(1-x^{-2})}
+\frac{\mathfrak{t}^\mathsf{p} x^{-\mathsf{p}}}{(1-\mathfrak{t}^2x^{-2})(1-x^{2})}
\nonumber\\
&=\frac{\mathfrak{t}^\mathsf{p} x^{-\mathsf{p}}(1-\mathfrak{t}^2x^2+\mathfrak{t}^2x^{2\mathsf{p}}-x^{2\mathsf{p}+2})}
{(1-x^2)(1-\mathfrak{t}^2x^2)(1-\mathfrak{t}^2x^{-2})}. 
\end{align}
On the other hand, the line defect index (\ref{SQED2V1_n}) for the vortex line is non-trivial in the Coulomb limit. 
It becomes
\begin{align}
\langle V_{\mathsf{q}}\rangle_{C}^{U(1)-[2]}(z;\mathfrak{t})
&=\frac{1}{1-\mathfrak{t}^2}\sum_{m\in \mathbb{Z}} 
\mathfrak{t}^{|m|+|m+\mathsf{q}|}z^{2m+\mathsf{q}}
\nonumber\\
&=\frac{1}{1-\mathfrak{t}^2}
\left[
\sum_{m=-\infty}^{-\mathsf{q}}
+\sum_{m=-(\mathsf{q}-1)}^{-1}
+\sum_{m=0}^{\infty}
\right]\mathfrak{t}^{|m|+|m+\mathsf{q}|}z^{2m+1}
\nonumber\\
&=\frac{1}{1-\mathfrak{t}^2}
\left[
\frac{\mathfrak{t}^\mathsf{q}z^{-\mathsf{q}}}{1-\mathfrak{t}^2z^{-2}}
+\frac{\mathfrak{t}^\mathsf{q} z^{-\mathsf{q}}(z^2-z^{2\mathsf{q}})}{1-z^2}
+\frac{\mathfrak{t}^\mathsf{q} z^\mathsf{q}}{1-\mathfrak{t}^2z^2}
\right]
\nonumber\\
&=\frac{\mathfrak{t}^\mathsf{q} z^{-\mathsf{q}}(1-\mathfrak{t}^2z^2+\mathfrak{t}^2z^{2\mathsf{q}}-z^{2\mathsf{q}+2})}
{(1-z^2)(1-\mathfrak{t}^2z^2)(1-\mathfrak{t}^2z^{-2})}. 
\end{align}
Thus
\begin{align}
\langle W_{\mathsf{p}=n} \rangle^{U(1)-[2]}_H(x;\mathfrak{t})
&=\langle V_{\mathsf{q}=n}\rangle_{C}^{U(1)-[2]}(x;\mathfrak{t}). 
\end{align}

Furthermore, we can consider the one-point function of the vortex-Wilson line operator $L_{\mathsf{p};\mathsf{q}}$ 
of electric gauge charge $\mathsf{p}$ and of the vortex number $\mathsf{q}$ that takes the form 
\begin{align}
\label{SQED2WnVn}
&
\langle L_{\mathsf{p};\mathsf{q}}\rangle^{U(1)-[2]}(t,x,z;q)
\nonumber\\
&=\frac{(q^{\frac12}t^{2};q)_{\infty}}{(q^{\frac12}t^{-2};q)_{\infty}}
\sum_{m\in \mathbb{Z}}
\oint \frac{ds}{2\pi is}
\frac{(q^{\frac34+\frac{|m+\mathsf{q}|}{2}}t^{-1}s^{\mp}x^{\mp};q)_{\infty}}
{(q^{\frac14+\frac{|m+\mathsf{q}|}{2}}ts^{\pm}x^{\pm};q)_{\infty}}
\frac{(q^{\frac34+\frac{|m|}{2}}t^{-1}s^{\mp}x^{\pm};q)_{\infty}}
{(q^{\frac14+\frac{|m|}{2}}ts^{\pm}x^{\mp};q)_{\infty}}
\nonumber\\
&\times 
s^{\mathsf{p}}
q^{\frac{|m|+|m+\mathsf{q}|}{4}}
t^{-|m|-|m+\mathsf{q}|}
z^{2m+\mathsf{q}}. 
\end{align}
It follows that when $\mathsf{p}=\mathsf{q}$, the line defect index (\ref{SQED2WnVn}) is invariant under the mirror transformation (\ref{mirror_trans}). 
This would imply that the vortex-Wilson line $L_{\mathsf{p}=n;\mathsf{q}=n}$ is self-mirror.

\subsection{SQED$_3$}
For the 3d $\mathcal{N}=4$ SQED$_3$, i.e. the $U(1)$ gauge theory with three fundamental hypers, 
the theory is mirror to the $U(1)\times U(1)$ linear quiver gauge theory with 
a bifundamental twisted hyper as well as a fundamental twisted hyper coupled to each of the gauge nodes. 

The one-point function of the charged Wilson line is 
\footnote{Here the flavor fugacities $z_a$ in the expression are redundant.
However, they are useful to enumerate the BPS local operators. }
\begin{align}
\label{SQED3Wn}
&
\langle W_\mathsf{p} \rangle^{U(1)-[3]}(t,x_\alpha,z_{a};q)
\nonumber\\
&=
\frac{(q^{\frac12}t^{2};q)_{\infty}}{(q^{\frac12}t^{-2};q)_{\infty}}
\sum_{m\in \mathbb{Z}}\oint \frac{ds}{2\pi is}
\prod_{\alpha=1}^3
\frac{(q^{\frac34+\frac{|m|}{2}}t^{-1}s^{\mp}x_{\alpha}^{\mp};q)_{\infty}}
{(q^{\frac14+\frac{|m|}{2}}ts^{\pm}x_{\alpha}^{\pm};q)_{\infty}}
s^\mathsf{p} 
q^{\frac{3|m|}{4}}t^{-3|m|}
\left(\frac{z_1}{z_2}\right)^{m}, 
\end{align}
with $x_1x_2x_3=1$. 

The mirror $U(1)\times U(1)$ linear quiver gauge theory has the $U(1)$ flavor symmetry rotating the a single twisted hypermultiplet (see Figure \ref{fig:Sdualbrane1}).
The one-point function of the flavor vortex line of the vortex number $\mathsf{q}$ can be written as the following form: 
\begin{align}
\label{mSQED3Vn}
&
\langle V_{\mathsf{q}}\rangle^{\widetilde{[1]-U(1)-U(1)-[1]}}(t,z_{a},x_\alpha;q)
\nonumber\\
&=\frac{(q^{\frac12}t^{-2};q)_{\infty}^2}{(q^{\frac12}t^2;q)_{\infty}^2}
\sum_{m^{(1)},m^{(2)}\in \mathbb{Z}}
\oint \frac{ds^{(1)}}{2\pi is^{(1)}}
\oint \frac{ds^{(2)}}{2\pi is^{(2)}}
\nonumber\\
&\times 
\frac{(q^{\frac34+\frac{|m^{(1)}|}{2}}ts^{(1)\mp}z_1^{\pm};q)_{\infty}}
{(q^{\frac14+\frac{|m^{(1)}|}{2}}t^{-1}s^{(1)\pm}z_1^{\mp};q)_{\infty}}
\frac{(q^{\frac34+\frac{|m^{(1)}-m^{(2)}|}{2}}ts^{(1)\mp}s^{(2)\pm};q)_{\infty}}
{(q^{\frac14+\frac{|m^{(1)}-m^{(2)}|}{2}}t^{-1}s^{(1)\pm}s^{(2)\mp};q)_{\infty}}
\nonumber\\
&\times 
\frac{(q^{\frac34+\frac{|m^{(2)}+\mathsf{q}|}{2}}ts^{(2)\mp}z_2^{\pm};q)_{\infty}}
{(q^{\frac14+\frac{|m^{(2)}+\mathsf{q}|}{2}}t^{-1}s^{(2)\pm}z_2^{\mp};q)_{\infty}}
\nonumber\\
&\times 
q^{\frac{|m^{(1)}|+|m^{(1)}-m^{(2)}|+|m^{(2)}+\mathsf{q}|}{4}}
t^{|m^{(1)}|+|m^{(1)}-m^{(2)}|+|m^{(2)}+\mathsf{q}|}
\left(\frac{x_1}{x_2}\right)^{m^{(1)}}
\left(\frac{x_2}{x_3}\right)^{m^{(2)}}
x_3^{-\mathsf{q}}. 
\end{align} 

We find that the line defect indices (\ref{SQED3Wn}) and (\ref{mSQED3Vn}) agree with each other 
when $\mathsf{p}=\mathsf{q}=n$
\begin{align}
\langle W_{\mathsf{p}=n} \rangle^{U(1)-[3]}(t,x_\alpha,z_{a};q)
&=\langle V_{\mathsf{q}=n}\rangle^{\widetilde{[1]-U(1)-U(1)-[1]}}(t,z_{a},x_\alpha;q). 
\end{align}
This demonstrates the duality of the charged Wilson line in the SQED$_3$ 
and the flavor vortex line in the mirror $U(1)\times U(1)$ linear quiver gauge theory! 

In the Coulomb limit, the one-point function (\ref{SQED3Wn}) vanishes. 
In the Higgs limit, it reduces to 
\begin{align}
\langle W_\mathsf{p} \rangle^{U(1)-[3]}_{H}(x_{\alpha};\mathfrak{t})
&=(1-\mathfrak{t}^2)\oint \frac{ds}{2\pi is}
\prod_{\alpha=1}^3\frac{1}{(1-\mathfrak{t}s^{\pm}x_{\alpha}^{\pm})}s^\mathsf{p}. 
\end{align}
This can be computed as a sum over the residues at three poles 
$s=\mathfrak{t}y_{\alpha}^{-1}$, $\alpha=1,2,3$
\begin{align}
\langle W_\mathsf{p} \rangle^{U(1)-[3]}_{H}(x_{\alpha};\mathfrak{t})
&=\sum_{\alpha=1}^{3}\frac{\mathfrak{t}^\mathsf{p} x_{\alpha}^{-\mathsf{p}}}
{\prod_{\beta\neq \alpha}(1-\mathfrak{t}^2\frac{x_{\beta}}{x_{\alpha}})(1-\frac{x_{\alpha}}{x_{\beta}})}. 
\end{align}

Next consider the flavor vortex line in the SQED$_3$. 
As the theory has $SU(3)$ flavor symmetry rotating three hypermultiplets, 
it is labeled by a pair of the vortex numbers $(\mathsf{q}_1,\mathsf{q}_2)$ which encodes the orders of singularities of hypermultiplets. 
According to the prescription (\ref{vortex_shift}), 
the one-point function of the vortex line with $(\mathsf{q}_1,\mathsf{q}_2)$ $=$ $(n^{(1)}+n^{(2)},n^{(2)})$ is given by 
\begin{align}
\label{SQED3Vn1n2}
&
\langle V_{n^{(1)}+n^{(2)},n^{(2)}} \rangle^{U(1)-[3]}(t,x_\alpha,z_{a};q)
\nonumber\\
&=
\frac{(q^{\frac12}t^{2};q)_{\infty}}{(q^{\frac12}t^{-2};q)_{\infty}}
\sum_{m\in \mathbb{Z}}\oint \frac{ds}{2\pi is}
\frac{(q^{\frac34+\frac{|m+n^{(1)}+n^{(2)}|}{2}}t^{-1}s^{\mp}x_{1}^{\mp};q)_{\infty}}
{(q^{\frac14+\frac{|m+n^{(1)}+n^{(2)}|}{2}}ts^{\pm}x_{1}^{\pm};q)_{\infty}}
\frac{(q^{\frac34+\frac{|m+n^{(2)}|}{2}}t^{-1}s^{\mp}x_{2}^{\mp};q)_{\infty}}
{(q^{\frac14+\frac{|m+n^{(2)}|}{2}}ts^{\pm}x_{2}^{\pm};q)_{\infty}}
\nonumber\\
&\times 
\frac{(q^{\frac34+\frac{|m|}{2}}t^{-1}s^{\mp}x_{3}^{\mp};q)_{\infty}}
{(q^{\frac14+\frac{|m|}{2}}ts^{\pm}x_{3}^{\pm};q)_{\infty}}
q^{\frac{|m+n^{(1)}+n^{(2)}|+|m+n^{(2)}|+|m|}{4}}t^{-|m+n^{(1)}+n^{(2)}|-|m+n^{(2)}|-|m|}
z_1^{m+n^{(1)}+n^{(2)}}z_2^{-m}. 
\end{align}

For the mirror $U(1)\times U(1)$ linear quiver gauge theory, there is a Wilson line of charge $(\mathsf{p}^{(1)},\mathsf{p}^{(2)})$. 
The one-point function can be evaluated as
\begin{align}
\label{mSQED3Wn1n2}
&
\langle W_{\mathsf{p}^{(1)},\mathsf{p}^{(2)}}\rangle^{\widetilde{[1]-U(1)-U(1)-[1]}}(t,z_a,x_\alpha;q)
=\frac{(q^{\frac12}t^{-2};q)_{\infty}^2}{(q^{\frac12}t^2;q)_{\infty}^2}
\sum_{m^{(1)},m^{(2)}\in \mathbb{Z}}
\oint \frac{ds^{(1)}}{2\pi is^{(1)}}
\oint \frac{ds^{(2)}}{2\pi is^{(2)}}
\nonumber\\
&\times 
\frac{(q^{\frac34+\frac{|m^{(1)}|}{2}}ts^{(1)\mp}z_1^{\pm};q)_{\infty}}
{(q^{\frac14+\frac{|m^{(1)}|}{2}}t^{-1}s^{(1)\pm}z_1^{\mp};q)_{\infty}}
\frac{(q^{\frac34+\frac{|m^{(1)}-m^{(2)}|}{2}}ts^{(1)\mp}s^{(2)\pm};q)_{\infty}}
{(q^{\frac14+\frac{|m^{(1)}-m^{(2)}|}{2}}t^{-1}s^{(1)\pm}s^{(2)\mp};q)_{\infty}}
\frac{(q^{\frac34+\frac{|m^{(2)}|}{2}}ts^{(2)\mp}z_2^{\pm};q)_{\infty}}
{(q^{\frac14+\frac{|m^{(2)}|}{2}}t^{-1}s^{(2)\pm}z_2^{\mp};q)_{\infty}}
\nonumber\\
&\times 
s^{(1)\mathsf{p}^{(1)}} s^{(2)\mathsf{p}^{(2)}}
q^{\frac{|m^{(1)}|+|m^{(1)}-m^{(2)}|+|m^{(2)}|}{4}}
t^{|m^{(1)}|+|m^{(1)}-m^{(2)}|+|m^{(2)}|}
\left(\frac{x_1}{x_2}\right)^{m^{(1)}}
\left(\frac{x_2}{x_3}\right)^{m^{(2)}}. 
\end{align}

We find that the two expressions (\ref{SQED3Vn1n2}) and (\ref{mSQED3Wn1n2}) coincide 
when $(\mathsf{q}_1,\mathsf{q}_2)$ $=$ $(n^{(1)}+n^{(2)},n^{(2)})$ 
and $(\mathsf{p}^{(1)},\mathsf{p}^{(2)})$ $=$ $(n^{(1)},n^{(2)})$
\begin{align}
\langle V_{n^{(1)}+n^{(2)},n^{(2)}} \rangle^{U(1)-[3]}(t,x_\alpha,z_{a};q)
&=\langle W_{n^{(1)},n^{(2)}}\rangle^{\widetilde{[1]-U(1)-U(1)-[1]}}(t,z_a,x_\alpha;q). 
\label{SQED3V=mirrorW}
\end{align}
This equality states that 
the flavor vortex line with a pair of charges $(n^{(1)}+n^{(2)},n^{(2)})$ 
is dual to the Wilson line of a pair of electric charges $(n^{(1)},n^{(2)})$! 
See also appendix \ref{app_proofHClinear} for an analytic proof of the equality \eqref{SQED3V=mirrorW} in the Coulomb and Higgs limits.
In the Higgs limit both sides of \eqref{SQED3V=mirrorW} vanishes, while in the Coulomb limit we obtain by using the formula \eqref{SQEDCoulomb}
\begin{align}
&\langle V_{n^{(1)}+n^{(2)},n^{(2)}}\rangle^{U(1)\text{-}[3]}_C(z_\alpha;\mathfrak{t})=
\langle W_{n^{(1)},n^{(2)}}\rangle^{\widetilde{[1]-U(1)-U(1)-[1]}}_C(z_\alpha;\mathfrak{t})\nonumber \\
&=
\frac{
z_1^{n^{(1)}}z_2^{n^{(2)}}
\mathfrak{t}^{n^{(1)}+n^{(2)}}(1-\mathfrak{t}^2)}{\prod_\pm(1-z^{\pm 1}\mathfrak{t})}
\Bigl(
1
-\frac{z^{-n^{(1)}-1}\mathfrak{t}^{n^{(1)}+1}(1-z\mathfrak{t})}{1-z^{-1}\mathfrak{t}^3}
-\frac{z^{n^{(2)}+1}\mathfrak{t}^{n^{(2)}+1}(1-z^{-1}\mathfrak{t})}{1-z\mathfrak{t}^3}
\Bigr)
\end{align}
for $n^{(1)},n^{(2)}\ge 0$, where $z=\frac{z_1}{z_2}$.


\subsection{SQED$_l$}

Now consider the general 3d $\mathcal{N}=4$ SQED$_l$ 
with gauge group $U(1)$ and $l$ fundamental hypermultiplets $(I_{\alpha},J_{\alpha})$, $\alpha=1,\cdots,l$.\footnote{
In order to reproduce the convention for $l=1$ in section \ref{sec_SQED1},
we need to set $x_1=1$ and $\frac{z_1}{z_2}=z$.
Similarly, in order to reproduce the convention for $l=2$ in section \ref{sec_TSU2}, we need to set $x_1=x$, $x_2=x^{-1}$ and $z_1=z$, $z_2=z^{-1}$.
}
The theory is mirror to the Abelian linear quiver gauge theory of gauge group $U(1)^{\otimes l-1}$ 
and bifundamental twisted hypers between the adjacent gauge nodes 
as well as a single twisted hyper charged under each end of the gauge nodes as depicted in Figure \ref{fig:Sdualbrane1}.

One has the one-point function of the charged Wilson line of the form
\begin{align}
\label{SQEDlWn}
&
\langle W_\mathsf{p} \rangle^{U(1)-[l]}(t,x_\alpha,z_{a};q)
\nonumber\\
&=
\frac{(q^{\frac12}t^{2};q)_{\infty}}{(q^{\frac12}t^{-2};q)_{\infty}}
\sum_{m\in \mathbb{Z}}\oint \frac{ds}{2\pi is}
\prod_{\alpha=1}^l 
\frac{(q^{\frac34+\frac{|m|}{2}}t^{-1}s^{\mp}x_{\alpha}^{\mp};q)_{\infty}}
{(q^{\frac14+\frac{|m|}{2}}ts^{\pm}x_{\alpha}^{\pm};q)_{\infty}}
s^\mathsf{p} 
q^{\frac{l|m|}{4}}t^{-l|m|}
\left(\frac{z_1}{z_2}\right)^{m}, 
\end{align}
with $\prod_{\alpha} x_{\alpha}=1$. 
The first term in the expansion that counts the Higgs branch operator with gauge charge $-\mathsf{p}$ is given by
\begin{align}
h_\mathsf{p}(x_1^{-1},\cdots,x_l^{-1})q^{\frac{\mathsf{p}}{4}}t^\mathsf{p}, 
\end{align}
where $h_\mathsf{p}(x_1,\cdots,x_l)$ is the complete homogeneous symmetric function of degree $\mathsf{p}$ in $l$ variables. 
In other words, it is the $\left(\begin{smallmatrix}\mathsf{p}+l-1\\ \mathsf{p}\\ \end{smallmatrix}\right)$ Higgs branch operator as the $\mathsf{p}$-th symmetric tensor power of the $SU(l)$ fundamental hypermultiplets. 

In the mirror $U(1)^{\otimes l-1}$ linear quiver gauge theory 
there exist the flavor vortex lines for the $U(1)$ flavor symmetry that rotates the twisted hypermultiplet. 
As this can be realized by a stack of $\mathsf{q}$ D1-branes stretched between a D3-brane and a probe NS5-branes, it is conjectured to be dual to the charged Wilson line in the SQED$_l$. 
The one-point function of the flavor vortex line in the mirror theory reads
\begin{align}
\label{mSQEDlVn}
&
\langle V_{\mathsf{q}}\rangle^{\widetilde{[1]-U(1)^{\otimes l-1}-[1]}}(t,z_{a},x_\alpha;q)
\nonumber\\
&=\frac{(q^{\frac12}t^{-2};q)_{\infty}^{l-1}}{(q^{\frac12}t^2;q)_{\infty}^{l-1}}
\sum_{m^{(1)},\cdots, m^{(l-1)}\in \mathbb{Z}}
\prod_{I=1}^{l-1}
\oint \frac{ds^{(I)}}{2\pi is^{(I)}}
\frac{(q^{\frac34+\frac{|m^{(1)}|}{2}}ts^{(1)\mp}z_1^{\pm};q)_{\infty}}
{(q^{\frac14+\frac{|m^{(1)}|}{2}}t^{-1}s^{(1)\pm}z_1^{\mp};q)_{\infty}}\nonumber \\
&\quad\times \prod_{I=1}^{l-2}
\frac{(q^{\frac34+\frac{|m^{(I)}-m^{(I+1)}|}{2}}ts^{(I)\mp}s^{(I+1)\pm};q)_{\infty}}
{(q^{\frac14+\frac{|m^{(I)}-m^{(I+1)}|}{2}}t^{-1}s^{(I)\pm}s^{(I+1)\mp};q)_{\infty}}
\frac{(q^{\frac34+\frac{|m^{(l-1)}+\mathsf{q}|}{2}}ts^{(l)\mp}z_2^{\pm};q)_{\infty}}
{(q^{\frac14+\frac{|m^{(l-1)}+\mathsf{q}|}{2}}t^{-1}s^{(l)\pm}z_2^{\mp};q)_{\infty}}\nonumber \\
&\quad\times q^{\frac{|m^{(1)}|+\sum_{I=1}^{l-2}|m^{(I)}-m^{(I+1)}|+|m^{(l-1)}+\mathsf{q}|}{4}}
t^{|m^{(1)}|+\sum_{I=1}^{l-2}|m^{(I)}-m^{(I+1)}|+|m^{(l-1)}+\mathsf{q}|}
\prod_{I=1}^{l-1}
\left(\frac{x_{I}}{x_{I+1}}\right)^{m^{(I)}}
x_{l}^{-\mathsf{q}}. 
\end{align}

We conjecture that the correlators (\ref{SQEDlWn}) and (\ref{mSQEDlVn}) are identical 
when $\mathsf{p}$ $=$ $\mathsf{q}$ $=$ $n$ 
so that the following duality holds: 
\begin{align}
\begin{matrix}
\textrm{gauge $W_{\mathsf{p}=n}$}\\
 \\
\textrm{in $U(1)-[l]$}\\
\end{matrix}
\qquad 
\leftrightarrow 
\qquad 
\begin{matrix}
\textrm{flavor $V_{\mathsf{q}=n}$}\\ 
 \\ 
\textrm{in $\widetilde{[1]-U(1)^{\otimes l-1}-[1]}$}\\
\end{matrix}. 
\end{align}

Next, we propose a generalization of the vortex line defect indices for the SQED$_l$. 
The theory has an $SU(l)$ flavor symmetry.  
The flavor vortex lines are indexed by $(l-1)$ vortex numbers $\{\mathsf{q}_{\alpha}\}_{\alpha=1}^{l-1}$ $\in$ $\mathbb{Z}^{l-1}$. 
When we relabel the charges as $\mathsf{q}_{\alpha}$ $=$ $\sum_{I=\alpha}^{l-1}n^{(I)}$, the one-point function of the vortex line is given by
\begin{align}
\label{SQEDlVnI}
&
\langle V_{\mathsf{q}_{\alpha}=\sum_{I=\alpha}^{l-1}n^{(I)}} \rangle^{U(1)-[l]}(t,x_\alpha,z_{a};q)
\nonumber\\
&=
\frac{(q^{\frac12}t^{2};q)_{\infty}}{(q^{\frac12}t^{-2};q)_{\infty}}
\sum_{m\in \mathbb{Z}}\oint \frac{ds}{2\pi is}
\prod_{\alpha=1}^{l-1}
\frac{(q^{\frac34+\frac{|m+\sum_{I=\alpha}^{l-1}n^{(I)}|}{2}}t^{-1}s^{\mp}x_{\alpha}^{\mp};q)_{\infty}}
{(q^{\frac14+\frac{|m+\sum_{I=\alpha}^{l-1}n^{(I)}|}{2}}ts^{\pm}x_{\alpha}^{\pm};q)_{\infty}}
\frac{(q^{\frac34+\frac{|m|}{2}}t^{-1}s^{\mp}x_{l}^{\mp};q)_{\infty}}
{(q^{\frac14+\frac{|m|}{2}}ts^{\pm}x_{l}^{\pm};q)_{\infty}}
\nonumber\\
&\quad\times 
q^{\frac{\sum_{\alpha=1}^{l-1}|m+\sum_{I=\alpha}^{l-1}n^{(I)}|+|m|}{4}}
t^{-\sum_{\alpha=1}^{l-1}|m+\sum_{I=\alpha}^{l-1}n^{(I)}|-|m|}
z_1^{m+\sum_{I=1}^{l-1}n^{(I)}}z_2^{-m}. 
\end{align}

On the other hand, 
the mirror $U(1)^{\otimes l-1}$ linear quiver gauge theory admits the charged Wilson lines 
which carry a set of the electric charges $(\mathsf{p}^{(1)},\cdots, \mathsf{p}^{(l-1)})$. 
The one-point function of the Wilson line can be computed as
\begin{align}
\label{mSQEDlWnI}
&
\langle W_{\mathsf{p}^{(I)}}\rangle^{\widetilde{[1]-U(1)^{\otimes l-1}-[1]}}(t,z_a,x_\alpha;q)\nonumber \\
&=\frac{(q^{\frac12}t^{-2};q)_{\infty}^{l-1}}{(q^{\frac12}t^2;q)_{\infty}^{l-1}}
\sum_{m^{(1)},\cdots, m^{(l-1)}\in \mathbb{Z}}
\prod_{I=1}^{l-1}
\oint \frac{ds^{(I)}}{2\pi is^{(I)}}
\frac{(q^{\frac34+\frac{|m^{(1)}|}{2}}ts^{(1)\mp}z_1^{\pm};q)_{\infty}}
{(q^{\frac14+\frac{|m^{(1)}|}{2}}t^{-1}s^{(1)\pm}z_1^{\mp};q)_{\infty}}\nonumber \\
&\quad\times \prod_{I=1}^{l-2}
\frac{(q^{\frac34+\frac{|m^{(I)}-m^{(I+1)}|}{2}}ts^{(I)\mp}s^{(I+1)\pm};q)_{\infty}}
{(q^{\frac14+\frac{|m^{(I)}-m^{(I+1)}|}{2}}t^{-1}s^{(I)\pm}s^{(I+1)\mp};q)_{\infty}}
\frac{(q^{\frac34+\frac{|m^{(l-1)}|}{2}}ts^{(l-1)\mp}z_2^{\pm};q)_{\infty}}
{(q^{\frac14+\frac{|m^{(l-1)}|}{2}}t^{-1}s^{(l-1)\pm}z_2^{\mp};q)_{\infty}}
\prod_{I=1}^{l-1}
s^{(I)\mathsf{p}^{(I)}}\nonumber \\
&\quad\times q^{\frac{|m^{(1)}|+\sum_{I=1}^{l-2}|m^{(I)}-m^{(I+1)}|+|m^{(l-1)}|}{4}}
t^{|m^{(1)}|+\sum_{I=1}^{l-2}|m^{(I)}-m^{(I+1)}|+|m^{(l-1)}|}
\prod_{I=1}^{l-1}
\left(
\frac{x_I}{x_{I+1}}
\right)^{m^{(I)}}. 
\end{align}

We conjecture that the two expressions (\ref{SQEDlVnI}) and (\ref{mSQEDlWnI}) are equal 
when $\mathsf{q}_{\alpha}$ $=$ $\sum_{I=\alpha}^{l-1}n^{(I)}$ 
and $\mathsf{p}^{(I)}$ $=$ $n^{(I)}$ 
as a consequence of the following vortex-Wilson duality: 
\begin{align}
\begin{matrix}
\textrm{flavor $V_{\mathsf{q}_{\alpha}=\sum_{I=\alpha}^{l-1}n^{(I)}}$}\\
 \\
\textrm{in $U(1)-[l]$}\\
\end{matrix}
\qquad 
\leftrightarrow 
\qquad 
\begin{matrix}
\textrm{gauge $W_{\mathsf{p}^{(I)}=n^{(I)}}$}\\ 
 \\ 
\textrm{in $\widetilde{[1]-U(1)^{\otimes l-1}-[1]}$}\\
\end{matrix}. 
\end{align}

Combining (\ref{SQEDlWn}) and (\ref{SQEDlVnI}), 
the vortex-Wilson line defect index for the SQED$_l$ takes the form
\begin{align}
\label{SQEDlLnm}
&
\langle L_{\mathsf{p};\mathsf{q}_{\alpha}=\sum_{I=\alpha}^{l-1}n^{(I)}} \rangle^{U(1)-[l]}(t,x_\alpha,z_{a};q)
\nonumber\\
&=
\frac{(q^{\frac12}t^{2};q)_{\infty}}{(q^{\frac12}t^{-2};q)_{\infty}}
\sum_{m\in \mathbb{Z}}\oint \frac{ds}{2\pi is}
\prod_{\alpha=1}^{l-1}
\frac{(q^{\frac34+\frac{|m+\sum_{I=\alpha}^{l-1}n^{(I)}|}{2}}t^{-1}s^{\mp}x_{\alpha}^{\mp};q)_{\infty}}
{(q^{\frac14+\frac{|m+\sum_{I=\alpha}^{l-1}n^{(I)}|}{2}}ts^{\pm}x_{\alpha}^{\pm};q)_{\infty}}
\frac{(q^{\frac34+\frac{|m|}{2}}t^{-1}s^{\mp}x_{l}^{\mp};q)_{\infty}}
{(q^{\frac14+\frac{|m|}{2}}ts^{\pm}x_{l}^{\pm};q)_{\infty}}
\nonumber\\
&\quad\times 
s^{\mathsf{p}} 
q^{\frac{\sum_{\alpha=1}^{l-1}|m+\sum_{I=\alpha}^{l-1}n^{(I)}|+|m|}{4}}
t^{-\sum_{\alpha=1}^{l-1}|m+\sum_{I=\alpha}^{l-1}n^{(I)}|-|m|}
z_1^{m+\sum_{I=1}^{l-1}n^{(I)}}z_2^{-m},
\end{align}
while the vortex-Wilson line defect index of the mirror theory is
\begin{align}
\label{mSQEDlLnm}
&
\langle L_{\mathsf{p}^{(I)};\mathsf{q}}\rangle^{\widetilde{[1]-U(1)^{\otimes l-1}-[1]}}(t,z_{a},x_\alpha;q)
\nonumber\\
&=\frac{(q^{\frac12}t^{-2};q)_{\infty}^{l-1}}{(q^{\frac12}t^2;q)_{\infty}^{l-1}}
\sum_{m^{(1)},\cdots, m^{(l-1)}\in \mathbb{Z}}
\prod_{I=1}^{l-1}
\oint \frac{ds^{(I)}}{2\pi is^{(I)}}
\frac{(q^{\frac34+\frac{|m^{(1)}|}{2}}ts^{(1)\mp}z_1^{\pm};q)_{\infty}}
{(q^{\frac14+\frac{|m^{(1)}|}{2}}t^{-1}s^{(1)\pm}z_1^{\mp};q)_{\infty}}
\nonumber \\
&\quad\times
\prod_{I=1}^{l-2}
\frac{(q^{\frac34+\frac{|m^{(I)}-m^{(I+1)}|}{2}}ts^{(I)\mp}s^{(I+1)\pm};q)_{\infty}}
{(q^{\frac14+\frac{|m^{(I)}-m^{(I+1)}|}{2}}t^{-1}s^{(I)\pm}s^{(I+1)\mp};q)_{\infty}}
\frac{(q^{\frac34+\frac{|m^{(l-1)}+\mathsf{q}|}{2}}ts^{(l-1)\mp}z_2^{\pm};q)_{\infty}}
{(q^{\frac14+\frac{|m^{(l-1)}+\mathsf{q}|}{2}}t^{-1}s^{(l-1)\pm}z_2^{\mp};q)_{\infty}}
\nonumber \\
&\quad\times
\prod_{I=1}^{l-1}
s^{(I)\mathsf{p}^{(I)}}
q^{\frac{|m^{(1)}|+\sum_{I=1}^{l-2}|m^{(I)}-m^{(I+1)}|+|m^{(l-1)}+\mathsf{q}|}{4}}
t^{|m^{(1)}|+\sum_{I=1}^{l-2}|m^{(I)}-m^{(I+1)}|+|m^{(l-1)}+\mathsf{q}|}
\nonumber \\
&\quad\times
\prod_{I=1}^{l-1}
\left(\frac{x_{I}}{x_{I+1}}\right)^{m^{(I)}}
x_{l}^{-\mathsf{q}}.
\end{align}
For these vortex-Wilson line defect indices, we conjecture
\begin{align}
\langle L_{\mathsf{p}=n;\mathsf{q}_\alpha=\sum_{I=\alpha}^{l-1}n^{(I)}} \rangle^{U(1)-[l]}(t,x_\alpha,z_{a};q)=\langle L_{\mathsf{p}^{(I)}=n^{(I)};\mathsf{q}=n}\rangle^{\widetilde{[1]-U(1)^{\otimes l-1}-[1]}}(t,z_{a},x_\alpha;q).
\end{align}

\subsection{$U(1)-U(1)-[1]$}
\label{sec_u1u1h1}
There is another type of the dualities, the Seiberg-like dualities in 3d $\mathcal{N}=4$ gauge theories. 
Let us consider the $\mathcal{N}=4$ quiver gauge theory with $U(1)\times U(1)$ gauge group and a single hyper in the fundamental representation under one of the $U(1)$ gauge nodes. 
The index is given by
\begin{align}
\label{u1u1-1_ind}
&
\mathcal{I}^{U(1)-U(1)-[1]}(t,x_\alpha;q)
\nonumber\\
&=
\frac{(q^{\frac12}t^2;q)_{\infty}^2}{(q^{\frac12}t^{-2};q)_{\infty}^2}
\sum_{m^{(1)},m^{(2)}\in \mathbb{Z}}
\oint \frac{ds^{(1)}}{2\pi is^{(1)}}
\oint \frac{ds^{(2)}}{2\pi is_2} 
\frac{(q^{\frac34+\frac{|m^{(1)}-m^{(2)}|}{2}}t^{-1}s^{(1)\mp}s^{(2)\pm};q)_{\infty}}
{(q^{\frac14+\frac{|m^{(1)}-m^{(2)}|}{2}}ts^{(1)\pm}s^{(2)\mp};q)_{\infty}}
\nonumber\\
&\times \frac{(q^{\frac34+\frac{|m^{(1)}|}{2}}t^{-1}s^{(1)\mp};q)_{\infty}}
{(q^{\frac14+\frac{|m^{(1)}|}{2}}ts^{(1)\pm};q)_{\infty}}
q^{\frac{|m^{(1)}-m^{(2)}|+|m^{(1)}|}{4}}t^{-|m^{(1)}-m^{(2)}|-|m^{(1)}|}x_1^{m^{(1)}}x_2^{m^{(2)}}. 
\end{align}
We find that the index (\ref{u1u1-1_ind}) is given as
\begin{align}
\mathcal{I}^{U(1)-U(1)-[1]}(t,x_\alpha;q)=
\frac{(q^{\frac34}tx_1^{\mp}x_2^{\mp};q)_{\infty} (q^{\frac34}tx_2^{\mp};q)_{\infty}}
{(q^{\frac14}t^{-1}x_1^{\pm}x_2^{\pm};q)_{\infty} (q^{\frac14}t^{-1}x_2^{\pm};q)_{\infty}}. 
\end{align}
This is the index for a free theory of a pair of the twisted hypermultiplets carrying the charges $(+,+)$ and $(0,+)$ 
under the $U(1)\times U(1)$ topological symmetry. 
This indicates that the $\mathcal{N}=4$ $U(1)\times U(1)$ quiver gauge theory with a fundamental hyper coupled to one of the gauge nodes is dual 
to a theory of a pair of the twisted hypermultiplets. 

The expression (\ref{u1u1-1_ind}) contains a tower of poles at $\frac{s^{(2)}}{s^{(1)}}$ $=$ $q^{\frac14+\frac{|m^{(1)}-m^{(2)}|}{2}+l}t$ with $l=0,1,\cdots$, corresponding to the bosonic modes of the bifundamental hypermultiplet. 
By picking up the residues at these poles, we can expand the index (\ref{u1u1-1_ind}) as
\begin{align}
&
\mathcal{I}^{U(1)-U(1)-[1]}(t,x_\alpha;q)
\nonumber\\
&=\frac{(q^{\frac12}t^2;q)_{\infty}^3}{(q)_{\infty}^2 (q^{\frac12}t^{-2};q)_{\infty}}
\sum_{\mathsf{q}\ge 0}
\sum_{\substack{m^{(1)},m^{(2)}\in\mathbb{Z},l\ge 0\\ (|m^{(1)}-m^{(2)}|+2l=\mathsf{q})}}
q^{\frac{|m^{(1)}-m^{(2)}|}{4}+\frac{|m^{(1)}|}{4}}t^{-|m^{(1)}-m^{(2)}|-|m^{(1)}|}x_1^{m^{(1)}}x_2^{m^{(2)}}
\nonumber\\
&\times 
\frac{(q^{1+l};q)_{\infty}(q^{1+|m^{(1)}-m^{(2)}|+l};q)_{\infty}}
{(q^{\frac12+|m^{(1)}-m^{(2)}|+l}t^2;q)_{\infty} (q^{\frac12+l}t^2;q)_{\infty}}
q^{\frac{l}{2}}t^{-2l}\oint \frac{ds}{2\pi is}
\frac{(q^{\frac{3}{4}+\frac{|m^{(1)}|}{2}}t^{-1}s^{\mp};q)_{\infty}}
{(q^{\frac{1}{4}+\frac{|m^{(1)}|}{2}}ts^{\pm};q)_{\infty}},
\end{align}
where we have separated the summation over the residues by the location of the poles $\frac{s(2)}{s(1)}=q^{\frac{1}{4}+\frac{\mathsf{q}}{2}}$ with $\mathsf{q}=0,1,\cdots$.
As argued in \cite{Gaiotto:2019jvo}, this expansion is associated with the Higgsing process that gives rise to the vev of the bifundamental hypermultiplet in such a way that the NS5-brane is removed from a single D3-brane. 
As a consequence, the $U(1)\times U(1)$ gauge group breaks down to $U(1)$ and a whole 3d $\mathcal{N}=4$ $U(1)$ with one flavor is left.
Picking the residues only at $\mathsf{q}=0$, one should set $l=0$ and reduce the sum over $m^{(1)}$ and $m^{(2)}$ to the sum over a single magnetic flux $m^{(1)}=m^{(2)}=m$.
As a result we find
\begin{align}
\mathcal{I}^{U(1)-U(1)-[1]}_{\textrm{Higgsed $\mathsf{q}=0$}}
&=
\frac{(q^{\frac12}t^2;q)_{\infty}}{(q^{\frac12}t^{-2};q)_{\infty}}
\sum_{m\in \mathbb{Z}}
\oint \frac{ds}{2\pi is}
\frac{(q^{\frac{3}{4}+\frac{|m|}{2}}t^{-1}s^{\mp};q)_{\infty}}
{(q^{\frac{1}{4}+\frac{|m|}{2}}ts^{\pm};q)_{\infty}}
q^{\frac{|m|}{4}}t^{-|m|}z^{m}. 
\end{align}
This is identified with the index of the SQED$_1$.
The other terms with $\mathsf{q}>0$ are expected to have an interpretation as contributions from configuration involving the BPS vortices.
One can introduce the Wilson line with charge $(\mathsf{p}^{(1)}, \mathsf{p}^{(2)})$ $=$ $(n^{(1)}, n^{(2)})$ under the $U(1)\times U(1)$ gauge group in the theory. 
We find that the one-point function can be written as
\begin{align}
&
\langle W_{n^{(1)},n^{(2)}}\rangle^{U(1)-U(1)-[1]}(t,x_\alpha;q)
\nonumber\\
&=q^{\frac{|n^{(1)}+2n^{(2)}|}{4}}
t^{|n^{(1)}+2n^{(2)}|}
\frac{(q^{\frac34+\frac{|n^{(1)}+n^{(2)}|}{2}}tx_1^{\mp}x_2^{\mp};q)_{\infty}}
{(q^{\frac14+\frac{|n^{(1)}+n^{(2)}|}{2}}t^{-1}x_1^{\pm}x_2^{\pm};q)_{\infty}}
\frac{(q^{\frac34+\frac{|n^{(2)}|}{2}}tx_2^{\mp};q)_{\infty}}
{(q^{\frac14+\frac{|n^{(2)}|}{2}}t^{-1}x_2^{\pm};q)_{\infty}}. 
\end{align}
This can be viewed as the line defect index of the flavor vortex line $V_{\mathsf{q}_1,\mathsf{q}_2}$ 
with charges $(\mathsf{q}_1,\mathsf{q}_2)$ $=$ $(n^{(1)}+n^{(2)},n^{(2)})$ 
under the flavor symmetries $U(1)\times U(1)$ attached with the flavor monopole of charge $n^{(1)}+2n^{(2)}$ 
for the theory of a pair of two free hypers. 

\subsection{$U(2)-[3]$}
\label{sec_u2h3}
The 3d $\mathcal{N}=4$ $U(N)$ gauge theory with $(2N-1)$ flavors is conjecturally 
equivalent to a product of the $U(N-1)$ gauge theory with $(2N-1)$ hypers and the theory of a free twisted hyper \cite{Gaiotto:2008ak}. 
This can be viewed as the Seiberg-like duality of the 3d $\mathcal{N}=4$ gauge theories. 
When $N=2$, the dual description is simply given by the SQED$_3$ and the free twisted hyper. 
Here we examine the Seiberg-like dualities of the line defects. 

Consider the Wilson line $W_{(n,n)}$ transforming in the irreducible representation 
whose highest weight is labeled by the $(n\times 2)$ rectangular Young diagram corresponding to the partition $(n,n)$ for the $U(2)$ gauge theory with three flavors. 
The one-point function of the Wilson line in the rectangular representation is given by
\begin{align}
\label{u2Nf3Wnn}
&
\langle W_{(n,n)}\rangle^{U(2)-[3]}(t,x_\alpha,z;q)
\nonumber\\
&=\frac12 \frac{(q^{\frac12}t^2;q)_{\infty}^2}{(q^{\frac12}t^{-2};q)_{\infty}^2}
\sum_{m_1,m_2\in \mathbb{Z}}
\oint \prod_{i=1}^2\frac{ds_i}{2\pi is_i}
(1-q^{\frac{|m_1-m_2|}{2}}s_1^{\pm}s_2^{\mp})
\frac{(q^{\frac12+\frac{|m_1-m_2|}{2}}t^2s_1^{\mp}s_2^{\pm};q)_{\infty}}
{(q^{\frac12+\frac{|m_1-m_2|}{2}}t^{-2}s_1^{\pm}s_2^{\mp};q)_{\infty}}
\nonumber\\
&\times 
\prod_{i=1}^{2}\prod_{\alpha=1}^{3}
\frac{(q^{\frac34+\frac{|m_i|}{2}}t^{-1}s_i^{\mp}x_{\alpha}^{\mp};q)_{\infty}}
{(q^{\frac14+\frac{|m_i|}{2}}ts_i^{\pm}x_{\alpha}^{\pm};q)_{\infty}}
s_i^{n}
q^{\frac34\sum_{i=1}^{2}|m_i|-\frac{|m_1-m_2|}{2}}
t^{-3\sum_{i=1}^2|m_i|+2|m_1-m_2|}z^{\sum_{i=1}^2m_i}. 
\end{align}
We find that 
\begin{align}
\langle W_{(n,n)}\rangle^{U(2)-[3]}(t,x_\alpha,z;q)
&=\langle W_{-n} \rangle^{U(1)-[3]}(t,x_\alpha,z^{\frac12};q)
\times 
\langle V_n \rangle^{\textrm{thyper}}(t,z;q), 
\end{align}
where $\langle W_{n} \rangle^{U(1)-[3]}$ and $\langle V_n \rangle^{\textrm{thyper}}$ 
are the SQED$_3$ Wilson line defect index (\ref{SQED3Wn}) and the twisted hyper vortex line defect index (\ref{thypVn}). 
This generalizes the equality of the indices in \cite{Okazaki:2019ony} 
and indicates the Seiberg-like duality of the line defect operators. 
In order to figure out more general Seiberg-like dualities of line operators, 
one needs to work on gauge theories with higher rank gauge groups. 
We leave more detailed analysis of the line defect indices for the non-Abelian gauge theories to future work.

\section{Circular quiver gauge theories}
\label{sec_ADHM}

\subsection{$U(1)$ ADHM$-[1]$}
As we have seen in section \ref{sec_branesetup}, when a single D3-brane wrapped on $S^1$ intersects with an NS5-brane and $l$ D5-branes, 
the low-energy effective theory on the D3-brane is given by the $U(1)$ ADHM theory, 
i.e. the 3d $\mathcal{N}=4$ supersymmetric gauge theory with gauge group $U(1)$, $l$ hypermultiplets of charge $+1$ and an neutral hypermultiplet. 
The theory also describes a single M2-brane probing $\mathbb{C}^2\times \mathbb{C}^2/\mathbb{Z}_l$ in M-theory. 

We begin with the case with $l=1$, for which the theory is self-mirror. 
The theory has enhanced $\mathcal{N}=8$ supersymmetry. 
The one-point function of the charged Wilson line $W_\mathsf{p}$ takes the form
\begin{align}
\label{ADHM1_Wn}
&
\langle W_\mathsf{p}\rangle^{\textrm{$U(1)$ ADHM-$[1]$}}(t,x,z;q)
\nonumber\\
&=
\frac{(q^{\frac12}t^2;q)_{\infty}}
{(q^{\frac12}t^{-2};q)_{\infty}}
\frac{(q^{\frac34}t^{-1}x^{\mp};q)_{\infty}}{(q^{\frac14}tx^{\pm};q)_{\infty}}
\sum_{m\in \mathbb{Z}}
\oint \frac{ds}{2\pi is}
\frac{(q^{\frac34+\frac{|m|}{2}}t^{-1}s^{\mp};q)_{\infty}}
{(q^{\frac14+\frac{|m|}{2}}ts^{\pm};q)_{\infty}}
s^{\mathsf{p}} q^{\frac{|m|}{4}}t^{-|m|}z^{m}. 
\end{align}
While the expression (\ref{ADHM1_Wn}) for the one-point function of the charged Wilson line for the $U(1)$ ADHM theory 
only differs from that for the SQED by the factor of the single neutral hypermultiplet, 
the associated duality of the line operators are distinguished from that for the SQED. 

Let us also consider
\begin{align}
\label{ADHM1_Vn}
&
\langle V_{\mathsf{q}_0}\rangle^{\widetilde{\textrm{$U(1)$ ADHM-$[1]$}}}(t,z,x;q)
\nonumber\\
&=
\frac{(q^{\frac12}t^{-2};q)_{\infty}}
{(q^{\frac12}t^{2};q)_{\infty}}
\frac{(q^{\frac34+\frac{|\mathsf{q}_0|}{2}}tz^{\mp};q)_{\infty}}{(q^{\frac14+\frac{|\mathsf{q}_0|}{2}}t^{-1}z^{\pm};q)_{\infty}}
\sum_{m\in \mathbb{Z}}
\oint \frac{ds}{2\pi is}
\frac{(q^{\frac34+\frac{|m|}{2}}ts^{\mp};q)_{\infty}}
{(q^{\frac14+\frac{|m|}{2}}t^{-1}s^{\pm};q)_{\infty}}
q^{\frac{|m|}{4}+\frac{|\mathsf{q}_0|}{4}}t^{|m|+|\mathsf{q}_0|}x^{m},
\end{align}
which is identified with the one-point function of the flavor vortex line $V_{\mathsf{q}_0}$ of charge $\mathsf{q}_0$ 
for the Higgs branch symmetry rotating the neutral hyper in the $U(1)$ ADHM theory with one flavor.
Then we find that the expression \eqref{ADHM1_Wn} agrees with \eqref{ADHM1_Vn} 
for $\mathsf{p}=\mathsf{q}_0=n$, both of which is evaluated as
\begin{align}
\label{ADHM1_Vn2}
&\langle W_{\mathsf{p}=n}\rangle^{\textrm{$U(1)$ ADHM-$[1]$}}(t,x,z;q)=
\langle V_{\mathsf{q}_0=n}\rangle^{\widetilde{\textrm{$U(1)$ ADHM-$[1]$}}}(t,z,x;q)\nonumber \\
&=
q^{\frac{|n|}{4}}t^{|n|}
\frac{(q^{\frac34}t^{-1}x^{\mp};q)_{\infty}}{(q^{\frac14}tx^{\pm};q)_{\infty}}
\frac{(q^{\frac34+\frac{|n|}{2}}tz^{\pm};q)_{\infty}}{(q^{\frac14+\frac{|n|}{2}}t^{-1}z^{\pm};q)_{\infty}}. 
\end{align}
This can be viewed as the one-point function of the flavor vortex line of charge $n$ 
associated with the flavor symmetry rotating the twisted hypers in the dual free theory. 

It is tempting to consider the one-point function of the mixed vortex-Wilson line operator $L_{\mathsf{p};\mathsf{q}_0}$ with electric gauge charge $\mathsf{p}$ and the vortex number $\mathsf{q}_0$ of the following form: 
\begin{align}
\label{ADHM1_VWn}
&\langle L_{\mathsf{p};\mathsf{q}_0}\rangle^{\textrm{$U(1)$ ADHM-$[1]$}}(t,x,z;q)\nonumber \\
&=
\frac{(q^{\frac12}t^2;q)_{\infty}}
{(q^{\frac12}t^{-2};q)_{\infty}}
\sum_{m\in \mathbb{Z}}
\frac{(q^{\frac34+\frac{|\mathsf{q}_0|}{2}}t^{-1}x^{\mp};q)_{\infty}}{(q^{\frac14+\frac{|\mathsf{q}_0|}{2}}tx^{\pm};q)_{\infty}}
\oint \frac{ds}{2\pi is}
\frac{(q^{\frac34+\frac{|m|}{2}}t^{-1}s^{\mp};q)_{\infty}}
{(q^{\frac14+\frac{|m|}{2}}ts^{\pm};q)_{\infty}}
s^{\mathsf{p}} q^{\frac{|m|}{4}+\frac{|\mathsf{q}_0|}{4}}t^{-|m|-|\mathsf{q}_0|}z^{m}. 
\end{align}
We find that the expression (\ref{ADHM1_VWn}) agrees with 
\begin{align}
\label{ADHM1_VWn2}
\langle L_{\mathsf{p};\mathsf{q}_0}\rangle^{\textrm{$U(1)$ ADHM-$[1]$}}(t,x,z;q)=q^{\frac{|\mathsf{p}|+|\mathsf{q}_0|}{4}}t^{|\mathsf{p}|-|\mathsf{q}_0|}
\frac{(q^{\frac34+\frac{|\mathsf{p}|}{2}}t^{-1}x^{\mp};q)_{\infty}}{(q^{\frac14+\frac{|\mathsf{p}|}{2}}tx^{\pm};q)_{\infty}}
\frac{(q^{\frac34+\frac{|\mathsf{q}_0|}{2}}tz^{\pm};q)_{\infty}}{(q^{\frac14+\frac{|\mathsf{q}_0|}{2}}t^{-1}z^{\pm};q)_{\infty}}. 
\end{align}
When $\mathsf{p}$ $=$ $\mathsf{q}_0$, the expression (\ref{ADHM1_VWn}) is invariant under the mirror transformation (\ref{mirror_trans}). 
This indicates that the corresponding line defect operator $L_{\mathsf{p}=n;\mathsf{q}_0=n}$ for the $U(1)$ ADHM with one flavor is self-mirror. 

\subsection{$U(1)$ ADHM$-[2]$}
For $l=2$, the $U(1)$ ADHM theory is mirror to 
the circular quiver gauge theory of $U(1)\times U(1)$ gauge group coupled 
to a pair of bifundamental twisted hypermultiplets and a single charged twisted hypermultiplet at one of the $U(1)$ gauge nodes (see Figure \ref{fig:Sdualbrane2}).

The line defect index of the charged Wilson line in the ADHM theory is 
\begin{align}
\label{ADHM2_Wn}
&
\langle W_{\mathsf{p}}\rangle^{\textrm{$U(1)$ ADHM-$[2]$}}(t,x,y_{\alpha},z;q)
\nonumber\\
&=
\frac{(q^{\frac12}t^2;q)_{\infty}}
{(q^{\frac12}t^{-2};q)_{\infty}}
\sum_{m\in \mathbb{Z}}
\frac{(q^{\frac34}t^{-1}x^{\mp};q)_{\infty}}{(q^{\frac14}tx^{\pm};q)_{\infty}}
\oint \frac{ds}{2\pi is}
\prod_{\alpha=1}^{2}
\frac{(q^{\frac34+\frac{|m|}{2}}t^{-1}s^{\mp}y_{\alpha}^{\mp};q)_{\infty}}
{(q^{\frac14+\frac{|m|}{2}}ts^{\pm}y_{\alpha}^{\pm};q)_{\infty}}
s^{\mathsf{p}} q^{\frac{|m|}{2}}t^{-2|m|}z^{2m}, 
\end{align}
with $y_1y_2$ $=$ $1$. 

The mirror circular $U(1)\times U(1)$ quiver gauge theory has the global symmetry rotating the bifundamental twisted hyper. 
This admits a non-trivial flavor vortex line. 
The one-point function of the flavor vortex line with charge $\mathsf{q}$ is evaluated as
\begin{align}
\label{mADHM2_Vn}
&{\langle V_\mathsf{q} \rangle^{\widetilde{U(1)^{\otimes2}\textrm{mADHM-}[1]}}}(t,y_{\alpha},z,x;q)
\nonumber\\
&=\frac{(q^{\frac12}t^{-2};q)_{\infty}^2}{(q^{\frac12}t^2;q)_{\infty}^2}
\sum_{m^{(1)},m^{(2)}\in \mathbb{Z}}
\oint \frac{ds^{(1)}}{2\pi is^{(1)}}
\oint \frac{ds^{(2)}}{2\pi is^{(2)}}
\frac{(q^{\frac34+\frac{|m^{(1)}|}{2}}ts^{(1)\mp};q)_{\infty}}
{(q^{\frac14+\frac{|m^{(1)}|}{2}}t^{-1}s^{(1)\pm};q)_{\infty}}
\nonumber\\
&\quad \times 
\frac{
(q^{\frac34+\frac{|m^{(1)}-m^{(2)}|}{2}}ts^{(1)\mp} s^{(2)\pm} z^{\mp};q)_{\infty}
}{
(q^{\frac14+\frac{|m^{(1)}-m^{(2)}|}{2}}t^{-1}s^{(1)\pm} s^{(2)\mp} z^{\pm};q)_{\infty}
}
\frac{
(q^{\frac34+\frac{|m^{(2)}-m^{(1)}+\mathsf{q}|}{2}}ts^{(2)\mp} s^{(1)\pm} z^{\mp};q)_{\infty}
}{
(q^{\frac14+\frac{|m^{(2)}-m^{(1)}+\mathsf{q}|}{2}}t^{-1}s^{(2)\pm} s^{(1)\mp} z^{\pm};q)_{\infty}
}\nonumber \\
&\quad\times 
q^{\frac{|m^{(1)}|}{4}+\frac{|m^{(1)}-m^{(2)}|}{4}+\frac{|m^{(2)}-m^{(1)}+\mathsf{q}|}{4}}
t^{|m^{(1)}|+|m^{(1)}-m^{(2)}|+|m^{(2)}-m^{(1)}+\mathsf{q}|}
x^{m^{(1)}}
y_1^{-(m^{(2)}-m^{(1)}+\mathsf{q})}\nonumber \\
&\quad\times y_2^{m^{(2)}-m^{(1)}}.
\end{align}

We find that the vortex line defect correlator (\ref{ADHM2_Wn}) is in agreement with the Wilson line defect correlator (\ref{mADHM2_Vn}) 
when $\mathsf{p}=\mathsf{q}=n$
\begin{align}
\langle W_{\mathsf{p}=n}\rangle^{\textrm{$U(1)$ ADHM-$[2]$}}(t,x,y_{\alpha},z;q)
&={\langle V_{\mathsf{q}=n} \rangle^{\widetilde{U(1)^{\otimes2}\textrm{mADHM-}[1]}}}(t,y_{\alpha},z,x;q). 
\end{align}
This supports the duality of the charged Wilson line in the $U(1)$ ADHM theory with two flavors 
and the flavor vortex line in the mirror $U(1)\times U(1)$ quiver. 

Next consider the vortex lines in the $U(1)$ ADHM theory with two flavors. 
The theory has two types of the flavor symmetry rotating the neutral hyper and the fundamental hypers 
so that the flavor vortex lines are characterized by a set of charges 
$(\mathsf{q}_0,\mathsf{q}_1,\mathsf{q}_2)$. 
Here $\mathsf{q}_0$ stands for the vortex charge for the neutral hyper and $\mathsf{q}_\alpha$, $\alpha=1,2$ 
stands for the vortex charge for the $\alpha$-th fundamental hyper.
However, one of the charges is screened by the gauge vortex 
so that we fix $\mathsf{q}_1$ to zero without loss of generality. 
The vortex line defect index reads
\begin{align}
\label{ADHM2_Vn}
&
\langle V_{\mathsf{q}_0,\mathsf{q}_\alpha}\rangle^{\textrm{$U(1)$ ADHM-$[2]$}}(t,x,y_{\alpha},z;q)
\nonumber\\
&=
\frac{(q^{\frac12}t^2;q)_{\infty}}
{(q^{\frac12}t^{-2};q)_{\infty}}
\sum_{m\in \mathbb{Z}}
\frac{(q^{\frac{3+|\mathsf{q}_0|}{4}}t^{-1}x^{\mp};q)_{\infty}}
{(q^{\frac{1+|\mathsf{q}_0|}{4}}tx^{\pm};q)_{\infty}}
\oint \frac{ds}{2\pi is}
\frac{(q^{\frac34+\frac{|m|}{2}}t^{-1}s^{\mp}y_{1}^{\mp};q)_{\infty}}
{(q^{\frac14+\frac{|m|}{2}}ts^{\pm}y_{1}^{\pm};q)_{\infty}}
\nonumber \\
&\quad\times 
\frac{(q^{\frac34+\frac{|m+\mathsf{q}_2|}{2}}t^{-1}s^{\mp}y_{2}^{\mp};q)_{\infty}}
{(q^{\frac14+\frac{|m+\mathsf{q}_2|}{2}}ts^{\pm}y_{2}^{\pm};q)_{\infty}}
q^{\frac{|m|}{4}+\frac{|m+\mathsf{q}_2|}{4}+\frac{|\mathsf{q}_0|}{4}}
t^{-|m|-|m+\mathsf{q}_2|-|\mathsf{q}_0|}
z^{2m+\mathsf{q}_2}. 
\end{align}

For the mirror circular quiver gauge theory 
there exist the Wilson lines carrying a pair of electric charges $(\mathsf{p}^{(1)},\mathsf{p}^{(2)})$ under the $U(1)\times U(1)$ gauge group. 
The one-point function of the Wilson line is evaluated as
\begin{align}
\label{mADHM2_Wn}
&{\langle W_{\mathsf{p}^{(1)},\mathsf{p}^{(2)}}\rangle^{\widetilde{U(1)^{\otimes2}\textrm{mADHM-}[1]}}}(t,y_{\alpha},z,x;q)
\nonumber\\
&=\frac{(q^{\frac12}t^{-2};q)_{\infty}^2}{(q^{\frac12}t^2;q)_{\infty}^2}
\sum_{m^{(1)},m^{(2)}\in \mathbb{Z}}
\oint \frac{ds^{(1)}}{2\pi is^{(1)}}
\oint \frac{ds^{(2)}}{2\pi is^{(2)}}
{s^{(1)}}^{\mathsf{p}^{(1)}} {s^{(2)}}^{\mathsf{p}^{(2)}}
\frac{(q^{\frac34+\frac{|m^{(1)}|}{2}}ts^{(1)\mp};q)_{\infty}}
{(q^{\frac14+\frac{|m^{(1)}|}{2}}t^{-1}s^{(1)\pm};q)_{\infty}}\nonumber \\
&\quad\times \frac{(q^{\frac34+\frac{|m^{(1)}-m^{(2)}|}{2}}ts^{(1)\mp}s^{(2)\pm}z^{\mp};q)_{\infty}}
{(q^{\frac14+\frac{|m^{(1)}-m^{(2)}|}{2}}t^{-1}s^{(1)\pm}s^{(2)\mp}z^{\pm};q)_{\infty}}
\frac{(q^{\frac34+\frac{|m^{(2)}-m^{(1)}|}{2}}ts^{(2)\mp}s^{(1)\pm}z^{\mp};q)_{\infty}}
{(q^{\frac14+\frac{|m^{(2)}-m^{(1)}|}{2}}t^{-1}s^{(2)\pm}s^{(1)\mp}z^{\pm};q)_{\infty}}\nonumber \\
&\quad\times q^{\frac{|m^{(1)}|}{4}+\frac{|m^{(1)}-m^{(2)}|}{2}}
t^{|m^{(1)}|+2|m^{(1)}-m^{(2)}|}
x^{m^{(1)}}
\left(\frac{y_1}{y_2}\right)^{m^{(1)}}
\left(\frac{y_2}{y_1}\right)^{m^{(2)}}.
\end{align}

Again we find the agreement of the vortex line defect index (\ref{ADHM2_Vn}) for the $U(1)$ ADHM with two flavors 
and the Wilson line defect index (\ref{mADHM2_Wn}) for its mirror 
when $(\mathsf{q}_0,\mathsf{q}_1,\mathsf{q}_2)$ $=$ $(n^{(1)}+n^{(2)},0,n^{(2)})$ 
and $(\mathsf{p}^{(1)},\mathsf{p}^{(2)})$ $=$ $(n^{(1)},n^{(2)})$
\begin{align}
&\langle V_{\mathsf{q}_0=n^{(1)}+n^{(2)},\mathsf{q}_1=0,\mathsf{q}_2=n^{(2)}}\rangle^{\textrm{$U(1)$ ADHM-$[2]$}}(t,x,y_{\alpha},z;q)\nonumber \\
&={\langle W_{\mathsf{p}_1=n^{(1)},\mathsf{p}_2=n^{(2)}}\rangle^{\widetilde{U(1)^{\otimes2}\textrm{mADHM-}[1]}}}(t,y_{\alpha},z,x;q). 
\end{align}
This supports the opposite type of vortex-Wilson duality under mirror symmetry! 

\subsection{$U(1)$ ADHM$-[l]$}

Now we are ready to discuss the general $U(1)$ ADHM theory. 
The theory is mirror to the circular $U(1)^{\otimes l}$ quiver gauge theory 
with the bifundamental twisted hypers between the adjacent gauge nodes and a single twisted hyper charged under the first gauge node as depicted in Figure \ref{fig:Sdualbrane2}.

The one-point function of the charged Wilson line in the $U(1)$ ADHM theory with $l$ flavors is given by
\begin{align}
\label{ADHMl_Wn}
&
\langle W_{\mathsf{p}}\rangle^{\textrm{$U(1)$ ADHM-$[l]$}}(t,x,y_{\alpha},z;q)
\nonumber\\
&=
\frac{(q^{\frac12}t^2;q)_{\infty}}
{(q^{\frac12}t^{-2};q)_{\infty}}
\sum_{m\in \mathbb{Z}}
\frac{(q^{\frac34}t^{-1}x^{\mp};q)_{\infty}}{(q^{\frac14}tx^{\pm};q)_{\infty}}
\oint \frac{ds}{2\pi is}
\prod_{\alpha=1}^{l}
\frac{(q^{\frac34+\frac{|m|}{2}}t^{-1}s^{\mp}y_{\alpha}^{\mp};q)_{\infty}}
{(q^{\frac14+\frac{|m|}{2}}ts^{\pm}y_{\alpha}^{\pm};q)_{\infty}}
s^{\mathsf{p}} q^{\frac{l|m|}{4}}t^{-l|m|}z^{lm}, 
\end{align}
with $\prod_{\alpha}y_{\alpha}$ $=$ $1$. 

In the mirror circular $U(1)^{\otimes l}$ quiver gauge theory 
there is the global symmetry rotating the charged bifundamental twisted hyper that allows for the associated flavor vortex lines. 
According to the prescription (\ref{vortex_shift}), 
the one-point function of the flavor vortex line with charge $\mathsf{q}$ takes the form
\begin{align}
\label{mADHMl_Vn}
&{\langle V_\mathsf{q} \rangle^{\widetilde{U(1)^{\otimes l}\textrm{mADHM-}[1]}}}(t,y_{\alpha},z,x;q)
\nonumber\\
&=\frac{(q^{\frac12}t^{-2};q)_{\infty}^l}{(q^{\frac12}t^2;q)_{\infty}^l}
\sum_{m^{(1)},\cdots,m^{(l)}\in \mathbb{Z}}
\prod_{I=1}^{l}
\oint \frac{ds^{(I)}}{2\pi is^{(I)}}
\frac{(q^{\frac34+\frac{|m^{(1)}|}{2}}ts^{(1)\mp};q)_{\infty}}
{(q^{\frac14+\frac{|m^{(1)}|}{2}}t^{-1}s^{(1)\pm};q)_{\infty}}
\nonumber\\
&\quad\times 
\prod_{I=1}^{l-1}
\frac{(q^{\frac34+\frac{|m^{(I)}-m^{(I+1)}|}{2}}ts^{(I)\mp}s^{(I+1)\pm}z^{\mp};q)_{\infty}}
{(q^{\frac14+\frac{|m^{(I)}-m^{(I+1)}|}{2}}t^{-1}s^{(I)\pm}s^{(I+1)\mp}z^{\pm};q)_{\infty}}
\frac{(q^{\frac34+\frac{|m^{(l)}-m^{(1)}+\mathsf{q}|}{2}}ts^{(l)\mp}s^{(1)\pm}z^{\mp};q)_{\infty}}
{(q^{\frac14+\frac{|m^{(l)}-m^{(1)}+\mathsf{q}|}{2}}t^{-1}s^{(l)\pm}s^{(1)\mp}z^{\pm};q)_{\infty}}
\nonumber\\
&\quad\times 
q^{\frac{|m^{(1)}|}{4}+\sum_{I=1}^{l-1}\frac{|m^{(I)}-m^{(I+1)}|}{4}+\frac{|m^{(l)}-m^{(1)}+\mathsf{q}|}{4}}
t^{|m^{(1)}|+\sum_{I=1}^{l-1}|m^{(I)}-m^{(I+1)}|+|m^{(l)}-m^{(1)}+\mathsf{q}|}
x^{m^{(1)}}\nonumber \\
&\quad\times
y_1^{-(m^{(l)}-m^{(1)}+\mathsf{q})}
\prod_{\alpha=2}^ly_\alpha^{m^{(\alpha)}-m^{(\alpha-1)}}.
\end{align}

We propose matching of the line defect indices (\ref{ADHMl_Wn}) and (\ref{mADHMl_Vn}) 
for $\mathsf{p}$ $=$ $\mathsf{q}$ $=$ $n$
which states the following duality of line operators: 
\begin{align}
\begin{matrix}
\textrm{gauge $W_{\mathsf{p}=n}$}\\
 \\
\textrm{in $U(1)$ ADHM-$[l]$}\\
\end{matrix}
\qquad 
\leftrightarrow 
\qquad 
\begin{matrix}
\textrm{flavor $V_{\mathsf{q}=n}$}\\ 
 \\ 
\textrm{in $\widetilde{U(1)^{\otimes l}\textrm{mADHM-}[1]}$}\\
\end{matrix}. 
\label{ADHMWmirrorVduality}
\end{align}

Let us turn to the opposite type of the line operators for a pair of the $U(1)$ ADHM theory and its mirror. 
In the $U(1)$ ADHM theory we have the global symmetries rotating the neutral hyper and the $(l-1)$ charged hypers. 
When we introduce the vortex charges $\mathsf{q}_{\alpha}$, $\alpha=1,\cdots, l$ for the latter, 
we can set $\mathsf{q}_1$ to zero, corresponding to the one screened by dynamical vortices. 
Then the flavor vortex lines can be labeled by a set of $l$ charges 
$(\mathsf{q}_0,\mathsf{q}_2,\cdots,\mathsf{q}_l)$ $\in$ $\mathbb{Z}^{l}$.
Relabeling the charges as
\begin{align}
\mathsf{q}_0=\sum_{I=1}^{l}n^{(I)}, \quad
\mathsf{q}_{\alpha}=\sum_{I=\alpha}^{l}n^{(I)}\quad (2\le\alpha\le l),
\label{qADHMwritteninn}
\end{align}
we have
\begin{align}
\label{ADHMl_Vn}
&
\langle V_{\mathsf{q}_0=\sum_{I=1}^{l}n^{(I)},\mathsf{q}_1=0,\mathsf{q}_{\alpha\ge 2}=\sum_{I=\alpha}^{l}n^{(I)}}\rangle^{\textrm{$U(1)$ ADHM-$[l]$}}(t,x,y_{\alpha},z;q)
\nonumber\\
&=
\frac{(q^{\frac12}t^2;q)_{\infty}}
{(q^{\frac12}t^{-2};q)_{\infty}}
\sum_{m\in \mathbb{Z}}
\frac{(q^{\frac{3+|\sum_{I=1}^{l}n^{(I)}|}{4}}t^{-1}x^{\mp};q)_{\infty}}
{(q^{\frac{1+|\sum_{I=1}^{l}n^{(I)}|}{4}}tx^{\pm};q)_{\infty}}
\oint \frac{ds}{2\pi is}
\frac{(q^{\frac34+\frac{|m|}{2}}t^{-1}s^{\mp}y_{1}^{\mp};q)_{\infty}}
{(q^{\frac14+\frac{|m|}{2}}ts^{\pm}y_{1}^{\pm};q)_{\infty}}
\nonumber \\
&\quad\times
\prod_{\alpha=1}^{l-1}
\frac{(q^{\frac34+\frac{|m+\sum_{I=\alpha+1}^{l}n^{(I)}|}{2}}t^{-1}s^{\mp}y_{\alpha+1}^{\mp};q)_{\infty}}
{(q^{\frac14+\frac{|m+\sum_{I=\alpha+1}^{l}n^{(I)}|}{2}}ts^{\pm}y_{\alpha+1}^{\pm};q)_{\infty}}
q^{\frac{|m|}{4}+\frac{\sum_{\alpha=1}^{l-1}|m+\sum_{I=\alpha+1}^{l}n^{(I)}|}{4}+\frac{|\sum_{I=1}^{l}n^{(I)}|}{4}}
\nonumber \\
&\quad\times t^{-|m|-\sum_{\alpha=1}^{l-1}|m+\sum_{I=\alpha+1}^{l}n^{(I)}|-|\sum_{I=1}^{l}n^{(I)}|}
z^{lm+\sum_{\alpha=2}^l\sum_{I=\alpha}^ln^{(I)}}. 
\end{align}

On the other hand, one has the charged Wilson lines carrying electric charges $(\mathsf{p}^{(1)},\cdots, \mathsf{p}^{(l)})$ $\in$ $\mathbb{Z}^l$ 
in the mirror $U(1)^{\otimes l}$ quiver gauge theory. 
The Wilson line defect index for the mirror theory takes the form
\begin{align}
\label{mADHMl_Wn}
&{\langle W_{\mathsf{p}^{(I)}}\rangle^{\widetilde{U(1)^{\otimes l}\textrm{mADHM-}[1]}}}(t,y_{\alpha},z,x;q)
\nonumber\\
&=\frac{(q^{\frac12}t^{-2};q)_{\infty}^l}{(q^{\frac12}t^2;q)_{\infty}^l}
\sum_{m^{(1)},\cdots, m^{(l)}\in \mathbb{Z}}
\prod_{I=1}^{l}
\oint \frac{ds^{(I)}}{2\pi is^{(I)}}
{s^{(I)}}^{\mathsf{p}^{(I)}}
\frac{(q^{\frac34+\frac{|m^{(1)}|}{2}}ts^{(1)\mp};q)_{\infty}}
{(q^{\frac14+\frac{|m^{(1)}|}{2}}t^{-1}s^{(1)\pm};q)_{\infty}}
\nonumber\\
&\quad\times 
\prod_{I=1}^{l}
\frac{(q^{\frac34+\frac{|m^{(I)}-m^{(I+1)}|}{2}}ts^{(I)\mp}s^{(I+1)\pm}z^{\mp};q)_{\infty}}
{(q^{\frac14+\frac{|m^{(I)}-m^{(I+1)}|}{2}}t^{-1}s^{(I)\pm}s^{(I+1)\mp}z^{\pm};q)_{\infty}}
q^{\frac{|m^{(1)}|+\sum_{I=1}^l |m^{(I)}-m^{(I+1)}|}{4}}
\nonumber\\
&\quad\times 
t^{|m^{(1)}|+\sum_{I=1}^l |m^{(I)}-m^{(I+1)}|}
x^{m^{(1)}}
\prod_{\alpha=1}^{l}
\left(\frac{y_{\alpha}}{y_{\alpha+1}}\right)^{m^{(\alpha)}},
\end{align}
with $y_{l+1}=y_1$.

We expect that the line defect indices (\ref{ADHMl_Vn}) and (\ref{mADHMl_Wn}) are equal if we identify the parameters $n^{(I)}$ \eqref{qADHMwritteninn} in the vortex charges in the ADHM theory with $\mathsf{p}^{(I)}$ in the mirror theory. 
Accordingly, we propose the following duality of the flavor vortex line with 
$(\mathsf{q}_0,\mathsf{q}_{\alpha\ge 2})$ $=$ $(\sum_{I=1}^l n^{(I)},\sum_{I=\alpha}^{l}n^{(I)})$
for the ADHM theory and the gauge Wilson line for the mirror theory with $p^{(I)}$ $=$ $n^{I}$: 
\begin{align}
\begin{matrix}
\textrm{flavor $V_{\mathsf{q}_0=\sum_{I=1}^{l}n^{(I)},\mathsf{q}_1=0,\mathsf{q}_{\alpha\ge 2}=\sum_{I=\alpha}^{l}n^{(I)}}$}\\
 \\
\textrm{in $U(1)$ ADHM-$[l]$}\\
\end{matrix}
\qquad 
\leftrightarrow 
\qquad 
\begin{matrix}
\textrm{gauge $W_{\mathsf{p}^{(I)}=n^{(I)}}$}\\ 
 \\ 
\textrm{in $\widetilde{U(1)^{\otimes l}\textrm{mADHM-}[1]}$}\\
\end{matrix}.
\label{ADHMVmirrorWduality}
\end{align}

Lastly, let us further consider the vortex-Wilson line index in the ADHM theory 
with charges $(\mathsf{p};\mathsf{q}_0,\mathsf{q}_\alpha)$, $\alpha=1,\cdots,l$, with $\mathsf{q}_1\equiv 0$
\begin{align}
&
\langle L_{\mathsf{p};\mathsf{q}_0,\mathsf{q}_{\alpha}}\rangle^{\textrm{$U(1)$ ADHM-$[l]$}}(t,x,y_{\alpha},z;q)
\nonumber\\
&=
\frac{(q^{\frac12}t^2;q)_{\infty}}
{(q^{\frac12}t^{-2};q)_{\infty}}
\sum_{m\in \mathbb{Z}}
\frac{
(q^{\frac{3+|\mathsf{q}_0|}{4}}t^{-1}x^{\mp};q)_{\infty}
}{
(q^{\frac{1+|\mathsf{q}_0|}{4}}tx^{\pm};q)_{\infty}
}
\oint \frac{ds}{2\pi is} s^{\mathsf{p}}
\frac{(q^{\frac34+\frac{|m|}{2}}t^{-1}s^{\mp}y_{1}^{\mp};q)_{\infty}}
{(q^{\frac14+\frac{|m|}{2}}ts^{\pm}y_{1}^{\pm};q)_{\infty}}
\nonumber \\
&\quad\times \prod_{\alpha=1}^{l-1}
\frac{(q^{\frac34+\frac{|m+\mathsf{q}_{\alpha+1}|}{2}}t^{-1}s^{\mp}y_{\alpha+1}^{\mp};q)_{\infty}}
{(q^{\frac14+\frac{|m+\mathsf{q}_{\alpha+1}|}{2}}ts^{\pm}y_{\alpha+1}^{\pm};q)_{\infty}}
q^{\frac{|m|}{4}+\frac{\sum_{\alpha=1}^{l-1}|m+\mathsf{q}_{\alpha+1}|}{4}+\frac{|\mathsf{q}_0|}{4}}
t^{-|m|-\sum_{\alpha=1}^{l-1}|m+\mathsf{q}_{\alpha+1}|-|\mathsf{q}_0|}\nonumber \\
&\quad\times z^{lm+\sum_{\alpha=2}^{l}\mathsf{q}_\alpha},
\label{ADHMlLpq}
\end{align}
and the vortex-Wilson line index in the mirror theory
\begin{align}
&{\langle L_{\mathsf{p}^{(I)};\mathsf{q}}\rangle^{\widetilde{U(1)^{\otimes l}\textrm{mADHM-}[1]}}}(t,y_{\alpha},z,x;q)
\nonumber\\
&=\frac{(q^{\frac12}t^{-2};q)_{\infty}^l}{(q^{\frac12}t^2;q)_{\infty}^l}
\sum_{m^{(1)},\cdots, m^{(l)}\in \mathbb{Z}}
\prod_{I=1}^{l}
\oint \frac{ds^{(I)}}{2\pi is^{(I)}}
{s^{(I)}}^{\mathsf{p}^{(I)}}
\frac{(q^{\frac34+\frac{|m^{(1)}|}{2}}ts^{(1)\mp};q)_{\infty}}
{(q^{\frac14+\frac{|m^{(1)}|}{2}}t^{-1}s^{(1)\pm};q)_{\infty}}
\nonumber\\
&\quad\times 
\prod_{I=1}^{l-1}
\frac{(q^{\frac34+\frac{|m^{(I)}-m^{(I+1)}|}{2}}ts^{(I)\mp}s^{(I+1)\pm}z^{\mp};q)_{\infty}}
{(q^{\frac14+\frac{|m^{(I)}-m^{(I+1)}|}{2}}t^{-1}s^{(I)\pm}s^{(I+1)\mp}z^{\pm};q)_{\infty}}
\frac{(q^{\frac34+\frac{|m^{(l)}-m^{(1)}+\mathsf{q}|}{2}}ts^{(l)\mp}s^{(1)\pm}z^{\mp};q)_{\infty}}
{(q^{\frac14+\frac{|m^{(l)}-m^{(1)}+\mathsf{q}|}{2}}t^{-1}s^{(l)\pm}s^{(1)\mp}z^{\pm};q)_{\infty}}
\nonumber\\
&\quad\times 
q^{\frac{|m^{(1)}|+\sum_{I=1}^{l-1} |m^{(I)}-m^{(I+1)}|+|m^{(l)}-m^{(1)}+\mathsf{q}|}{4}}
t^{|m^{(1)}|+\sum_{I=1}^{l-1} |m^{(I)}-m^{(I+1)}|+|m^{(l)}-m^{(1)}+\mathsf{q}|}
x^{m^{(1)}}
y_1^{m^{(1)}-m^{(l)}-\mathsf{q}}\nonumber \\
&\quad\times 
\prod_{\alpha=2}^{l}y_\alpha^{m^{(\alpha)}-m^{(\alpha-1)}}.
\label{mirrorADHMlLpq}
\end{align}
By combining \eqref{ADHMWmirrorVduality} and \eqref{ADHMVmirrorWduality}, 
we propose the duality of the vortex-Wilson line indices 
\begin{align}
\begin{matrix}
\textrm{$L_{\mathsf{p}=n;\mathsf{q}_0=\sum_{I=1}^l n^{(I)},\mathsf{q}_1=0,\mathsf{q}_{\alpha\ge 2}=\sum_{I=\alpha}^l n^{(I)}}$}\\
 \\
\textrm{in $U(1)$ ADHM-$[l]$}\\
\end{matrix}
\qquad 
\leftrightarrow 
\qquad 
\begin{matrix}
\textrm{$L_{\mathsf{p}^{(I)}=n^{(I)};\mathsf{q}=n}$}\\ 
 \\ 
\textrm{in $\widetilde{U(1)^{\otimes l}\textrm{mADHM-}[1]}$}\\
\end{matrix}. 
\label{ADHMlmirrorADHMlLpqduality}
\end{align}

\section{Linear quiver CS theory}
\label{sec_linearCS}
\subsection{$U(1)_1\times U(1)_{-1}$ CS theory}
\label{sec_U1_1xU1_-1}

We begin with the quiver CS theory consisting of the $\mathcal{N}=2$ $U(1)_1\times U(1)_{-1}$ CS theory 
coupled to the twisted hypermultiplet transforming as the bifundamental representation under the $U(1)\times U(1)$ gauge group. 
The theory is dual to the SQED$_1$ and the theory of a free twisted hyper \cite{Jafferis:2008em}. 

For the $I$-th gauge node we can introduce a charged Wilson line $W_{\mathsf{p}^{(I)}}$, with $I=1,2$. 
The correlation function of the Wilson lines takes the form
\begin{align}
\langle W_{\mathsf{p}^{(I)}}\rangle^{\textrm{$U(1)_1\times U(1)_{-1}$}}(t,z;q)
&=\sum_{m^{(1)},m^{(2)}\in \mathbb{Z}}
\oint \frac{ds^{(1)}}{2\pi is^{(1)}}
\oint \frac{ds^{(2)}}{2\pi is^{(2)}}
(s^{(1)})^{m^{(1)}+\mathsf{p}^{(1)}}(s^{(2)})^{-m^{(2)}+\mathsf{p}^{(2)}}
\nonumber\\
&\quad \times 
\frac{(q^{\frac34+\frac{|m^{(1)}-m^{(2)}|}{2}}t(z\frac{s^{(1)}}{s^{(2)}})^{\mp 1};q)_{\infty}}
{(q^{\frac14+\frac{|m^{(1)}-m^{(2)}|}{2}}t^{-1}(z\frac{s^{(1)}}{s^{(2)}})^{\pm 1};q)_{\infty}}
q^{\frac{|m^{(1)}-m^{(2)}|}{4}}t^{|m^{(1)}-m^{(2)}|}.
\label{u1^l+1CSWn}
\end{align}
We find that 
\begin{align}
\langle W_{\mathsf{p}^{(I)}}\rangle^{\textrm{$U(1)_1\times U(1)_{-1}$}}(t,z;q)
=q^{\frac{|\mathsf{p}^{(1)}+\mathsf{p}^{(2)}|}{4}}t^{|\mathsf{p}^{(1)}+\mathsf{p}^{(2)}|}
\frac{(q^{\frac34+\frac{|\mathsf{p}^{(1)}+\mathsf{p}^{(2)}|}{2}}tz^{\pm};q)_{\infty}}
{(q^{\frac14+\frac{|\mathsf{p}^{(1)}+\mathsf{p}^{(2)}|}{2}}t^{-1}z^{\pm};q)_{\infty}}. 
\end{align}
Therefore we have 
\begin{align}
\langle W_{\mathsf{p}^{(I)}=n^{(I)}}\rangle^{\textrm{$U(1)_1\times U(1)_{-1}$}}(t,z;q)
=\langle W_{\mathsf{p}=n^{(1)}+n^{(2)}} \rangle^{U(1)-[1]}(t,z;q).
\end{align}
where $\langle W_{\mathsf{p}} \rangle^{U(1)-[1]}(t,z;q)$ is the line defect index of electric charge $\mathsf{p}$ in the $\text{SQED}_1$ given in the convention of \eqref{SQED1Wn}.

\subsection{$U(1)_1\times U(1)_0 \times U(1)_{-1}$ CS theory}
\label{sec_3nodelinearquiverCS}
Next consider the quiver $U(1)_1\times U(1)_0\times U(1)_{-1}$ CS theory 
with the bifundamental twisted hypermultiplets between the adjacent gauge nodes 
where the $U(1)_0$ gauge node corresponds to the $\mathcal{N}=4$ twisted vector multiplet 
whereas the other gauge nodes to $\mathcal{N}=2$ vector multiplet (see Figure \ref{fig:CSbrane1}).
The theory is dual to the $T[SU(2)]$. 

The theory has the Wilson line of charges $(\mathsf{p}^{(1)},\mathsf{p}^{(2)},\mathsf{p}^{(3)})$ under the $U(1)\times U(1)\times U(1)$ gauge group. 
The correlation function of the charged Wilson line can be written as 
\begin{align}
\label{u1u1u1CSWn}
&
\langle W_{\mathsf{p}^{(I)}}\rangle^{\textrm{$U(1)_1\times U(1)_0\times U(1)_{-1}$}}(t,x,z;q)
\nonumber\\
&=
\frac{(q^{\frac12}t^{-2};q)_{\infty}}{(q^{\frac12}t^2;q)_{\infty}}
\sum_{m^{(1)},m^{(2)},m^{(3)}\in \mathbb{Z}}
\prod_{I=1}^{3}
\oint \frac{ds^{(I)}}{2\pi is^{(I)}}
(s^{(1)})^{m^{(1)}+\mathsf{p}^{(1)}}
(s^{(2)})^{n^{(2)}}
(s^{(3)})^{-m^{(3)}+\mathsf{p}^{(3)}}
\nonumber\\
&\times 
\prod_{I=1}^2
\frac{(q^{\frac34+\frac{|m^{(I)}-m^{(I+1)}|}{2}}t(z\frac{s^{(I)}}{s^{(I+1)}})^{\mp 1};q)_{\infty}}
{(q^{\frac14+\frac{|m^{(I)}-m^{(I+1)}|}{2}}t^{-1}(z\frac{s^{(I)}}{s^{(I+1)}})^{\pm 1};q)_{\infty}}
\nonumber\\
&\times 
q^{\frac{\sum_{I=1}^2 |m^{(I)}-m^{(I+1)}|}{4}}
t^{\sum_{I=1}^2|m^{(I)}-m^{(I+1)}|}
x^{m^{(1)}-2m^{(2)}+m^{(3)}}.
\end{align}

We find that 
\begin{align}
\langle W_{\mathsf{p}^{(I)}=n^{(I)}}\rangle^{\textrm{$U(1)_1\times U(1)_0\times U(1)_{-1}$}}(t,x,z;q)
&=\langle L_{\mathsf{p}=n^{(1)}+n^{(2)}+n^{(3)};\mathsf{q}=n^{(2)}}\rangle^{U(1)-[2]}(t,x,z;q),
\end{align}
where $\langle L_{\mathsf{p};\mathsf{q}}\rangle^{U(1)-[2]}(t,x,z;q)$ is the line defect index of the vortex-Wilson line of electric charge $\mathsf{p}$ and the vortex number $\mathsf{q}$ in the $T[SU(2)]$ given by (\ref{SQED2WnVn}).

In particular, when $\mathsf{p}^{(1)}=\mathsf{p}^{(3)}=0$, 
the charged Wilson line in the quiver $U(1)_1\times U(1)_0\times U(1)_{-1}$ CS theory is self-mirror. 

\subsection{$U(1)_1 \times U(1)_0^{l-1}\times U(1)_{-1}$ CS theory}
Now we propose a generalization of the dualities of the line defects for the linear quiver CS theory.
In the same way, the $\text{SQED}_l$ and its mirror are dual to the $U(1)_1 \times U(1)_0^{l-1}\times U(1)_{-1}$ linear quiver CS theory. 
Here the linear quiver CS theory consists of two ${\cal N}=2$ vector multiplets on the nodes $U(1)_{\pm 1}$, 
${\cal N}=4$ twisted vector multiplet on each node $U(1)_0$ and the bifundamental twisted hypermultiplets between adjacent nodes (see Figure \ref{fig:CSbrane1}).

The linear quiver CS theory has the Wilson line carrying a set of $(l+1)$ electric charges $\mathsf{p}^{(I)}$, with $I=1,\cdots,l+1$. 
The correlation function of the Wilson line takes the form\footnote{
In order to reproduce the convention for $l=1$ in section \ref{sec_U1_1xU1_-1}, we need to set $x_1=1$ and $\frac{z_1}{z_2}=z$.
Similarly, in order to reproduce the convention for $l=2$ in section \ref{sec_3nodelinearquiverCS}, we need to set $x_1=x_2^{-1}=x$ and $z_1=z_2^{-1}=z$.
}
\begin{align}
\label{u1^l1+1CSWn}
&\langle W_{\mathsf{p}^{(1)},\mathsf{p}^{(2)}}\rangle^{U(1)_1\times U(1)_{-1}}(t,x_{\alpha},z_a;q)
\nonumber \\
&=
\sum_{m^{(1)},m^{(2)}\in \mathbb{Z}}
\prod_{I=1}^{2}
\oint \frac{ds^{(I)}}{2\pi is^{(I)}}
(s^{(1)})^{m^{(1)}+\mathsf{p}^{(1)}}
(s^{(2)})^{-m^{(2)}+\mathsf{p}^{(2)}}
\frac{(q^{\frac34+\frac{|m^{(I)}-m^{(I+1)}|}{2}}t(\frac{z_1}{z_2}\frac{s^{(1)}}{s^{(2)}})^{\mp 1};q)_{\infty}}
{(q^{\frac14+\frac{|m^{(I)}-m^{(I+1)}|}{2}}t^{-1}(\frac{z_1}{z_2}\frac{s^{(1)}}{s^{(2)}})^{\pm 1};q)_{\infty}}\nonumber \\
&\quad\times q^{\frac{|m^{(1)}-m^{(2)}|}{4}}
t^{|m^{(1)}-m^{(2)}|}
x_1^{m^{(1)}-m^{(2)}}
\end{align}
for $l=1$ and
\begin{align}
\label{u1^lge2+1CSWn}
&\langle W_{\mathsf{p}^{(I)}}\rangle^{U(1)_1\times U(1)_0^{l-1}\times U(1)_{-1}}(t,x_{\alpha},z_a;q)
\nonumber \\
&=
\frac{(q^{\frac12}t^{-2};q)_{\infty}^{l-1}}{(q^{\frac12}t^2;q)_{\infty}^{l-1}}
\sum_{m^{(1)},\cdots,m^{(l+1)}\in \mathbb{Z}}
\prod_{I=1}^{l+1}
\oint \frac{ds^{(I)}}{2\pi is^{(I)}}
(s^{(1)})^{m^{(1)}+\mathsf{p}^{(1)}}
\prod_{I=2}^{l} (s^{(I)})^{\mathsf{p}^{(I)}}
(s^{(l+1)})^{-m^{(l+1)}+\mathsf{p}^{(l+1)}}\nonumber \\
&\quad\times \frac{(q^{\frac34+\frac{|m^{(1)}-m^{(2)}|}{2}}t(z_1\frac{s^{(1)}}{s^{(2)}})^{\mp 1};q)_{\infty}}
{(q^{\frac14+\frac{|m^{(1)}-m^{(2)}|}{2}}t^{-1}(z_1\frac{s^{(1)}}{s^{(2)}})^{\pm 1};q)_{\infty}}
\prod_{I=1}^{l-1}
\frac{(q^{\frac34+\frac{|m^{(I)}-m^{(I+1)}|}{2}}ts^{(I)\mp}s^{(I+1)\pm};q)_{\infty}}
{(q^{\frac14+\frac{|m^{(I)}-m^{(I+1)}|}{2}}t^{-1}s^{(I)\pm}s^{(I+1)\mp};q)_{\infty}}\nonumber \\
&\quad\times \frac{(q^{\frac34+\frac{|m^{(I)}-m^{(I+1)}|}{2}}t(\frac{1}{z_2}\frac{s^{(l)}}{s^{(l+1)}})^{\mp 1};q)_{\infty}}
{(q^{\frac14+\frac{|m^{(I)}-m^{(I+1)}|}{2}}t^{-1}(\frac{1}{z_2}\frac{s^{(l)}}{s^{(l+1)}})^{\pm 1};q)_{\infty}}
q^{\frac{\sum_{I=1}^l |m^{(I)}-m^{(I+1)}|}{4}}
t^{\sum_{I=1}^l |m^{(I)}-m^{(I+1)}|}
x_1^{m^{(1)}}\nonumber \\
&\quad\times \prod_{\alpha=1}^{l-1}\left(\frac{x_{\alpha+1}}{x_\alpha}\right)^{m^{(\alpha+1)}}
x_l^{-m^{(l+1)}}
\end{align}
for $l\ge 2$.

We conjectured that 
\begin{align}
\label{triality_SQED}
&\langle W_{\mathsf{p}^{(I)}=n^{(I)}}\rangle^{U(1)_1\times U(1)_0^{l-1}\times U(1)_{-1}}(t,x_{\alpha},z_a;q)\nonumber \\
&=\langle L_{
\mathsf{p}=\sum_{I=1}^{l+1}n^{(I)};
\mathsf{q}_\alpha=\sum_{I=\alpha}^{l-1}n^{(I+1)}
} \rangle^{U(1)-[l]}(t,x_\alpha,z_{a};q)
\nonumber\\
&=\langle L_{
\mathsf{p}^{(I)}=
n^{(I+1)};
\mathsf{q}=\sum_{I=1}^{l+1}n^{(I)}
}\rangle^{\widetilde{[1]-U(1)^{\otimes l-1}-[1]}}(t,z_{a},x_\alpha;q), 
\end{align}
where $\langle L_{\mathsf{p}=n;\mathsf{q}_\alpha=\sum_{I=\alpha}^{l-1}\ell^{(I)}} \rangle^{U(1)-[l]}$ and $\langle L_{\mathsf{p}^{(I)}=\ell^{(I)};\mathsf{q}=n}\rangle^{\widetilde{[1]-U(1)^{\otimes l-1}-[1]}}$ are the vortex-Wilson line defect indices (\ref{SQEDlLnm}) for the SQED$_l$ and (\ref{mSQEDlLnm}) for its mirror respectively. 
The relation (\ref{triality_SQED}) indicates the following triality of the line operators: 
\begin{align}
\begin{tikzpicture}[scale=0.3]
\draw [<->] (12,8)--(16,11);
\draw [<->] (38,8)--(34,11);
\draw [<->] (16,3)--(35,3);
\node [above] at (25,17) {\textrm{Wilson}};
\node [above] at (25,14) {\textrm{$W_{\mathsf{p}^{(I)}=n^{(I)}}$}};
\node [above] at (25,11) {\textrm{in $U(1)_1\times U(1)_0^{l-1}\times U(1)_{-1}$}};
\node [above] at (8,6) {\textrm{vortex-Wilson}};
\node [above] at (8,3) {$L_{\mathsf{p}=\sum_{I=1}^{l+1}n^{(I)};\mathsf{q}_\alpha=\sum_{I=\alpha}^{l-1}n^{(I+1)}}$};
\node [above] at (8,0) {\textrm{in $U(1)-[l]$}};
\node [above] at (43,6) {\textrm{vortex-Wilson}};
\node [above] at (43,3) {$L_{\mathsf{p}^{(I)}=n^{(I+1)};\mathsf{q}=\sum_{I=1}^{l+1}n^{(I)}}$};
\node [above] at (43,0) {\textrm{in $\widetilde{[1]-U(1)^{\otimes l-1}-[1]}$}};
\end{tikzpicture}.
\end{align}

It is also interesting to examine the line defect indices of the charged Wilson lines 
for the $U(1)_k \times U(1)_0^{l-1}\times U(1)_{-k}$ linear quiver CS theory.
Let us define the Wilson line defect index for the $U(1)_k \times U(1)_0^{l-1}\times U(1)_{-k}$ linear quiver CS theory as
\begin{align}
\label{u1^l1+1CSWngenk}
&\langle W_{\mathsf{p}^{(1)},\mathsf{p}^{(2)}}\rangle^{U(1)_k\times U(1)_{-k}}(t,x_{\alpha},z_a;q)
\nonumber \\
&=
\sum_{m^{(1)},m^{(2)}\in \mathbb{Z}}
\prod_{I=1}^{2}
\oint \frac{ds^{(I)}}{2\pi is^{(I)}}
(s^{(1)})^{km^{(1)}+\mathsf{p}^{(1)}}
(s^{(2)})^{-km^{(2)}+\mathsf{p}^{(2)}}
\frac{(q^{\frac34+\frac{|m^{(I)}-m^{(I+1)}|}{2}}t(\frac{z_1}{z_2}\frac{s^{(1)}}{s^{(2)}})^{\mp 1};q)_{\infty}}
{(q^{\frac14+\frac{|m^{(I)}-m^{(I+1)}|}{2}}t^{-1}(\frac{z_1}{z_2}\frac{s^{(1)}}{s^{(2)}})^{\pm 1};q)_{\infty}}\nonumber \\
&\quad\times q^{\frac{|m^{(1)}-m^{(2)}|}{4}}
t^{|m^{(1)}-m^{(2)}|}
x_1^{k(m^{(1)}-m^{(2)})}
\end{align}
for $l=1$ and
\begin{align}
\label{u1^lge2+1CSWngenk}
&\langle W_{\mathsf{p}^{(I)}}\rangle^{U(1)_k\times U(1)_0^{l-1}\times U(1)_{-k}}(t,x_{\alpha},z_a;q)
\nonumber \\
&=
\frac{(q^{\frac12}t^{-2};q)_{\infty}^{l-1}}{(q^{\frac12}t^2;q)_{\infty}^{l-1}}
\sum_{m^{(1)},\cdots,m^{(l+1)}\in \mathbb{Z}}
\prod_{I=1}^{l+1}
\oint \frac{ds^{(I)}}{2\pi is^{(I)}}
(s^{(1)})^{km^{(1)}+\mathsf{p}^{(1)}}
\prod_{I=2}^{l} (s^{(I)})^{\mathsf{p}^{(I)}}
(s^{(l+1)})^{-km^{(l+1)}+\mathsf{p}^{(l+1)}}\nonumber \\
&\quad\times \frac{(q^{\frac34+\frac{|m^{(1)}-m^{(2)}|}{2}}t(z_1\frac{s^{(1)}}{s^{(2)}})^{\mp 1};q)_{\infty}}
{(q^{\frac14+\frac{|m^{(1)}-m^{(2)}|}{2}}t^{-1}(z_1\frac{s^{(1)}}{s^{(2)}})^{\pm 1};q)_{\infty}}
\prod_{I=1}^{l-1}
\frac{(q^{\frac34+\frac{|m^{(I)}-m^{(I+1)}|}{2}}ts^{(I)\mp}s^{(I+1)\pm};q)_{\infty}}
{(q^{\frac14+\frac{|m^{(I)}-m^{(I+1)}|}{2}}t^{-1}s^{(I)\pm}s^{(I+1)\mp};q)_{\infty}}\nonumber \\
&\quad\times \frac{(q^{\frac34+\frac{|m^{(I)}-m^{(I+1)}|}{2}}t(\frac{1}{z_2}\frac{s^{(l)}}{s^{(l+1)}})^{\mp 1};q)_{\infty}}
{(q^{\frac14+\frac{|m^{(I)}-m^{(I+1)}|}{2}}t^{-1}(\frac{1}{z_2}\frac{s^{(l)}}{s^{(l+1)}})^{\pm 1};q)_{\infty}}
q^{\frac{\sum_{I=1}^l |m^{(I)}-m^{(I+1)}|}{4}}
t^{\sum_{I=1}^l |m^{(I)}-m^{(I+1)}|}
x_1^{km^{(1)}}\nonumber \\
&\quad\times \prod_{\alpha=1}^{l-1}\left(\frac{x_{\alpha+1}}{x_\alpha}\right)^{km^{(\alpha+1)}}
x_l^{-km^{(l+1)}}
\end{align}
for $l\ge 2$.
Then we observe
\begin{align}
&\langle W_{\mathsf{p}^{(I)}=(kn,0,\cdots,0)}\rangle^{U(1)_k\times U(1)_0^{l-1}\times U(1)_{-k}}(t,x_\alpha,z_a;q)\nonumber \\
&=\frac{1}{k}\sum_{j=0}^{k-1}\langle W_{\mathsf{p}^{(I)}=(n,0,\cdots,0)}\rangle^{U(1)_1\times U(1)_0^{l-1}\times U(1)_{-1}}(t,x_\alpha,z_a;q)\Bigr|_{z_2\rightarrow e^{\frac{2\pi ij}{k}}z_2,x_\alpha\rightarrow x_\alpha^k}.
\end{align}



\subsection{$U(2)_{2,-2}$ CS theory}
In section \ref{sec_u1u1h1} and \ref{sec_u2h3}, 
we examined the Seiberg-like dualities of line operators in 3d $\mathcal{N}=4$ gauge theories. 
Let us investigate similar dualities for the CS theories. 
It is argued that \cite{Kubo:2021ecs} 
the SQED$_1$ is dual to the $U(2)_{2,-2}$ CS theory coupled to a fundamental hyper. 

The one-point function of the Wilson line $W_n$ of charge $n$ in the $U(2)_{2,-2}$ CS theory with a fundamental hyper is given by
\begin{align}
\langle W_{n}\rangle^{U(2)_{2,-2}-[1]}
&=\frac12 \sum_{m_1, m_2\in \mathbb{Z}}
\oint \prod_{i=1}^2 \frac{ds_i}{2\pi is_i}
(1-q^{\frac{|m_1-m_2|}{2}}s_1^{\pm}s_2^{\mp})
s_1^{2m_1+n}s_2^{2m_2+n}(s_1s_2)^{-2m_1-2m_2}
\nonumber\\
&\times 
\frac{(q^{\frac34}ts_i^{\mp};q)_{\infty}}
{(q^{\frac14}t^{-1}s_i^{\pm};q)_{\infty}}
q^{-\frac{|m_1-m_2|}{2}+\sum_{i=1}^2 \frac{|m_i|}{4}}t^{\sum_{i=1}^2|m_i|}. 
\end{align}
We find that it agrees with 
\begin{align}
(-1)^n z^n
q^{\frac{|n|}{4}}t^{|n|}
\frac{(q^{\frac{3}{4}+\frac{|n|}{2}}tz^{\mp};q)_{\infty}}
{(q^{\frac14+\frac{|n|}{2}}t^{-1}z^{\pm};q)_{\infty}}. 
\end{align}
Hence 
\begin{align}
(-1)^n z^{-n} \langle W_{n}\rangle^{U(2)_{2,-2}-[1]}&
=\langle W_n \rangle^{U(1)-[1]}. 
\end{align}

\section{Circular quiver CS theory}
\label{sec_ABJM}

\subsection{$U(1)_1 \times U(1)_{-1}$ ABJM theory}
Let us start from the $U(1)_1\times U(1)_{-1}$ ABJM theory. 
The theory has enhanced $\mathcal{N}=8$ supersymmetry and it is dual to the $U(1)$ ADHM theory with one flavor \cite{Bashkirov:2011pt}. 

There exist the $1/6$-BPS Wilson lines of charge $(\mathsf{p}^{(1)},\mathsf{p}^{(2)})$ for the $U(1)\times U(1)$ gauge group \cite{Drukker:2008zx,Chen:2008bp,Rey:2008bh}. 
The Wilson line defect index is given by
\begin{align}
\label{ABJMu1k1W}
&
{\langle W_{\mathsf{p}^{(1)},\mathsf{p}^{(2)}}\rangle^{U(1)_{1}\times U(1)_{-1}}}(t,x,z;q)
\nonumber\\
&=
\sum_{m^{(1)},m^{(2)}\in \mathbb{Z}}
\oint \frac{ds^{(1)}}{2\pi is^{(1)}} 
\oint \frac{ds^{(2)}}{2\pi is^{(2)}} 
\nonumber\\
&\times 
\frac{(q^{\frac34+\frac{|m^{(1)}-m^{(2)}|}{2}}t^{-1}s^{(1)\mp}s^{(2)\pm}x^{\mp};q)_{\infty}
(q^{\frac34+\frac{|m^{(1)}-m^{(2)}|}{2}}ts^{(1)\mp}s^{(2)\pm}z^{\pm};q)_{\infty}
}
{
(q^{\frac14+\frac{|m^{(1)}-m^{(2)}|}{2}}ts^{(1)\pm}s^{(2)\mp}x^{\pm};q)_{\infty}
(q^{\frac14+\frac{|m^{(1)}-m^{(2)}|}{2}}t^{-1}s^{(1)\pm}s^{(2)\mp}z^{\mp};q)_{\infty}
}
\nonumber\\
&\times 
{s^{(1)}}^{m^{(1)}+\mathsf{p}^{(1)}} {s^{(2)}}^{-m^{(2)}+\mathsf{p}^{(2)}}
q^{\frac{|m^{(1)}-m^{(2)}|}{2}}. 
\end{align}
We find that 
\begin{align}
\label{ABJMu1k1W2}
&
{\langle W_{\mathsf{p}^{(1)},\mathsf{p}^{(2)}}\rangle^{U(1)_{1}\times U(1)_{-1}}}(t,x,z;q)
\nonumber\\
&=q^{\frac{|\mathsf{p}^{(1)}+\mathsf{p}^{(2)}|}{2}}
\frac{(q^{\frac34+\frac{|\mathsf{p}^{(1)}+\mathsf{p}^{(2)}|}{2}}t^{-1}x^{\mp};q)_{\infty}}
{(q^{\frac14+\frac{|\mathsf{p}^{(1)}+\mathsf{p}^{(2)}|}{2}}tx^{\pm};q)_{\infty}}
\frac{(q^{\frac34+\frac{|\mathsf{p}^{(1)}+\mathsf{p}^{(2)}|}{2}}tz^{\mp};q)_{\infty}}
{(q^{\frac14+\frac{|\mathsf{p}^{(1)}+\mathsf{p}^{(2)}|}{2}}t^{-1}z^{\pm};q)_{\infty}}. 
\end{align}
We see that the line defect index (\ref{ABJMu1k1W2}) only depends on the total charge 
$\mathsf{p}^{(1)}+\mathsf{p}^{(2)}$ carried by the Wilson line. 
This implies that the Wilson lines with any pair of gauge charges for each of gauge nodes 
are not independent as they involve some overcounting \cite{Drukker:2008zx,Rey:2008bh}. 
In particular, when $\mathsf{p}^{(1)}=-\mathsf{p}^{(2)}$, the line defect index is identical to the original index 
because such Wilson lines can be screened by matter fields.  
The expression (\ref{ABJMu1k1W}) or equivalently (\ref{ABJMu1k1W2}) is invariant under the mirror transformation (\ref{mirror_trans}) 
so that the charged Wilson line in the ABJM theory is self-mirror. 

The theory is dual to the $U(1)$ ADHM theory with one flavor. 
From (\ref{ADHM1_VWn2}) and (\ref{ABJMu1k1W2}) we find that 
\begin{align}
\langle L_{\mathsf{p}=n^{(1)}+n^{(2)};\mathsf{q}_0=n^{(1)}+n^{(2)}}\rangle^{\textrm{$U(1)$ ADHM-$[1]$}}(t,x,z;q)
&=
{\langle W_{\mathsf{p}^{(1)}=n^{(1)},\mathsf{p}^{(2)}=n^{(2)}}\rangle^{U(1)_{1}\times U(1)_{-1}}}(t,x,z;q)
\end{align}
This supports the duality of the line operators
\begin{align}
\begin{matrix}
\textrm{vortex-Wilson $L_{\mathsf{p}=n^{(1)}+n^{(2)};\mathsf{q}_0=n^{(1)}+n^{(2)}}$}\\
 \\
\textrm{in $U(1)$ ADHM-$[1]$}\\
\end{matrix}
\qquad 
\leftrightarrow 
\qquad 
\begin{matrix}
\textrm{gauge $W_{\mathsf{p}^{(1)}=n^{(1)},\mathsf{p}^{(2)}=n^{(2)}}$}\\ 
 \\ 
\textrm{in $U(1)_1\times U(1)_{-1}$ ABJM}\\
\end{matrix}. 
\end{align}
This is consistent with the fact that the Wilson line in the $U(1)_1\times U(1)_{-1}$ ABJM theory is realized by the fundamental strings ending on a D5-brane in Type IIB string theory which is obtainable from the $(1,-1)$ strings ending on a $(1,-1)$ 5-brane upon the $STS$ transformation. 

The vortex line in the ABJM theory can be constructed by requiring a certain singular profile of matter fields. 
As the theory has a pair of the global symmetries rotating the hypermultiplet and that rotating the twisted hypermultiplet, 
we can introduce the associated flavor vortex lines carrying pairs $(\mathsf{q}_1,\mathsf{q}_2)$ of charges. 
The vortex line defect index takes the form
\begin{align}
\label{ABJMu1k1V}
&
{\langle V_{\mathsf{q}_1,\mathsf{q}_2}\rangle^{U(1)_{1}\times U(1)_{-1}}}(t,x,z;q)
\nonumber\\
&=
\sum_{m^{(1)},m^{(2)}\in \mathbb{Z}}
\oint \frac{ds^{(1)}}{2\pi is^{(1)}}
\oint \frac{ds^{(2)}}{2\pi is^{(2)}} 
\nonumber\\
&\times 
\frac{(q^{\frac34+\frac{|m^{(1)}-m^{(2)}+\mathsf{q}_1|}{2}}t^{-1}s^{(1)\mp}s^{(2)\pm}x^{\mp};q)_{\infty}
(q^{\frac34+\frac{|m^{(1)}-m^{(2)}+\mathsf{q}_2|}{2}}ts^{(1)\mp}s^{(2)\pm}z^{\pm};q)_{\infty}
}
{
(q^{\frac14+\frac{|m^{(1)}-m^{(2)}+\mathsf{q}_1|}{2}}ts^{(1)\pm}s^{(2)\mp}x^{\pm};q)_{\infty}
(q^{\frac14+\frac{|m^{(1)}-m^{(2)}+\mathsf{q}_2|}{2}}t^{-1}s^{(1)\pm}s^{(2)\mp}z^{\mp};q)_{\infty}
}
\nonumber\\
&\times 
{s^{(1)}}^{m^{(1)}} {s^{(2)}}^{-m^{(2)}}
q^{\frac{|m^{(1)}-m^{(2)}+\mathsf{q}_1|}{4}+\frac{|m^{(1)}-m^{(2)}+\mathsf{q}_2|}{4}}
t^{|m^{(1)}-m^{(2)}+\mathsf{q}_2|-|m^{(1)}-m^{(2)}+\mathsf{q}_1|}. 
\end{align}
We get
\begin{align}
\label{ABJMu1k1V2}
&
{\langle V_{\mathsf{q}_1,\mathsf{q}_2}\rangle^{U(1)_{1}\times U(1)_{-1}}}(t,x,z;q)
\nonumber\\
&=q^{\frac{|\mathsf{q}_1|+|\mathsf{q}_2|}{4}}
t^{|\mathsf{q}_2|-|\mathsf{q}_1|}
\frac{(q^{\frac34+\frac{|\mathsf{q}_1|}{2}}t^{-1}x^{\mp};q)_{\infty}}
{(q^{\frac14+\frac{|\mathsf{q}_1|}{2}}tx^{\pm};q)_{\infty}}
\frac{(q^{\frac34+\frac{|\mathsf{q}_2|}{2}}tz^{\mp};q)_{\infty}}
{(q^{\frac14+\frac{|\mathsf{q}_2|}{2}}t^{-1}z^{\pm};q)_{\infty}}. 
\end{align}
When we turn off $\mathsf{q}_1$ (resp. $\mathsf{q}_2$), 
we excite two complex scalar fields in the bifundamental hypermultiplet (resp. twisted hypermultiplet). 
Then we generically obtain the $1/3$-BPS vortex line \cite{Drukker:2008jm}. 

Comparing the expression (\ref{ABJMu1k1V2}) with (\ref{ADHM1_Vn2}), 
we get
\begin{align}
\langle V_{\mathsf{q}_0=n}\rangle^{\textrm{$U(1)$ ADHM-$[1]$}}(t,z,x;q)
&=
{\langle V_{\mathsf{q}_1=n,\mathsf{q}_2=0}\rangle^{U(1)_{1}\times U(1)_{-1}}}(t,x,z;q). 
\end{align}
This indicates the following duality: 
\begin{align}
\begin{matrix}
\textrm{flavor $V_{\mathsf{q}_0=n}$}\\
 \\
\textrm{in $U(1)$ ADHM-$[1]$}\\
\end{matrix}
\qquad 
\leftrightarrow 
\qquad 
\begin{matrix}
\textrm{flavor $V_{\mathsf{q}_1=n,\mathsf{q}_2=0}$}\\ 
 \\ 
\textrm{in $U(1)_1\times U(1)_{-1}$ ABJM}\\
\end{matrix}. 
\end{align}

On the other hand, when we allow all the complex scalar fields 
in the bifundamental hyper and twisted hyper to acquire singular profiles, 
the vortex lines are the $1/6$-BPS operators \cite{Drukker:2008jm}. 
In particular, when the two charges are equal, 
the vortex line defect index (\ref{ABJMu1k1V2}) is identical to the Wilson line defect index (\ref{ABJMu1k1W2})
\begin{align}
{\langle V_{\mathsf{q}_1=n^{(1)}+n^{(2)},\mathsf{q}_2=n^{(1)}+n^{(2)}}\rangle^{U(1)_{1}\times U(1)_{-1}}}(t,x,z;q)
&={\langle W_{\mathsf{p}^{(1)}=n^{(1)},\mathsf{p}^{(2)}=n^{(2)}}\rangle^{U(1)_{1}\times U(1)_{-1}}}(t,x,z;q). 
\end{align}
Thus we propose the following duality: 
\begin{align}
\begin{matrix}
\textrm{flavor $V_{\mathsf{q}_1=n^{(1)}+n^{(2)},\mathsf{q}_2=n^{(1)}+n^{(2)}}$}\\
 \\
\textrm{in $U(1)_1\times U(1)_{-1}$ ABJM}\\
\end{matrix}
\qquad 
\leftrightarrow 
\qquad 
\begin{matrix}
\textrm{gauge $W_{\mathsf{p}^{(1)}=n^{(1)},\mathsf{p}^{(2)}=n^{(2)}}$}\\ 
 \\ 
\textrm{in $U(1)_1\times U(1)_{-1}$ ABJM}\\
\end{matrix}. 
\end{align}
In other words, the $1/6$-BPS dynamical Wilson line and the $1/6$-BPS flavor vortex line 
in the $U(1)_1\times U(1)_{-1}$ ABJM theory are equivalent. 

More generally, the line defect index of the mixed vorex-Wilson line carrying  
a pair $(\mathsf{p}^{(1)},\mathsf{p}^{(2)})$ of electric charges 
and a pair $(\mathsf{q}_1,\mathsf{q}_2)$ of the vortex numbers takes the form 
\begin{align}
\label{ABJMu1k1VW}
&
{\langle L_{\mathsf{p}^{(1)},\mathsf{p}^{(2)};\mathsf{q}_1,\mathsf{q}_2}\rangle^{U(1)_{1}\times U(1)_{-1}}}(t,x,z;q)
\nonumber\\
&=
\sum_{m^{(1)},m^{(2)}\in \mathbb{Z}}
\oint \frac{ds^{(1)}}{2\pi is^{(1)}}
\oint \frac{ds^{(2)}}{2\pi is^{(2)}} 
\nonumber\\
&\times 
\frac{(q^{\frac34+\frac{|m^{(1)}-m^{(2)}+\mathsf{q}_1|}{2}}t^{-1}s^{(1)\mp}s^{(2)\pm}x^{\mp};q)_{\infty}
(q^{\frac34+\frac{|m^{(1)}-m^{(2)}+\mathsf{q}_2|}{2}}ts^{(1)\mp}s^{(2)\pm}z^{\pm};q)_{\infty}
}
{
(q^{\frac14+\frac{|m^{(1)}-m^{(2)}+\mathsf{q}_1|}{2}}ts^{(1)\pm}s^{(2)\mp}x^{\pm};q)_{\infty}
(q^{\frac14+\frac{|m^{(1)}-m^{(2)}+\mathsf{q}_2|}{2}}t^{-1}s^{(1)\pm}s^{(2)\mp}z^{\mp};q)_{\infty}
}
\nonumber\\
&\times 
{s^{(1)}}^{m^{(1)}+\mathsf{p}^{(1)}} {s^{(2)}}^{-m^{(2)}+\mathsf{p}^{(2)}}
q^{\frac{|m^{(1)}-m^{(2)}+\mathsf{q}_1|}{4}+\frac{|m^{(1)}-m^{(2)}+\mathsf{q}_2|}{4}}
t^{|m^{(1)}-m^{(2)}+\mathsf{q}_2|-|m^{(1)}-m^{(2)}+\mathsf{q}_1|}.
\end{align}
It can be evaluated as
\begin{align}
\label{ABJMu1k1VW2}
&
{\langle L_{\mathsf{p}^{(1)},\mathsf{p}^{(2)};\mathsf{q}_1,\mathsf{q}_2}\rangle^{U(1)_{1}\times U(1)_{-1}}}(t,x,z;q)
\nonumber\\
&=q^{\frac{
|\mathsf{p}^{(1)}+\mathsf{p}^{(2)}-\mathsf{q}_1|
+|\mathsf{p}^{(1)}+\mathsf{p}^{(2)}-\mathsf{q}_2|
}{4}}
t^{|\mathsf{q}_2|-|\mathsf{q}_1|}
\nonumber\\
&\times 
\frac{(q^{\frac34+\frac{|\mathsf{p}^{(1)}+\mathsf{p}^{(2)}-\mathsf{q}_1|}{2}}t^{-1}x^{\mp};q)_{\infty}}
{(q^{\frac14+\frac{|\mathsf{p}^{(1)}+\mathsf{p}^{(2)}-\mathsf{q}_2|}{2}}tx^{\pm};q)_{\infty}}
\frac{(q^{\frac34+\frac{|\mathsf{p}^{(1)}+\mathsf{p}^{(2)}-\mathsf{q}_2|}{2}}tz^{\mp};q)_{\infty}}
{(q^{\frac14+\frac{|\mathsf{p}^{(1)}+\mathsf{p}^{(2)}-\mathsf{q}_2|}{2}}t^{-1}z^{\pm};q)_{\infty}}. 
\end{align}

Similarly, from (\ref{ADHM1_Vn2}) and (\ref{ABJMu1k1VW2}) 
we find that 
\begin{align}
\langle W_{\mathsf{p}=n}\rangle^{\textrm{$U(1)$ ADHM-$[1]$}}(t,z,x;q)
&={\langle L_{\mathsf{p}^{(1)}=n^{(1)},\mathsf{p}^{(2)}=-n^{(1)};\mathsf{q}_1=0,\mathsf{q}_2=n}\rangle^{U(1)_{1}\times U(1)_{-1}}}(t,x,z;q)
\nonumber\\
&={\langle V_{\mathsf{q}_1=0,\mathsf{q}_2=n}\rangle^{U(1)_{1}\times U(1)_{-1}}}(t,x,z;q). 
\end{align}
This indicates the dualities 
\begin{align}
\begin{matrix}
\textrm{gauge $W_{\mathsf{p}=n}$}\\
 \\
\textrm{in $U(1)$ ADHM-$[1]$}\\
\end{matrix}
\qquad 
&\leftrightarrow 
\qquad 
\begin{matrix}
\textrm{vortex-Wilson $L_{\mathsf{p}^{(1)}=n^{(1)},\mathsf{p}^{(2)}=-n^{(1)};\mathsf{q}_1=0,\mathsf{q}_2=n}$}\\ 
 \\ 
\textrm{in $U(1)_1\times U(1)_{-1}$ ABJM}\\
\end{matrix}\\ \nonumber\\
&\leftrightarrow 
\qquad 
\begin{matrix}
\textrm{flavor $V_{\mathsf{q}_1=0,\mathsf{q}_2=n}$}\\ 
 \\ 
\textrm{in $U(1)_1\times U(1)_{-1}$ ABJM}\\
\end{matrix}. 
\end{align}
The mixed vortex-Wilson line $L_{\mathsf{p}^{(1)}=n^{(1)},\mathsf{p}^{(2)}=-n^{(1)};\mathsf{q}_1=0,\mathsf{q}_2=n}$ is equivalent to the $1/3$-BPS flavor vortex line $V_{\mathsf{q}_1=0,\mathsf{q}_2=n}$ 
in the $U(1)_1\times U(1)_{-1}$ ABJM theory as the Wilson line can be screened by the matter fields. 
It is dual to the gauge Wilson line in the $U(1)$ ADHM theory with one flavor.

\subsection{$U(1)_2 \times U(1)_{-2}$ ABJM theory}
For $k=2$ the $U(1)_k\times U(1)_{-k}$ ABJM theory has enhanced $\mathcal{N}=8$ supersymmetry. 
Although it is not dual to the ADHM theory, it has a dual description as the $\mathbb{Z}_2$ gauge theory coupled to a pair of the hyper and twisted hyper \cite{Hayashi:2022ldo}. 
\footnote{The basic idea of such dualities can be found in \cite{Kapustin:2014gua}. } 

Similarly, the theory has the charged $1/6$-BPS Wilson line operators with a pair of electric charges. 
The Wilson line defect index takes the form
\begin{align}
\label{ABJMu1k2W}
&
{\langle W_{\mathsf{p}^{(1)},\mathsf{p}^{(2)}}\rangle^{U(1)_{2}\times U(1)_{-2}}}(t,x,z;q)
\nonumber\\
&=
\sum_{m^{(1)},m^{(2)}\in \mathbb{Z}}
\oint \frac{ds^{(1)}}{2\pi is^{(1)}}
\oint \frac{ds^{(2)}}{2\pi is^{(2)}} 
\nonumber\\
&\times 
\frac{(q^{\frac34+\frac{|m^{(1)}-m^{(2)}|}{2}}t^{-1}s^{(1)\mp}s^{(2)\pm}x^{\mp};q)_{\infty}
(q^{\frac34+\frac{|m^{(1)}-m^{(2)}|}{2}}ts^{(1)\mp}s^{(2)\pm}z^{\pm};q)_{\infty}
}
{
(q^{\frac14+\frac{|m^{(1)}-m^{(2)}|}{2}}ts^{(1)\pm}s^{(2)\mp}x^{\pm};q)_{\infty}
(q^{\frac14+\frac{|m^{(1)}-m^{(2)}|}{2}}t^{-1}s^{(1)\pm}s^{(2)\mp}z^{\mp};q)_{\infty}
}
\nonumber\\
&\times 
{s^{(1)}}^{2m^{(1)}+\mathsf{p}^{(1)}} {s^{(2)}}^{-2m^{(2)}+\mathsf{p}^{(2)}}
q^{\frac{|m^{(1)}-m^{(2)}|}{2}}. 
\end{align}
It vanishes for a pair of even and odd charges. 
When both charges are even, 
we find that the Wilson line defect index is evaluated as
\begin{align}
\label{ABJMu1k2Weven}
&
{\langle W_{\mathsf{p}^{(1)}=2n^{(1)},\mathsf{p}^{(2)}=2n^{(2)}}\rangle^{U(1)_{2}\times U(1)_{-2}}}(t,x,z;q)
\nonumber\\
&=
\frac12 
q^{\frac{|n^{(1)}+n^{(2)}|}{2}}
\Biggl[
\frac{(q^{\frac34+\frac{|n^{(1)}+n^{(2)}|}{2}}t^{-1}x^{\mp};q)_{\infty}}
{(q^{\frac14+\frac{|n^{(1)}+n^{(2)}|}{2}}tx^{\pm};q)_{\infty}}
\frac{(q^{\frac34+\frac{|n^{(1)}+n^{(2)}|}{2}}tz^{\mp};q)_{\infty}}
{(q^{\frac14+\frac{|n^{(1)}+n^{(2)}|}{2}}t^{-1}z^{\pm};q)_{\infty}}
\nonumber\\
&+\frac{(-q^{\frac34+\frac{|n^{(1)}+n^{(2)}|}{2}}t^{-1}x^{\mp};q)_{\infty}}
{(-q^{\frac14+\frac{|n^{(1)}+n^{(2)}|}{2}}tx^{\pm};q)_{\infty}}
\frac{(-q^{\frac34+\frac{|n^{(1)}+n^{(2)}|}{2}}tz^{\mp};q)_{\infty}}
{(-q^{\frac14+\frac{|n^{(1)}+n^{(2)}|}{2}}t^{-1}z^{\pm};q)_{\infty}}
\Biggr]. 
\end{align}
This can be identified with the line defect index for the flavor vortex line of charges 
$(\mathsf{q}_1,\mathsf{q}_2)$ $=$
$(n^{(1)}+n^{(2)},n^{(1)}+n^{(2)})$ 
for the flavor symmetry rotating the hyper and twisted hyper 
in the dual $\mathbb{Z}_2$ even gauge theory. 
The overall factor $q^{\frac{|n^{(1)}+n^{(2)}|}{2}}$ corresponds to the contributions from 
a pair of the flavor monopoles of charges $n^{(1)}+n^{(2)}$. 
So we have the duality
\begin{align}
\begin{matrix}
\textrm{gauge $W_{\mathsf{p}^{(1)}=2n^{(1)},\mathsf{p}^{(2)}=2n^{(2)}}$}\\ 
 \\ 
\textrm{in $U(1)_2\times U(1)_{-2}$ ABJM}\\
\end{matrix}
\qquad 
&\leftrightarrow 
\qquad 
\begin{matrix}
\textrm{flavor $V_{\mathsf{q}_1=n^{(1)}+n^{(2)},\mathsf{q}_2=n^{(1)}+n^{(2)}}$}\\
 \\
\textrm{in $\mathbb{Z}_2^+$ gauge theory}\\
\end{matrix}
\end{align}
With both odd charges we find
\begin{align}
\label{ABJMu1k2Wodd}
&
{\langle W_{\mathsf{p}^{(1)}=2n^{(1)}-1,\mathsf{p}^{(2)}=2n^{(2)}-1}\rangle^{U(1)_{2}\times U(1)_{-2}}}(t,x,z;q)
\nonumber\\
&=
\frac12 
q^{\frac{|n^{(1)}+n^{(2)}|}{2}}
\Biggl[
\frac{(q^{\frac34+\frac{|n^{(1)}+n^{(2)}|}{2}}t^{-1}x^{\mp};q)_{\infty}}
{(q^{\frac14+\frac{|n^{(1)}+n^{(2)}|}{2}}tx^{\pm};q)_{\infty}}
\frac{(q^{\frac34+\frac{|n^{(1)}+n^{(2)}|}{2}}tz^{\mp};q)_{\infty}}
{(q^{\frac14+\frac{|n^{(1)}+n^{(2)}|}{2}}t^{-1}z^{\pm};q)_{\infty}}
\nonumber\\
&-\frac{(-q^{\frac34+\frac{|n^{(1)}+n^{(2)}|}{2}}t^{-1}x^{\mp};q)_{\infty}}
{(-q^{\frac14+\frac{|n^{(1)}+n^{(2)}|}{2}}tx^{\pm};q)_{\infty}}
\frac{(-q^{\frac34+\frac{|n^{(1)}+n^{(2)}|}{2}}tz^{\mp};q)_{\infty}}
{(-q^{\frac14+\frac{|n^{(1)}+n^{(2)}|}{2}}t^{-1}z^{\pm};q)_{\infty}}
\Biggr]. 
\end{align} 
This is the line defect index for the flavor vortex line of charges $(n^{(1)}+n^{(2)},n^{(1)}+n^{(2)})$ 
in the $\mathbb{Z}_2$ odd gauge theory. 
Thus when we introduce a pair of Wilson lines with odd charges in the $U(1)_2\times U(1)_{-2}$ ABJM theory, 
the dual description as the $\mathbb{Z}_2$ even gauge theory transforms onto the $\mathbb{Z}_2$ odd gauge theory
\begin{align}
\begin{matrix}
\textrm{gauge $W_{\mathsf{p}^{(1)}=2n^{(1)}-1,\mathsf{p}^{(2)}=2n^{(2)}-1}$}\\ 
 \\ 
\textrm{in $U(1)_2\times U(1)_{-2}$ ABJM}\\
\end{matrix}
\qquad 
&\leftrightarrow 
\qquad 
\begin{matrix}
\textrm{flavor $V_{\mathsf{q}_1=n^{(1)}+n^{(2)},\mathsf{q}_2=n^{(1)}+n^{(2)}}$}\\
 \\
\textrm{in $\mathbb{Z}_2^-$ gauge theory}\\
\end{matrix}.
\end{align}

Next consider the vortex lines of the $U(1)_2\times U(1)_{-2}$ ABJM theory. 
The theory has the flavor vortex line of charges $(\mathsf{q}_1,\mathsf{q}_2)$ 
associated with the global symmetry rotating the hyper and twisted hyper. 
The one-point function of the flavor vortex line is
\begin{align}
\label{ABJMu1k2V}
&
{\langle V_{\mathsf{q}_1,\mathsf{q}_2}\rangle^{U(1)_{2}\times U(1)_{-2}}}(t,x,z;q)
\nonumber\\
&=
\sum_{m^{(1)},m^{(2)}\in \mathbb{Z}}
\oint \frac{ds^{(1)}}{2\pi is^{(1)}} 
\oint \frac{ds^{(2)}}{2\pi is^{(2)}} 
\nonumber\\
&\times 
\frac{(q^{\frac34+\frac{|m^{(1)}-m^{(2)}+\mathsf{q}_1|}{2}}t^{-1}s^{(1)\mp}s^{(2)\pm}x^{\mp};q)_{\infty}
(q^{\frac34+\frac{|m^{(1)}-m^{(2)}+\mathsf{q}_2|}{2}}ts^{(1)\mp}s^{(2)\pm}z^{\pm};q)_{\infty}
}
{
(q^{\frac14+\frac{|m^{(1)}-m^{(2)}+\mathsf{q}_1|}{2}}ts^{(1)\pm}s^{(2)\mp}x^{\pm};q)_{\infty}
(q^{\frac14+\frac{|m^{(1)}-m^{(2)}+\mathsf{q}_2|}{2}}t^{-1}s^{(1)\pm}s^{(2)\mp}z^{\mp};q)_{\infty}
}
\nonumber\\
&\times 
{s^{(1)}}^{2m^{(1)}} {s^{(2)}}^{-2m^{(2)}}
q^{\frac{|m^{(1)}-m^{(2)}+\mathsf{q}_1|}{4}+\frac{|m^{(1)}-m^{(2)}+\mathsf{q}_2|}{4}}
t^{|m^{(1)}-m^{(2)}+\mathsf{q}_2|-|m^{(1)}-m^{(2)}+\mathsf{q}_1|}.
\end{align}
This can be evaluated as
\begin{align}
\label{ABJMu1k2V2}
&
{\langle V_{\mathsf{q}_1,\mathsf{q}_2}\rangle^{U(1)_{2}\times U(1)_{-2}}}(t,x,z;q)
\nonumber\\
&=
\frac12 
q^{\frac{|\mathsf{q}_1|+|\mathsf{q}_2|}{4}}
t^{\mathsf{q}_2-\mathsf{q}_1}
\Biggl[
\frac{(q^{\frac34+\frac{|\mathsf{q}_1|}{2}}t^{-1}x^{\mp};q)_{\infty}}
{(q^{\frac14+\frac{|\mathsf{q}_1|}{2}}tx^{\pm};q)_{\infty}}
\frac{(q^{\frac34+\frac{|\mathsf{q}_2|}{2}}tz^{\mp};q)_{\infty}}
{(q^{\frac14+\frac{|\mathsf{q}_2|}{2}}t^{-1}z^{\pm};q)_{\infty}}
\nonumber\\
&+\frac{(-q^{\frac34+\frac{|\mathsf{q}_1|}{2}}t^{-1}x^{\mp};q)_{\infty}}
{(-q^{\frac14+\frac{|\mathsf{q}_1|}{2}}tx^{\pm};q)_{\infty}}
\frac{(-q^{\frac34+\frac{|\mathsf{q}_2|}{2}}tz^{\mp};q)_{\infty}}
{(-q^{\frac14+\frac{|\mathsf{q}_2|}{2}}t^{-1}z^{\pm};q)_{\infty}}
\Biggr]. 
\end{align}
Here the overall factor $q^{\frac{|\mathsf{q}_1|+|\mathsf{q}_2|}{4}} t^{\mathsf{q}_2-\mathsf{q}_1}$ stems from 
the monopole of charges $\mathsf{q}_1$ for the flavor symmetry rotating the hyper 
and that of charge $\mathsf{q}_2$ for the twisted hyper. 
Again this can be understood as the line defect index for the flavor vortex line of charges $(\mathsf{q}_1,\mathsf{q}_2)$ in the $\mathbb{Z}_2$ even gauge theory. 
Thus
\begin{align}
\begin{matrix}
\textrm{flavor $V_{\mathsf{q}_1,\mathsf{q}_2}$}\\ 
 \\ 
\textrm{in $U(1)_2\times U(1)_{-2}$ ABJM}\\
\end{matrix}
\qquad 
&\leftrightarrow 
\qquad 
\begin{matrix}
\textrm{flavor $V_{\mathsf{q}_1,\mathsf{q}_2}$}\\
 \\
\textrm{in $\mathbb{Z}_2^+$ gauge theory}\\
\end{matrix}.
\end{align}
Also when the two vortex charges are equal $\mathsf{q}_1=\mathsf{q}_2=n^{(1)}+n^{(2)}$, 
the expression (\ref{ABJMu1k2V2}) becomes the correlation function (\ref{ABJMu1k2Weven}) of the Wilson lines of even charges. 
This indicates the following duality: 
\begin{align}
\begin{matrix}
\textrm{flavor $V_{\mathsf{q}_1=n^{(1)}+n^{(2)},\mathsf{q}_2=n^{(1)}+n^{(2)}}$}\\
 \\
\textrm{in $U(1)_2\times U(1)_{-2}$ ABJM}\\
\end{matrix}
\qquad 
\leftrightarrow 
\qquad 
\begin{matrix}
\textrm{gauge $W_{\mathsf{p}^{(1)}=2n^{(1)},\mathsf{p}^{(2)}=2n^{(2)}}$}\\ 
 \\ 
\textrm{in $U(1)_2\times U(1)_{-2}$ ABJM}\\
\end{matrix}. 
\end{align}

\subsection{$U(1)_k \times U(1)_{-k}$ ABJM theory}
Now we propose the generalization for the $U(1)_k\times U(1)_{-k}$ ABJM theory. 
The theory is dual to the $\mathbb{Z}_k$ gauge theory coupled to the hypermultiplet and twisted hypermultiplet \cite{Hayashi:2022ldo}. 

According to the Gauss law constraint, the Wilson line is non-vanishing only if the total electric charge is a multiple of $k$. 
We can write a pair the electric charges for which the Wilson line defect index is non-trivial as 
$(\mathsf{p}^{(1)},\mathsf{p}^{(2)})$ $=$ $(kn^{(1)}+i, kn^{(2)}-i)$, 
where $n^{(1)}$, $n^{(2)}$ $\in$ $\mathbb{Z}$ and $i=0,1,\cdots, k-1$. 
Then the Wilson line defect index reads
\begin{align}
\label{ABJMu1kW}
&
{\langle W_{\mathsf{p}^{(1)}=kn^{(1)}+i,\mathsf{p}^{(2)}=kn^{(2)}-i}\rangle^{U(1)_{k}\times U(1)_{-k}}}(t,x,z;q)
\nonumber\\
&=
\sum_{m^{(1)},m^{(2)}\in \mathbb{Z}}
\oint \frac{ds^{(1)}}{2\pi is^{(1)}} 
\oint \frac{ds^{(2)}}{2\pi is^{(2)}} 
\nonumber\\
&\times 
\frac{(q^{\frac34+\frac{|m^{(1)}-m^{(2)}|}{2}}t^{-1}s^{(1)\mp}s^{(2)\pm}x^{\mp};q)_{\infty}
(q^{\frac34+\frac{|m^{(1)}-m^{(2)}|}{2}}ts^{(1)\mp}s^{(2)\pm}z^{\pm};q)_{\infty}
}
{
(q^{\frac14+\frac{|m^{(1)}-m^{(2)}|}{2}}ts^{(1)\pm}s^{(2)\mp}x^{\pm};q)_{\infty}
(q^{\frac14+\frac{|m^{(1)}-m^{(2)}|}{2}}t^{-1}s^{(1)\pm}s^{(2)\mp}z^{\mp};q)_{\infty}
}
\nonumber\\
&\times 
{s^{(1)}}^{km^{(1)}+kn^{(1)}+i} {s^{(2)}}^{-km^{(2)}+kn^{(2)}-i}
q^{\frac{|m^{(1)}-m^{(2)}|}{2}}. 
\end{align}
This is invariant under the mirror transformation (\ref{mirror_trans}). 
We find that it is given by
\begin{align}
\label{ABJMu1kW2}
&
{\langle W_{\mathsf{p}^{(1)}=kn^{(1)}+i,\mathsf{p}^{(2)}=kn^{(2)}-i}\rangle^{U(1)_{k}\times U(1)_{-k}}}(t,x,z;q)
\nonumber\\
&=
\frac{1}{k}
q^{\frac{|n^{(1)}+n^{(2)}|}{2}}
\sum_{l=0}^{k-1}
\Biggl[
\omega^{i(k-l)} 
\frac{(q^{\frac34+\frac{|n^{(1)}+n^{(2)}|}{2}}t^{-1}\omega^{\mp l}x^{\mp};q)_{\infty}}
{(q^{\frac14+\frac{|n^{(1)}+n^{(2)}|}{2}}t\omega^{\pm l}x^{\pm};q)_{\infty}}
\frac{(q^{\frac34+\frac{|n^{(1)}+n^{(2)}|}{2}}t\omega^{\mp l}z^{\pm};q)_{\infty}}
{(q^{\frac14+\frac{|n^{(1)}+n^{(2)}|}{2}}t^{-1}\omega^{\pm l}z^{\mp};q)_{\infty}}
\Biggr], 
\end{align}
where $\omega$ $=$ $e^{2\pi i/k}$ is the discrete gauge fugacity for the $\mathbb{Z}_k$ gauge group. 
The overall factor $q^{\frac{|n^{(1)}+n^{(2)}|}{2}}$ is a contribution from the background monopole. 
When $i=0$, i.e. each of gauge charges is a multiple of $k$, 
the expression (\ref{ABJMu1kW2}) is identified with the one-point function 
for the flavor vortex line of charges $(n^{(1)}+n^{(2)},n^{(1)}+n^{(2)} )$ 
in the dual $\mathbb{Z}_k$ gauge theory. 
In particular, when $n^{(1)}+n^{(2)}=0$, the line defect index is equal to the index without any insertion of the line operator. 
This simply implies that the Wilson line with a pair of opposite charges can be screened by the matter fields.  
\footnote{See \cite{Bergman:2020ifi} for related discussion. }
For $i\neq 0$, we can view the expression (\ref{ABJMu1kW2}) as 
the vortex line defect correlator in one of the $(k-1)$ different $\mathbb{Z}_k$ gauge theories labeled by $i=1,\cdots, k-1$, 
for which gauging the $\mathbb{Z}_k$ symmetry is performed  
by projecting the $\mathbb{Z}_k$ charged states onto the $\mathbb{Z}_k$ gauge invariant states 
with non-trivial phase factor $\omega^{i}$. 
So we find the duality 
\begin{align}
\begin{matrix}
\textrm{gauge $W_{\mathsf{p}^{(1)}=kn^{(1)}+i,\mathsf{p}^{(2)}=kn^{(2)}-i}$}\\ 
 \\ 
\textrm{in $U(1)_k\times U(1)_{-k}$ ABJM}\\
\end{matrix}
\qquad 
&\leftrightarrow 
\qquad 
\begin{matrix}
\textrm{flavor $V_{\mathsf{q}_1=n^{(1)}+n^{(2)},\mathsf{q}_2=n^{(1)}+n^{(2)} }$}\\
 \\
\textrm{in $\mathbb{Z}_k^{(i)}$ gauge theory}\\
\end{matrix}. 
\end{align}

For the $U(1)_k \times U(1)_{-k}$ ABJM model 
we can introduce the flavor vortex lines which admit singular profile for the bifundamental hyper or/and twisted hyper. 
The vortex line is characterized by two charges $\mathsf{q}_1$ and $\mathsf{q}_2$ for the global symmetries rotating the hyper and twisted hyper. 
The vortex line defect index takes the form
\begin{align}
\label{ABJMu1kV}
&
{\langle V_{\mathsf{q}_1,\mathsf{q}_2}\rangle^{U(1)_{k}\times U(1)_{-k}}}(t,x,z;q)
\nonumber\\
&=
\sum_{m^{(1)},m^{(2)}\in \mathbb{Z}}
\oint \frac{ds^{(1)}}{2\pi is^{(1)}} 
\oint \frac{ds^{(2)}}{2\pi is^{(2)}} 
\nonumber\\
&\times 
\frac{(q^{\frac34+\frac{|m^{(1)}-m^{(2)}+\mathsf{q}_1|}{2}}t^{-1}s^{(1)\mp}s^{(2)\pm}x^{\mp};q)_{\infty}
(q^{\frac34+\frac{|m^{(1)}-m^{(2)}+\mathsf{q}_2|}{2}}ts^{(1)\mp}s^{(2)\pm}z^{\pm};q)_{\infty}
}
{
(q^{\frac14+\frac{|m^{(1)}-m^{(2)}+\mathsf{q}_1|}{2}}ts^{(1)\pm}s^{(2)\mp}x^{\pm};q)_{\infty}
(q^{\frac14+\frac{|m^{(1)}-m^{(2)}+\mathsf{q}_2|}{2}}t^{-1}s^{(1)\pm}s^{(2)\mp}z^{\mp};q)_{\infty}
}
\nonumber\\
&\times 
{s^{(1)}}^{km^{(1)}} {s^{(2)}}^{-km^{(2)}}
q^{\frac{|m^{(1)}-m^{(2)}+\mathsf{q}_1|}{4}+\frac{|m^{(1)}-m^{(2)}+\mathsf{q}_2|}{4}}
t^{|m^{(1)}-m^{(2)}+\mathsf{q}_2|-|m^{(1)}-m^{(2)}+\mathsf{q}_1|}.
\end{align}
We find that it is evaluated as
\begin{align}
\label{ABJMu1kV2}
&
{\langle V_{\mathsf{q}_1,\mathsf{q}_2}\rangle^{U(1)_{k}\times U(1)_{-k}}}(t,x,z;q)
\nonumber\\
&=
\frac{1}{k}
q^{\frac{|\mathsf{q}_1|+|\mathsf{q}_2|}{4}}
t^{\mathsf{q}_2-\mathsf{q}_1}
\sum_{l=0}^{k-1}
\Biggl[
\frac{(q^{\frac34+\frac{|\mathsf{q}_1|}{2}}t^{-1}\omega^{\mp l}x^{\mp};q)_{\infty}}
{(q^{\frac14+\frac{|\mathsf{q}_1|}{2}}t\omega^{\pm l}x^{\pm};q)_{\infty}}
\frac{(q^{\frac34+\frac{|\mathsf{q}_2|}{2}}t\omega^{\mp l}z^{\pm};q)_{\infty}}
{(q^{\frac14+\frac{|\mathsf{q}_2|}{2}}t^{-1}\omega^{\pm l}z^{\mp};q)_{\infty}}
\Biggr], 
\end{align}
where the overall factor $q^{\frac{|\mathsf{q}_1|+|\mathsf{q}_2|}{4}} t^{\mathsf{q}_2-\mathsf{q}_1}$ is the index for 
the monopole of charges $\mathsf{q}_1$ for the flavor symmetry rotating the hyper 
and that of charge $\mathsf{q}_2$ for the twisted hyper. 
The expression (\ref{ABJMu1kV2}) agrees with the line defect index of the flavor vortex line of charges $(\mathsf{q}_1,\mathsf{q}_2)$ 
in the dual $\mathbb{Z}_k$ gauge theory. 
Thus we propose the following duality: 
\begin{align}
\begin{matrix}
\textrm{flavor $V_{\mathsf{q}_1,\mathsf{q}_2}$}\\ 
 \\ 
\textrm{in $U(1)_k\times U(1)_{-k}$ ABJM}\\
\end{matrix}
\qquad 
&\leftrightarrow 
\qquad 
\begin{matrix}
\textrm{flavor $V_{\mathsf{q}_1,\mathsf{q}_2}$}\\
 \\
\textrm{in $\mathbb{Z}_k^{(0)}$ gauge theory}\\
\end{matrix}.
\end{align}
For $\mathsf{q}_1=\mathsf{q}_2$ $=$ $n^{(1)}+n^{(2)}$, 
the vortex line defect index (\ref{ABJMu1kV2}) coincides with the line defect index (\ref{ABJMu1kW2}) 
of the Wilson line of charges $(\mathsf{p}^{(1)},\mathsf{p}^{(2)})$ $=$ $(kn^{(1)},kn^{(2)})$. 
This would imply the duality of the vortex and Wilson lines in the $U(1)_k\times U(1)_{-k}$ ABJM theory 
\begin{align}
\begin{matrix}
\textrm{flavor $V_{\mathsf{q}_1=n^{(1)}+n^{(2)},\mathsf{q}_2=n^{(1)}+n^{(2)}}$}\\
 \\
\textrm{in $U(1)_k\times U(1)_{-k}$ ABJM}\\
\end{matrix}
\qquad 
\leftrightarrow 
\qquad 
\begin{matrix}
\textrm{gauge $W_{\mathsf{p}^{(1)}=kn^{(1)},\mathsf{p}^{(2)}=kn^{(2)}}$}\\ 
 \\ 
\textrm{in $U(1)_k\times U(1)_{-k}$ ABJM}\\
\end{matrix}. 
\end{align}

\subsection{$U(2)_1 \times U(1)_{-1}$ ABJ theory}
The $U(2)_1\times U(1)_{-1}$ ABJ theory is dual to the $U(1)_1\times U(1)_{-1}$ ABJM theory. 
We propose a simple example of the duality of the Wilson line operators in these theories. 

Let $W_{\mathcal{R},\mathsf{p}}$ be the Wilson line operator in the $U(2)_1\times U(1)_{-1}$ ABJ theory 
that is characterized by the irreducible representation $\mathcal{R}$ of the $U(2)$ gauge group and the $U(1)$ charge $\mathsf{p}$. 
We find that the $k$-point function of the Wilson lines transforming in the rank-$2$ antisymmetric representation under the $U(2)$ gauge group 
and carrying charge $\mathsf{p}_i$ $=$ $n_i$, with $i=1,\cdots, k$
\begin{align}
&{\langle \underbrace{W_{\tiny \yng(1,1),n_1} \cdots W_{\tiny \yng(1,1),n_k}}_{k} \rangle^{U(2)_{1}\times U(1)_{-1}}}(t,x,z;q)\nonumber \\
&=
\sum_{m_i^{(1)},m^{(2)}\in \mathbb{Z}}
\prod_{i=1}^2\oint \frac{ds_i^{(1)}}{2\pi is_i^{(1)}} 
\oint \frac{ds^{(2)}}{2\pi is^{(2)}} 
\prod_{i=1}^2
\frac{(q^{\frac34+\frac{|m^{(1)}_i-m^{(2)}|}{2}}t^{-1}s^{(1)\mp}_is^{(2)\pm}x^{\mp};q)_{\infty}
}{
(q^{\frac14+\frac{|m^{(1)}_i-m^{(2)}|}{2}}ts^{(1)\pm}_is^{(2)\mp}x^{\pm};q)_{\infty}
}\nonumber \\
&\quad\times
\frac{
(q^{\frac34+\frac{|m^{(1)}_i-m^{(2)}|}{2}}ts^{(1)\mp}_is^{(2)\pm}z^{\pm};q)_{\infty}
}{
(q^{\frac14+\frac{|m^{(1)}_i-m^{(2)}|}{2}}t^{-1}s^{(1)\pm}_is^{(2)\mp}z^{\mp};q)_{\infty}
}
 \prod_{i=1}^2{s_i^{(1)}}^{m^{(1)}_i+k} {s^{(2)}}^{-m^{(2)}+\sum_{i=1}^kn_i}\nonumber \\
&\quad\times q^{-\frac{|m^{(1)}_1-m^{(1)}_2|}{2}+\sum_{i=1}^2\frac{|m^{(1)}_i-m^{(2)}|}{2}}
\end{align}
is given by
\begin{align}
\label{ABJu2u1k1_(W11Wn)^k}
&
{\langle \underbrace{W_{\tiny \yng(1,1),n_1} \cdots W_{\tiny \yng(1,1),n_k}}_{k} \rangle^{U(2)_{1}\times U(1)_{-1}}}(t,x,z;q)
\nonumber\\
&=q^{\frac{k+\sum_{i=1}^{k}n_i}{2}}
x^{k+\sum_{i=1}^{k}n_i}z^{-k-\sum_{i=1}^{k}n_i}
\frac{(q^{\frac34+\frac{|k+\sum_{i=1}^{k}n_i|}{2}}t^{-1}x^{\mp};q)_{\infty}}
{(q^{\frac14+\frac{|k+\sum_{i=1}^{k}n_i|}{2}}t x^{\pm};q)_{\infty}}
\frac{(q^{\frac34+\frac{|k+\sum_{i=1}^{k}n_i|}{2}}tz^{\mp};q)_{\infty}}
{(q^{\frac14+\frac{|k+\sum_{i=1}^{k}n_i|}{2}}t^{-1} z^{\pm};q)_{\infty}}. 
\end{align}
Note that this is equal to the one-point function of the Wilson line $W_{\emptyset, k+\sum_{i=1}^{k}n_i}$ 
of the trivial representation under the $U(2)$ and of $(k+\sum_{i=1}^{k}n_i)$ units of the $U(1)$ charge.
This follows from the fact that an insertion of the rank-$2$ antisymmetric Wilson line for $U(2)$ can be replaced with that of the Wilson line of charge $+1$ for $U(1)$ by shifting the monopole charges $m^{(1)}_i,m^{(2)}$ uniformly.
As the overall factor can be regarded as contributions from the flavor Wilson lines, we find the following duality: 
\begin{align}
&
\begin{matrix}
\textrm{gauge $W_{\emptyset,n^{(1)}+n^{(2)}}$ $+$ flavor$_x$ $W_{-n^{(1)}-n^{(2)}}$ $+$ flavor$_z$ $W_{n^{(1)}+n^{(2)}}$}\\
 \\
\textrm{in $U(2)_1\times U(1)_{-1}$ ABJ}\\
\end{matrix}
\nonumber\\
&
\nonumber\\
&\qquad 
\leftrightarrow 
\qquad \qquad 
\begin{matrix}
\textrm{gauge $W_{\mathsf{p}^{(1)}=n^{(1)},\mathsf{p}^{(2)}=n^{(2)}}$}\\ 
 \\ 
\textrm{in $U(1)_1\times U(1)_{-1}$ ABJM}\\
\end{matrix}. 
\end{align}
It would be intriguing to explore more dualities of line operators in the ABJ(M) theories. 

\subsection{$U(1)_1\times U(1)_0^{l-1}\times U(1)_{-1}$ circular quiver CS theory}
Let us consider a generalization of the ABJM theory obtained by adding $(l-1)$ $(1,k)$5-branes to the Type IIB brane construction of the ABJM theory 
\cite{Hosomichi:2008jd,Imamura:2008nn,Imamura:2008dt}.
This theory is an ${\cal N}=4$ $U(1)_k\times U(1)_0^{l-1}\times U(1)_{-k}$ circular quiver CS theory with a bifundamental twisted hypermultiplet between each pair of $I$-th gauge node and $(I+1)$-th gauge node ($I=1,\cdots,l$) and a bifundamental twisted hypermultiplet between the first and the last gauge nodes (see Figure \ref{fig:CSbrane2}).
To distinguish from the linear quiver CS theory in the superscript of the line defect index, let us refer to the circular quiver CS theory with $(l+1)$ gauge nodes and the CS levels $(k,0,\cdots,0,-k)$ as ``$(l,1)_k$ model''.
Let us define the line defect index of the $(l,1)_k$ model as\footnote{
The convention in \eqref{l1kmodelLpq} is related to the convention adopted in section \ref{sec_ABJM} \eqref{ABJMu1k1VW} as
\begin{align}
\langle L_{\mathsf{p}^{(1)},\mathsf{p}^{(2)}=n^{(2)};\mathsf{q}_1=\ell_1,\mathsf{q}_2=\ell_2}\rangle^{U(1)_1\times U(1)_{-1}}(t,x,z;q)
=\langle L_{\mathsf{p}^{(1)},\mathsf{p}^{(2)}=n^{(2)}-\ell_2;\mathsf{q}=\ell_2-\ell_1}\rangle^{(1,1)_1}(t,x,z;q).
\end{align}
}
\begin{align}
&\langle L_{\mathsf{p}^{(I)};\mathsf{q}}\rangle^{(l,1)_k}(t,x,y_\alpha,z;q)
=\sum_{m^{(1)},\cdots,m^{(l+1)}}
\frac{
(q^{\frac{1}{2}}t^{-2};q)_\infty^{l-1}
}{
(q^{\frac{1}{2}}t^2;q)_\infty^{l-1}
}
\prod_{I=1}^{l+1}\oint\frac{ds^{(I)}}{2\pi is^{(I)}}
(s^{(1)})^{km^{(1)}+\mathsf{p}^{(1)}}
\prod_{I=2}^l(s^{(I)})^{\mathsf{p}^{(I)}}\nonumber \\
&\quad\times (s^{(l+1)})^{-km^{(l+1)}+\mathsf{p}^{(l+1)}}
\prod_{I=1}^l
\frac{
(q^{\frac{3}{4}+\frac{|m^{(I)}-m^{(I+1)}|}{2}}t(z\frac{s^{(I)}}{s^{(I+1)}})^{\mp 1};q)_\infty
}{
(q^{\frac{1}{4}+\frac{|m^{(I)}-m^{(I+1)}|}{2}}t^{-1}(z\frac{s^{(I)}}{s^{(I+1)}})^{\pm 1};q)_\infty
}\nonumber \\
&\quad\times \frac{
(q^{\frac{3}{4}+\frac{|m^{(l+1)}-m^{(1)}+\mathsf{q}|}{2}}t^{-1}(x\frac{s^{(l+1)}}{s^{(1)}})^{\mp 1};q)_\infty
}{
(q^{\frac{1}{4}+\frac{|m^{(l+1)}-m^{(1)}+\mathsf{q}|}{2}}t(x\frac{s^{(l+1)}}{s^{(1)}})^{\pm 1};q)_\infty
}
q^{\frac{1}{4}(\sum_{I=1}^l|m^{(I)}-m^{(I+1)}|+|m^{(l+1)}-m^{(1)}+\mathsf{q}|)}\nonumber \\
&\quad\times t^{\sum_{I=1}^l|m^{(I)}-m^{(I+1)}|-|m^{(l+1)}-m^{(1)}+\mathsf{q}|}
y_2^{m^{(1)}}
\prod_{\alpha=2}^{l-1}\Bigl(\frac{y_{\alpha+1}}{y_{\alpha}}\Bigr)^{m^{(\alpha)}}
\Bigl(\frac{y_1}{y_l}\Bigr)^{m^{(l)}}
\Bigl(\frac{1}{y_1}\Bigr)^{m^{(l+1)}}.
\label{l1kmodelLpq}
\end{align}

For $k=1$, the brane construction suggests that the $(l,1)_1$ models is dual to the ADHM theory with $l$ fundamental hypermultiplet.
Indeed, we find that the line defect index of $(l,1)_1$ model satisfies the duality relation with the line defect index of the ADHM theory as
\begin{align}
&\langle L_{\mathsf{p}^{(I)}=n^{(I)};\mathsf{q}=n}\rangle^{(l,1)_1}(t,x,y_\alpha,z;q)\nonumber \\
&=
\langle
L_{\mathsf{p}=\sum_{I=1}^{l+1}n^{(I)};\mathsf{q}_0=\sum_{I=1}^{l+1}n^{(I)}+n,\mathsf{q}_1=0,\mathsf{q}_{\alpha\ge 2}=\sum_{I=\alpha}^ln^{(I)}}
\rangle^{U(1)\text{ADHM-}[l]}(t,x,y_\alpha,z;q)\nonumber \\
&=\langle L_{\mathsf{p}^{(1)}=n^{(1)}+n^{(l+1)}+n,\mathsf{p}^{(I\ge 2)}=n^{(I)};\mathsf{q}=\sum_{I=1}^{l+1}n^{(I)}}\rangle^{\widetilde{U(1)^{\otimes l}\textrm{mADHM-}[1]}}(t,y_\alpha,z,x;q),
\label{l1k-ADHMduality}
\end{align}
which indicates the following triality of the line operators
\begin{align}
\begin{tikzpicture}[scale=0.3]
\draw [<->] (12,8)--(16,11);
\draw [<->] (32,8)--(28,11);
\draw [<->] (20,3)--(27,3);
\node [above] at (23,17) {vortex-Wilson};
\node [above] at (23,14) {$L_{\mathsf{p}^{(I)}=n^{(I)};\mathsf{q}=n}$};
\node [above] at (23,11) {in $(l,1)_1$ model};
\node [above] at (8,6) {vortex-Wilson};
\node [above] at (8,3) {$L_{\mathsf{p}=\sum_{I=1}^{l+1}n^{(I)};\mathsf{q}_0=\sum_{I=1}^{l+1}n^{(I)}+n,\mathsf{q}_1=0,\mathsf{q}_{\alpha\ge 2}=\sum_{I=\alpha}^{l}n^{(I)}}$};
\node [above] at (8,0) {in $U(1)$ ADHM-[$l$]};
\node [above] at (37,6) {vortex-Wilson};
\node [above] at (37,3) {$L_{\mathsf{p}^{(I)}=n^{(I)}+(n^{(l+1)}+n)\delta^I_1;\mathsf{q}=\sum_{I=1}^{l+1}n^{(I)}}$};
\node [above] at (37,0) {in $\widetilde{U(1)^{\otimes l}\textrm{mADHM-}[1]}$};
\end{tikzpicture}. 
\end{align}

\section*{Acknowledgement}
The authors would like to thank Jean-Emile Bourgine, Luca Cassia and Douglas J.~Smith for useful discussions and comments.
The work of H.H.~is supported in part by JSPS KAKENHI Grand Number JP23K03396.
The work of T.N.~is supported by the Startup Funding no.~2302-SRFP-2024-0012 of Shanghai Institute for Mathematics and Interdisciplinary Sciences.
The work of T.O.~was supported by the Startup Funding no.~4007012317 of the Southeast University.

\appendix

\section{Proofs of identities in Coulomb/Higgs limit}
\label{app_proofHC}

In this section we prove the duality relation we proposed between the line defect indices \eqref{triality_SQED} and \eqref{l1k-ADHMduality} in the Coulomb limit and the Higgs limit \eqref{CoulombHiggslimit}.
The following two formulas play key roles for this purpose.
\begin{align}
&f(s)=\oint \frac{ds'}{2\pi is'}\sum_{n\in\mathbb{Z}}\Bigl(\frac{s}{s'}\Bigr)^nf(s'), \label{formuladelta} \\
&I_m=\oint\frac{ds}{2\pi is}s^m\prod_\pm\frac{1}{1-\mathfrak{t} s^{\pm 1}}=\frac{\mathfrak{t}^{|m|}}{1-\mathfrak{t}^2},\quad (m\in\mathbb{Z}),\label{formulaI}
\end{align}
where in the first formula $f(s)$ is any function which can be expanded as $f(s)=\sum_{n\in\mathbb{Z}}s^n f_n$.

\subsection{(mirror) SQED and linear quiver CS theory}
\label{app_proofHClinear}

First let us consider the duality \eqref{triality_SQED} among the $\text{SQED}_l$ \eqref{SQEDlLnm}, the mirror linear quiver theory \eqref{mSQEDlLnm} and linear quiver CS theory \eqref{u1^l+1CSWn} in the Coulomb/Higgs limit.

\subsubsection{$\text{SQED}_l$}

Let us consider the line defect index of $\text{SQED}_l$ \eqref{SQEDlLnm}.
To simplify the intermediate expressions, we use the expression before substituting $\mathsf{q}_\alpha=\sum_{I=\alpha}^{l-1}n^{(I)}$
\begin{align}
&
\langle L_{\mathsf{p};\mathsf{q}_{\alpha}} \rangle^{U(1)-[l]}(t,x_\alpha,z_{a};q)
\nonumber\\
&=
\frac{(q^{\frac12}t^{2};q)_{\infty}}{(q^{\frac12}t^{-2};q)_{\infty}}
\sum_{m\in \mathbb{Z}}\oint \frac{ds}{2\pi is}
\prod_{\alpha=1}^{l-1}
\frac{(q^{\frac34+\frac{|m+\mathsf{q}_\alpha|}{2}}t^{-1}s^{\mp}x_{\alpha}^{\mp};q)_{\infty}}
{(q^{\frac14+\frac{|m+\mathsf{q}_\alpha|}{2}}ts^{\pm}x_{\alpha}^{\pm};q)_{\infty}}
\frac{(q^{\frac34+\frac{|m|}{2}}t^{-1}s^{\mp}x_{l}^{\mp};q)_{\infty}}
{(q^{\frac14+\frac{|m|}{2}}ts^{\pm}x_{l}^{\pm};q)_{\infty}}
\nonumber\\
&\times 
s^\mathsf{p} 
q^{\frac{\sum_{\alpha=1}^{l-1}|m+\mathsf{q}_\alpha|+|m|}{4}}
t^{-\sum_{\alpha=1}^{l-1}|m+\mathsf{q}_\alpha|-|m|}
z_1^{m+\mathsf{q}_1}z_2^{-m}.
\label{SQEDlLnmwithmalphakept}
\end{align}
In the Coulomb limit the line defect index of the $\text{SQED}_l$ simplifies as
\begin{align}
\langle L_{\mathsf{p};\mathsf{q}_\alpha}\rangle^{U(1)\text{-}[l]}_C(z_a;\mathfrak{t})
&=\frac{1}{1-\mathfrak{t}^2}\sum_{m\in\mathbb{Z}}\oint \frac{ds}{2\pi is} s^\mathsf{p}\mathfrak{t}^{\sum_{\alpha=1}^l|m+\mathsf{q}_\alpha|}\Bigl(\frac{z_1}{z_2}\Bigr)^m z_1^{\mathsf{q}_1}\nonumber \\
&=\frac{1}{1-\mathfrak{t}^2} \delta_{\mathsf{p},0}\sum_{m\in\mathbb{Z}}\mathfrak{t}^{\sum_{\alpha=1}^l|m+\mathsf{q}_\alpha|}
z_1^{m+\mathsf{q}_1}z_2^{-m}.
\label{SQEDCoulomb}
\end{align}

In the Higgs limit, the monopole charge $m$ in the line defect index \eqref{SQEDlLnmwithmalphakept} is restricted to $m=0$ and hence we obtain
\begin{align}
&\langle L_{\mathsf{p};\mathsf{q}_\alpha}\rangle^{U(1)\text{-}[l]}_H(x_\alpha;\mathfrak{t})\nonumber \\
&=
(1-\mathfrak{t}^2)\prod_{\alpha=1}^{l-1}\delta_{\mathsf{q}_\alpha,0}\oint\frac{dx}{2\pi is}s^\mathsf{p}\prod_\pm\prod_{\alpha=1}^l\frac{1}{1-\mathfrak{t}(sx_\alpha)^{\pm 1}}.
\end{align}
Using the delta function formula \eqref{formuladelta} and the integration formula \eqref{formulaI}, we can rewrite this integration as
\begin{align}
&\langle L_{\mathsf{p};\mathsf{q}_\alpha}\rangle^{U(1)\text{-}[l]}_H(x_\alpha;\mathfrak{t})\nonumber \\
&=
(1-\mathfrak{t}^2)
\prod_{\alpha=1}^{l-1}\delta_{\mathsf{q}_\alpha,0}\sum_{n_1,\cdots,n_{l-1}\in\mathbb{Z}}\prod_{\alpha=1}^{l-1}
\Bigl(\frac{sx_\alpha}{\sigma_\alpha}\Bigr)^{n_\alpha}\oint\frac{d\sigma_\alpha}{2\pi i\sigma_\alpha}\prod_\pm\frac{1}{1-\mathfrak{t}\sigma_\alpha^{\pm 1}}
\oint\frac{ds}{2\pi is}s^\mathsf{p}\nonumber \\
&\quad\times \prod_\pm\frac{1}{1-\mathfrak{t}(sx_l)^{\pm 1}}\nonumber \\
&=\frac{1}{(1-\mathfrak{t}^2)^{l-1}}
\prod_{\alpha=1}^{l-1}\delta_{\mathsf{q}_\alpha,0}\sum_{n_1,\cdots,n_{l-1}\in\mathbb{Z}}
\prod_{\alpha=1}^{l-1}x_\alpha^{n_\alpha}
x_l^{-\mathsf{p}-\sum_{\alpha=1}^{l-1}n_\alpha}
\mathfrak{t}^{\sum_{\alpha=1}^{l-1}|n_\alpha|+|\mathsf{p}+\sum_{\alpha=1}^{l-1}n_\alpha|}.
\label{SQEDlLnmHiggs}
\end{align}


\subsubsection{$\widetilde{[1]-U(1)^{\otimes l-1}-[1]}$}

In the line defect index of the mirror linear quiver theory \eqref{mSQEDlLnm}, the monopole charges are restricted to $m^{(1)}=\cdots=m^{(l+1)}$ in the Coulomb limit and hence \eqref{mSQEDlLnm} reduces to
\begin{align}
&\langle L_{\mathsf{p}^{(I)};\mathsf{q}}\rangle^{\widetilde{[1]-U(1)^{\otimes l-1}-[1]}}_C(z_{a};\mathfrak{t})\nonumber \\
&=(1-\mathfrak{t})^{l-1}\delta_{\mathsf{q},0}\prod_{I=1}^{l-1}\oint\frac{ds^{(I)}}{2\pi is^{(I)}}
\prod_\pm
\frac{1}{1-\mathfrak{t}(\frac{s^{(1)}}{z_1})^{\pm 1}}
\prod_{I=1}^{l-2}\frac{1}{1-\mathfrak{t}(\frac{s^{(I+1)}}{s^{(I)}})^{\pm 1}}
\frac{1}{1-\mathfrak{t}(\frac{s^{(l-1)}}{z_2})^{\pm 1}}\nonumber \\
&\quad\times \prod_{I=1}^{l-1}(s^{(I)})^{\mathsf{p}^{(I)}}\nonumber \\
&=(1-\mathfrak{t})^{l-1}\delta_{\mathsf{q},0}z_1^{\sum_{I=1}^{l-1}\mathsf{p}^{(I)}}\prod_{I=1}^{l-1}\oint\frac{d\sigma^{(I)}}{2\pi i\sigma^{(I)}}
(\sigma^{(I)})^{\sum_{J=I}^{l-1}\mathsf{p}^{(J)}}
\prod_\pm
\frac{1}{1-\mathfrak{t}(\sigma^{(I)})^{\pm 1}}\nonumber \\
&\quad\times \prod_\pm 
\frac{1}{1-\mathfrak{t}(\frac{z_1}{z_2}\prod_{J=1}^{l-1}\sigma^{(J)})^{\pm1}},
\end{align}
where in the second line we have chosen new integration variables $\sigma^{(I)}$ as $s^{(1)}=z_1\sigma^{(1)}$ and $s^{(I)}=s^{(I)}\sigma^{(2)}$ ($I=2,\cdots,l-1$).
By using the formula for the delta function on the circle \eqref{formuladelta}, we can further rewrite the last factor to obtain
\begin{align}
&\langle L_{\mathsf{p}^{(I)};\mathsf{q}}\rangle^{\widetilde{[1]-U(1)^{\otimes l-1}-[1]}}_C(z_{a};\mathfrak{t})\nonumber \\
&=(1-\mathfrak{t})^{l-1}\delta_{\mathsf{q},0}z_1^{\sum_{I=1}^{l-1}\mathsf{p}^{(I)}}\sum_{n\in\mathbb{Z}}\Bigl(\frac{z_1}{z_2}\Bigr)^n\prod_{I=1}^{l-1}\oint\frac{d\sigma^{(I)}}{2\pi i\sigma^{(I)}}
(\sigma^{(I)})^{n+\sum_{J=I}^{l-1}\mathsf{p}^{(J)}}
\prod_\pm
\frac{1}{1-\mathfrak{t}(\sigma^{(I)})^{\pm 1}}\nonumber \\
&\quad\times
\oint \frac{d\rho}{2\pi i\rho}
\rho^{-n}
\prod_\pm 
\frac{1}{1-\mathfrak{t}\rho^{\pm 1}}.
\end{align}
Now we can perform the integrations over $\sigma^{(I)}$ and $\rho$ by the formula \eqref{formulaI} and obtain
\begin{align}
\langle L_{\mathsf{p}^{(I)};\mathsf{q}}\rangle^{\widetilde{[1]-U(1)^{\otimes l-1}-[1]}}_C(z_{a};\mathfrak{t})
=\frac{1}{1-\mathfrak{t}^2}
\delta_{\mathsf{q},0}
\sum_{n\in\mathbb{Z}}\mathfrak{t}^{\sum_{\alpha=1}^l|n+\sum_{I=\alpha}^{l-1}\mathsf{p}^{(I)}|}
z_1^{n+\sum_{I=1}^{l-1}\mathsf{p}^{(I)}}
z_2^{-n}.
\label{mirrorSQEDCoulomb}
\end{align}
By comparing \eqref{SQEDCoulomb} with \eqref{mirrorSQEDCoulomb}, we find that the line defect index of the mirror theory agree with the line defect index of $\text{SQED}_l$ in the Coulomb limit with the following identification of the charges
\begin{align}
\langle L_{\mathsf{p}=n;\mathsf{q}_\alpha=\sum_{I=\alpha}^{l-1}n^{(I)}} \rangle^{U(1)-[l]}_C(z_{a};\mathfrak{t})
=\langle L_{\mathsf{p}^{(I)}=n^{(I)};\mathsf{q}=n}\rangle^{\widetilde{[1]-U(1)^{\otimes l-1}-[1]}}_C(z_{a};\mathfrak{t}).
\label{SQEDCoulombmirrorparameteridentification}
\end{align}

On the other hand, the line defect index of the mirror linear quiver theory \eqref{mSQEDlLnm} simplifies in the Higgs limit as
\begin{align}
&\langle L_{\mathsf{p}^{(I)};\mathsf{q}}\rangle^{\widetilde{[1]-U(1)^{\otimes l-1}-[1]}}_H(x_\alpha;\mathfrak{t})\nonumber \\
&=\frac{1}{(1-\mathfrak{t}^2)^{l-1}}\prod_{I=1}^{l-1}\delta_{\mathsf{p}^{(I)},0}\sum_{m^{(1)},\cdots,m^{(l-1)}\in\mathbb{Z}}\prod_{I=1}^{l-1}\Bigl(\frac{x_I}{x_{I+1}}\Bigr)^{m^{(I)}}x_l^{-\mathsf{q}}\nonumber \\
&\quad\times \mathfrak{t}^{|m^{(1)}|+\sum_{I=1}^{l-2}|m^{(I)}-m^{(I+1)}|+|m^{(l-1)}+\mathsf{q}|}\nonumber \\
&=\frac{1}{(1-\mathfrak{t}^2)^{l-1}}
\prod_{I=1}^{l-1}\delta_{\mathsf{p}^{(I)},0}\sum_{n_1,\cdots,n_{l-1}\in\mathbb{Z}}
\prod_{I=1}^{l-1}x_I^{n_I}
x_l^{-\mathsf{q}-\sum_{I=1}^{l-1}n_I}
\mathfrak{t}^{\sum_{I=1}^{l-1}|n_I|+|\mathsf{q}+\sum_{I=1}^{l-1}n_I|}.
\label{mirrorSQEDHiggs}
\end{align}
Here in the final expression we have renamed the dummy variables $m^{(I)}$ as $n_1=m^{(1)}$ and $n_I=m^{(I)}-m^{(I-1)}$ ($I\ge 2$).
By comparing \eqref{SQEDlLnmHiggs} with \eqref{mirrorSQEDHiggs}, we find that the line defect index of the mirror theory agree with the line defect index of $\text{SQED}_l$ in the Higgs with the same parameter identification as \eqref{SQEDCoulombmirrorparameteridentification}
\begin{align}
\langle L_{\mathsf{p}=n;\mathsf{q}_\alpha=\sum_{I=\alpha}^{l-1}n^{(I)}} \rangle^{U(1)-[l]}_H(x_\alpha;\mathfrak{t})
=\langle L_{\mathsf{p}^{(I)}=n^{(I)};\mathsf{q}=n}\rangle^{\widetilde{[1]-U(1)^{\otimes l-1}-[1]}}_H(x_\alpha;\mathfrak{t}).
\end{align}


\subsubsection{Linear quiver CS theory}

Lastly let us calculate the Coulomb/Higgs limit of the line defect index of the $U(1)_1\times U(1)^{l-1}_0\times U(1)_{-1}$ linear quiver CS theory, \eqref{u1^l1+1CSWn} for $l=1$ and \eqref{u1^lge2+1CSWn} for $l\ge 2$.
In order to treat the line defect index for $l=1$ and $l\ge 2$ in the same manner, let us first rescale the integration variables $s^{(1)}$ and $s^{(l+1)}$ as $s^{(1)}\rightarrow z_1^{-1}s^{(1)}$ and $s^{(2)}\rightarrow z_2s^{(2)}$ to rewrite \eqref{u1^l1+1CSWn} and \eqref{u1^lge2+1CSWn} as
\begin{align}
&\langle W_{\mathsf{p}^{(I)}}\rangle^{U(1)_1\times U(1)_0^{l-1}\times U(1)_{-1}}(t,x_{\alpha},z_a;q)
\nonumber \\
&=
\frac{(q^{\frac12}t^{-2};q)_{\infty}^{l-1}}{(q^{\frac12}t^2;q)_{\infty}^{l-1}}
\sum_{m^{(1)},\cdots,m^{(l+1)}\in \mathbb{Z}}
\prod_{I=1}^{l+1}
\oint \frac{ds^{(I)}}{2\pi is^{(I)}}
\nonumber\\
&\quad \times 
(s^{(1)})^{m^{(1)}+\mathsf{p}^{(1)}}
\prod_{I=2}^{l} (s^{(I)})^{\mathsf{p}^{(I)}}
(s^{(l+1)})^{-m^{(l+1)}+\mathsf{p}^{(l+1)}}
\prod_{I=1}^{l}
\frac{(q^{\frac34+\frac{|m^{(I)}-m^{(I+1)}|}{2}}ts^{(I)\mp}s^{(I+1)\pm};q)_{\infty}}
{(q^{\frac14+\frac{|m^{(I)}-m^{(I+1)}|}{2}}t^{-1}s^{(I)\pm}s^{(I+1)\mp};q)_{\infty}}
\nonumber\\
&\quad \times 
q^{\frac{\sum_{I=1}^l |m^{(I)}-m^{(I+1)}|}{4}}
t^{\sum_{I=1}^l |m^{(I)}-m^{(I+1)}|}
\left( \frac{x_1}{z_1} \right)^{m^{(1)}}
\prod_{\alpha=1}^{l-1}
\left(\frac{x_{\alpha+1}}{x_{\alpha}}\right)^{m^{(\alpha+1)}}
\left( \frac{z_2}{x_l} \right)^{m^{(l+1)}}\nonumber \\
&\quad \times z_1^{-\mathsf{p}^{(1)}}z_2^{-\mathsf{p}^{(l+1)}}.
\label{u1^l+1CSWnforappendix}
\end{align}

In the Coulomb limit, the monopole charges in the line defect index of the linear quiver CS theory \eqref{u1^l+1CSWnforappendix} are restricted to $m^{(2)}=\cdots=m^{(l+1)}=m^{(1)}$ and hence \eqref{u1^l+1CSWnforappendix} reduces to
\begin{align}
&\langle W_{\mathsf{p}^{(I)}}\rangle^{U(1)_1\times U(1)_0^{l-1}\times U(1)_{-1}}_C(z_a;\mathfrak{t})\nonumber \\
&=(1-\mathfrak{t}^2)^{l-1}\sum_{m^{(1)}}\oint\frac{ds^{(I)}}{2\pi is^{(I)}}(s^{(1)})^{m^{(1)}+\mathsf{p}^{(1)}}\prod_{I=2}^l(s^{(I)})^{\mathsf{p}^{(I)}}(s^{(l+1)})^{-m^{(1)}+\mathsf{p}^{(l+1)}}
\prod_{I=1}^l\frac{1}{1-\mathfrak{t}(\frac{s^{(I+1)}}{s^{(I)}})^{\pm 1}}\nonumber \\
&\quad\times \Bigl(\frac{z_2}{z_1}\Bigr)^{m^{(1)}}
z_1^{-\mathsf{p}^{(1)}}z_2^{-\mathsf{p}^{(l+1)}}.
\end{align}
Redefining the integration variables as $s^{(1)}=\sigma^{(1)}$, $s^{(I)}=\sigma^{(I)}s^{(I-1)}$ ($I\ge 2$), we can evaluate the line defect index by using the integration formula \eqref{formulaI} as
\begin{align}
&\langle W_{\mathsf{p}^{(I)}}\rangle^{U(1)_1\times U(1)_0^{l-1}\times U(1)_{-1}}_C(z_a;\mathfrak{t})\nonumber \\
&=\frac{1}{1-\mathfrak{t}^2}\delta_{\sum_{I=1}^{l+1}\mathsf{p}^{(I)},0}\sum_{m^{(1)}}\mathfrak{t}^{\sum_{I=1}^{l}|\sum_{J=I}^{l-1}\mathsf{p}^{(J+1)}-m^{(1)}+\mathsf{p}^{(l+1)}|}
z_1^{-m^{(1)}-\mathsf{p}^{(1)}}z_2^{m^{(1)}-\mathsf{p}^{(l+1)}}.
\end{align}
If we denote $-m^{(1)}+\mathsf{p}^{(l+1)}=m$, we find
\begin{align}
&\langle W_{\mathsf{p}^{(I)}}\rangle^{U(1)_1\times U(1)_0^{l-1}\times U(1)_{-1}}_C(z_a;\mathfrak{t})\nonumber \\
&=
\frac{1}{1-\mathfrak{t}^2}\delta_{\sum_{I=1}^{l+1}\mathsf{p}^{(I)},0}\sum_m \mathfrak{t}^{\sum_{I=1}^l|m+\sum_{J=I}^{l-1}\mathsf{p}^{(J+1)}|}z_1^{m+\sum_{J=1}^{l-1}\mathsf{p}^{(J+1)}}z_2^{-m}.
\end{align}
By comparing this result with the Coulomb limit of the line defect index of $\text{SQED}_l$ \eqref{SQEDCoulomb}, we find they agree with the following parameter identification
\begin{align}
\langle W_{\mathsf{p}^{(I)}=n^{(I)}}\rangle^{U(1)_1\times U(1)_0^{l-1}\times U(1)_{-1}}_C(z_a;\mathfrak{t})
=
\langle L_{\mathsf{p}=\sum_{I=1}^{l+1}n^{(I)};\mathsf{q}_\alpha=\sum_{I=\alpha}^{l-1}n^{(I+1)}}\rangle^{U(1)\text{-}[l]}_C(z_a;\mathfrak{t}).
\label{SQEDCSchargedictionary}
\end{align}

In the Higgs limit the line defect index of the linear quiver CS theory \eqref{u1^l+1CSWnforappendix} simplifies as
\begin{align}
&\langle W_{\mathsf{p}^{(I)}}\rangle^{U(1)_1\times U(1)_0^{l-1}\times U(1)_{-1}}_H(x_\alpha;\mathfrak{t})\nonumber \\
&=\frac{1}{(1-\mathfrak{t}^2)^{l-1}}\sum_{m^{(1)},\cdots,m^{(l+1)}\in\mathbb{Z}}\prod_{I=1}^{l+1}\oint\frac{ds^{(I)}}{2\pi s^{(I)}}(s^{(1)})^{m^{(1)}+\mathsf{p}^{(1)}}
\prod_{I=2}^l(s^{(I)})^{\mathsf{p}^{(I)}}
(s^{(l+1)})^{-m^{(l+1)}+\mathsf{p}^{(l+1)}}\nonumber \\
&\quad\times \mathfrak{t}^{\sum_{I=1}^l|m^{(I)}-m^{(I+1)}|}
\Bigl(\frac{x_1}{z_1}\Bigr)^{m^{(1)}}
\prod_{\alpha=1}^{l-1}\Bigl(\frac{x_{\alpha+1}}{x_\alpha}\Bigr)^{m^{(\alpha+1)}}\Bigl(\frac{z_2}{x_l}\Bigr)^{m^{(l+1)}}
z_1^{-\mathsf{p}^{(1)}}z_2^{-\mathsf{p}^{(l+1)}}.
\end{align}
The integrations over $s^{(I)}$ gives $\delta_{m^{(1)}+\mathsf{p}^{(1)},0}$, $\delta_{\mathsf{p}^{(I)},0}$ ($I=2,\cdots,l$) and $\delta_{-m^{(l+1)}+\mathsf{p}^{(l+1)},0}$ and hence we obtain
\begin{align}
&\langle W_{\mathsf{p}^{(I)}}\rangle^{U(1)_1\times U(1)_0^{l-1}\times U(1)_{-1}}_H(x_\alpha;\mathfrak{t})\nonumber \\
&=\frac{1}{(1-\mathfrak{t}^2)^{l-1}}
x_1^{-\mathsf{p}^{(1)}}
x_l^{-\mathsf{p}^{(l+1)}}
\prod_{I=2}^{l}\delta_{\mathsf{p}^{(I)},0}
\sum_{m^{(2)},\cdots,m^{(l)}\in\mathbb{Z}}
\prod_{\alpha=1}^{l-1}
\Bigl(\frac{x_{\alpha+1}}{x_\alpha}\Bigr)^{m^{(\alpha+1)}}\nonumber \\
&\quad\times \mathfrak{t}^{|\mathsf{p}^{(1)}+m^{(2)}|+\sum_{I=2}^{l-1}|m^{(I)}-m^{(I+1)}|+|m^{(l)}-\mathsf{p}^{(l+1)}|}.
\end{align}
Comparing this expression with the Higgs index of $\text{SQED}_l$ \eqref{SQEDlLnmHiggs} after renaming the dummy variables $m^{(I)}$ ($I=2,\cdots,l$) as
\begin{align}
-\mathsf{p}^{(1)}-m^{(2)}=n_1,\quad
m^{(I)}-m^{(I+1)}=n_I\quad (I=2,\cdots,l-1),
\end{align}
we find the line defect indices agree with the same parameter identification as in the Coulomb limit \eqref{SQEDCSchargedictionary}
\begin{align}
\langle W_{\mathsf{p}^{(I)}=n^{(I)}}\rangle^{U(1)_1\times U(1)_0^{l-1}\times U(1)_{-1}}_H(x_\alpha;\mathfrak{t})
=
\langle L_{\mathsf{p}=\sum_{I=1}^{l+1}n^{(I)};\mathsf{q}_\alpha=\sum_{I=\alpha}^{l-1}n^{(I+1)}}\rangle^{U(1)\text{-}[l]}_H(x_\alpha;\mathfrak{t}).
\end{align}

\subsection{(mirror) ADHM theory and circular quiver CS theory}

Next let us consider the duality \eqref{l1k-ADHMduality} among the $U(N)$ ADHM theory \eqref{ADHMlLpq}, its mirror circular quiver theory \eqref{mirrorADHMlLpq} and the circular quiver CS theory \eqref{l1kmodelLpq} in the Coulomb/Higgs limit.

\subsubsection{$U(N)$-$[l]$ ADHM}

The line defect index of the ADHM theory simplifies in the Coulomb limit as
\begin{align}
\langle L_{\mathsf{p};\mathsf{q}_0,\mathsf{q}_{\alpha}}\rangle^{\textrm{$U(1)$ ADHM-$[l]$}}_C(z;\mathfrak{t})
=\frac{1}{1-\mathfrak{t}^2}\delta_{\mathsf{p},0}\mathfrak{t}^{|\mathsf{q}_0|}\sum_{m\in\mathbb{Z}}\mathfrak{t}^{\sum_{\alpha=1}^l|m+\mathsf{q}_\alpha|}z^{lm+\sum_{\alpha=2}^l\mathsf{q}_\alpha}.
\label{ADHMLCoulomb}
\end{align}

In the Higgs limit, the monopole charge in the line defect index of the ADHM theory is restricted to $m=0$ and we obtain
\begin{align}
&\langle L_{\mathsf{p};\mathsf{q}_0,\mathsf{q}_{\alpha}}\rangle^{\textrm{$U(1)$ ADHM-$[l]$}}_H(x,y_{\alpha};\mathfrak{t})\nonumber \\
&=(1-\mathfrak{t}^2)\delta_{\mathsf{q}_0,0}\prod_{\alpha=2}^l\delta_{\mathsf{q}_\alpha,0}\prod_\pm \frac{1}{1-\mathfrak{t}x^{\pm 1}}\oint\frac{ds}{2\pi is}s^\mathsf{p}
\prod_{\alpha=1}^l\prod_\pm\frac{1}{1-\mathfrak{t}s^{\pm 1}y_\alpha^{\pm 1}}.
\end{align}
By using the delta function formula \eqref{formuladelta} we can rewrite this as
\begin{align}
&\langle L_{\mathsf{p};\mathsf{q}_0,\mathsf{q}_{\alpha}}\rangle^{\textrm{$U(1)$ ADHM-$[l]$}}_H(x,y_{\alpha};\mathfrak{t})\nonumber \\
&=(1-\mathfrak{t}^2)\delta_{\mathsf{q}_0,0}\prod_{\alpha=2}^l\delta_{\mathsf{q}_\alpha,0}\prod_\pm \frac{1}{1-\mathfrak{t}x^{\pm 1}}
\sum_{p_2,\cdots,p_l}
\oint\frac{ds}{2\pi is}s^\mathsf{p}
\prod_{\alpha=2}^l\oint \frac{d\sigma^{(\alpha)}}{2\pi i\sigma^{(\alpha)}}\Bigl(\frac{s}{\sigma^{(\alpha)}}\Bigr)^{p_\alpha}\nonumber \\
&\quad\times \prod_\pm
\frac{1}{1-\mathfrak{t}s^{\pm 1}y_1^{\pm 1}}
\prod_{\alpha=2}^l
\prod_\pm
\frac{1}{1-\mathfrak{t}(\sigma^{(\alpha)})^{\pm 1}y_\alpha^{\pm 1}}.
\end{align}
Performing the integrations by the formula \eqref{formulaI} we obtain
\begin{align}
&\langle L_{\mathsf{p};\mathsf{q}_0,\mathsf{q}_{\alpha}}\rangle^{\textrm{$U(1)$ ADHM-$[l]$}}_H(x,y_{\alpha};\mathfrak{t})\nonumber \\
&=\frac{1}{(1-\mathfrak{t}^2)^{l-1}}\delta_{\mathsf{q}_0,0}\prod_{\alpha=2}^l\delta_{\mathsf{q}_\alpha,0}\prod_\pm\frac{1}{1-\mathfrak{t}x^{\pm 1}}\sum_{n_2,\cdots,n_l}\mathfrak{t}^{\sum_{\alpha=2}^l|n_\alpha|+|\mathsf{p}+\sum_{\alpha=2}^ln_\alpha|}y_1^{-\mathsf{p}-\sum_{\alpha=2}^ln_\alpha}\prod_{\alpha=2}^ly_\alpha^{n_\alpha}.
\label{ADHMLHiggs}
\end{align}

\subsubsection{$\widetilde{U(1)^{\otimes l}\textrm{mADHM-}[1]}$}

In the line defect index of the mirror circular quiver theory, the monopole charges are restricted to $m^{(I)}=0$ in the Coulomb limit and we obtain
\begin{align}
&{\langle L_{\mathsf{p}^{(I)};\mathsf{q}}\rangle^{\widetilde{U(1)^{\otimes l}\textrm{mADHM-}[1]}}_C}(z;\mathfrak{t})\nonumber \\
&=(1-\mathfrak{t}^2)^l\delta_{\mathsf{q},0}\prod_{I=1}^l\oint\frac{ds^{(I)}}{2\pi is^{(I)}}
(s^{(I)})^{\mathsf{p}^{(I)}}
\frac{1}{1-\mathfrak{t}(s^{(1)})^{\pm 1}}\prod_{I=1}^l\frac{1}{1-\mathfrak{t}(\frac{s^{(I)}z}{s^{(I+1)}})^{\pm 1}}.
\end{align}
By redefining the integration variables as $s^{(1)}=\sigma^{(1)}$, $s^{(I)}=zs^{(I-1)}\sigma^{(I)}$ ($I=2,\cdots,l$) we find
\begin{align}
&{\langle L_{\mathsf{p}^{(I)};\mathsf{q}}\rangle_C^{\widetilde{U(1)^{\otimes l}\textrm{mADHM-}[1]}}}(z;\mathfrak{t})\nonumber \\
&=(1-\mathfrak{t}^2)^l\delta_{\mathsf{q},0}\prod_{I=1}^l\oint\frac{d\sigma^{(I)}}{2\pi i\sigma^{(I)}}
z^{\sum_I(I-1)\mathsf{p}^{(I)}}\prod_{I=1}^l(\sigma^{(I)})^{\sum_{J=I}^l\mathsf{p}^{(J)}}
\prod_{I=1}^l\prod_\pm \frac{1}{1-\mathfrak{t}(\sigma^{(I)})^{\pm 1}}\nonumber \\
&\quad\times \prod_\pm \frac{1}{1-\mathfrak{t}(z^l\sigma^{(2)}\cdots\sigma^{(l)})^{\pm 1}}.
\end{align}
After rewriting the last factor using the delta function formula \eqref{formuladelta}, we can perform the integrations by using the formula \eqref{formulaI} as
\begin{align}
&{\langle L_{\mathsf{p}^{(I)};\mathsf{q}}\rangle_C^{\widetilde{U(1)^{\otimes l}\textrm{mADHM-}[1]}}}(z;\mathfrak{t})\nonumber \\
&=\frac{1}{1-\mathfrak{t}^2}\delta_{\mathsf{q},0}\sum_{n\in\mathbb{Z}}
z^{\sum_I(I-1)\mathsf{p}^{(I)}+ln}
\mathfrak{t}^{|\sum_{I=1}^l|\mathsf{p}^{(I)}|+\sum_{I=2}^l|n+\sum_{J=I}^l\mathsf{p}^{(J)}|+|n|}.
\end{align}
This coincides with the Coulomb limit of the line defect index of the ADHM theory \eqref{ADHMLCoulomb} under the following identification of the charges
\begin{align}
\langle L_{\mathsf{p}=n;\mathsf{q}_0=\sum_{I=1}^ln^{(I)},\mathsf{q}_1=0,\mathsf{q}_{\alpha\ge 2}=\sum_{I=\alpha}^l\mathsf{p}^{(I)}}\rangle^{\textrm{$U(1)$ ADHM-$[l]$}}_C(z;\mathfrak{t})
={\langle L_{\mathsf{p}^{(I)}=n^{(I)};\mathsf{q}=n}\rangle_C^{\widetilde{U(1)^{\otimes l}\textrm{mADHM-}[1]}}}(z;\mathfrak{t}).
\label{ADHMLmADHMLdictionary}
\end{align}

In the Higgs limit, line defect index of the mirror circular quiver theory simplifies as
\begin{align}
&{\langle L_{\mathsf{p}^{(I)};\mathsf{q}}\rangle_H^{\widetilde{U(1)^{\otimes l}\textrm{mADHM-}[1]}}}(y_{\alpha},x;\mathfrak{t})\nonumber \\
&=\frac{1}{(1-\mathfrak{t}^2)^l}\sum_{m^{(1)},\cdots,m^{(l)}}\prod_{I=1}^l\oint\frac{ds^{(I)}}{2\pi is^{(I)}}\prod_I(s^{(I)})^{\mathsf{p}^{(I)}}\mathfrak{t}^{|m^{(1)}|+\sum_{I=1}^{l-1}|m^{(I)}-m^{(I+1)}|+|m^{(l)}-m^{(1)}+\mathsf{q}|}
x^{m^{(1)}}\nonumber \\
&\quad\times y_1^{m^{(1)}-m^{(l)}-\mathsf{q}}
\prod_{\alpha=2}^ly_\alpha^{m^{(\alpha)}-m^{(\alpha-1)}}\nonumber \\
&=\frac{1}{(1-\mathfrak{t}^2)^l}\prod_{I=1}^l\delta_{\mathsf{p}^{(I)},0}\sum_{m^{(1)},\cdots,m^{(l)}}\mathfrak{t}^{|m^{(1)}|+\sum_{I=1}^{l-1}|m^{(I)}-m^{(I+1)}|+|m^{(l)}-m^{(1)}+\mathsf{q}|}
x^{m^{(1)}}\nonumber \\
&\quad\times y_1^{m^{(1)}-m^{(l)}-\mathsf{q}}
\prod_{\alpha=2}^ly_\alpha^{m^{(\alpha)}-m^{(\alpha-1)}}.
\end{align}
Changing the dummy variables as $m^{(1)}=n_0$ and $m^{(I)}-m^{(I-1)}=n_I$ ($I=2,\cdots,l$), we obtain
\begin{align}
&{\langle L_{\mathsf{p}^{(I)};\mathsf{q}}\rangle_H^{\widetilde{U(1)^{\otimes l}\textrm{mADHM-}[1]}}}(y_{\alpha},x;\mathfrak{t})\nonumber \\
&=\frac{1}{(1-\mathfrak{t}^2)^l}\prod_{I=1}^l\delta_{\mathsf{p}^{(I)},0}\sum_{n_0,n_2,\cdots,n_l}\mathfrak{t}^{|n_0|+\sum_{\alpha=2}^l|n_\alpha|+|\sum_{\alpha=2}^ln_\alpha+\mathsf{q}|}
x^{n_0}
y_1^{-\mathsf{q}-\sum_{\alpha=2}^ln_\alpha}
\prod_{\alpha=2}^ly_\alpha^{n_\alpha}.
\end{align}
Performing the summation over $n_0$ we are left with
\begin{align}
&{\langle L_{\mathsf{p}^{(I)};\mathsf{q}}\rangle_H^{\widetilde{U(1)^{\otimes l}\textrm{mADHM-}[1]}}}(y_{\alpha},x;\mathfrak{t})\nonumber \\
&=\frac{1}{(1-\mathfrak{t}^2)^{l-1}}\prod_{I=1}^l\delta_{\mathsf{p}^{(I)},0}\prod_\pm\frac{1}{1-\mathfrak{t}x^{\pm 1}}\sum_{n_2,\cdots,n_l}\mathfrak{t}^{\sum_{\alpha=2}^l|n_\alpha|+|\sum_{\alpha=2}^ln_\alpha+\mathsf{q}|}
y_1^{-\mathsf{q}-\sum_{\alpha=2}^ln_\alpha}
\prod_{\alpha=2}^ly_\alpha^{n_\alpha}.
\end{align}
This coincides with the Higgs limit in the ADHM theory \eqref{ADHMLHiggs} under the same identification of the charges as in the Coulomb limit \eqref{ADHMLmADHMLdictionary}
\begin{align}
&\langle L_{\mathsf{p}=n;\mathsf{q}_0=\sum_{I=1}^l\mathsf{p}^{(I)},\mathsf{q}_1=0,\mathsf{q}_{\alpha\ge 2}=\sum_{I=\alpha}^ln^{(I)}}\rangle^{\textrm{$U(1)$ ADHM-$[l]$}}_H(x,y_{\alpha};\mathfrak{t})\nonumber \\
&={\langle L_{\mathsf{p}^{(I)}=n^{(I)};\mathsf{q}=n}\rangle_H^{\widetilde{U(1)^{\otimes l}\textrm{mADHM-}[1]}}}(y_{\alpha},x;\mathfrak{t}).
\end{align}

\subsubsection{Circular quiver CS theory}

Lastly let us consider the line defect index of the circular quiver CS theory which we call $(l,1)_1$ model \eqref{l1kmodelLpq}.

In the Coulomb limit, monopole charges in the line defect index \eqref{l1kmodelLpq} are restricted to $m^{(I)}=m^{(1)}$ for $I=2,\cdots,l+1$, and hence the line defect index simplifies as
\begin{align}
&\langle L_{\mathsf{p}^{(I)};\mathsf{q}}\rangle^{(l,1)_1}_C(z;\mathfrak{t})=\sum_{m^{(1)}}(1-\mathfrak{t}^2)^{l-1}\prod_{I=1}^{l+1}\oint\frac{ds^{(I)}}{2\pi is^{(I)}}
(s^{(1)})^{m^{(1)}+\mathsf{p}^{(1)}}
\prod_{I=2}^l(s^{(I)})^{\mathsf{p}^{(I)}}
(s^{(l+1)})^{-m^{(1)}+\mathsf{p}^{(l+1)}}\nonumber \\
&\quad\times \prod_{I=1}^l\frac{1}{1-\mathfrak{t}(z\frac{s^{(I)}}{s^{(I+1)}})^{\pm 1}}
\mathfrak{t}^{|\mathsf{q}|}
z^{-lm^{(1)}}.
\end{align}
By changing the integration variables as $\frac{s^{(I)}}{zs^{(I-1)}}=\sigma_I$ ($I=2,\cdots,l+1$), we can perform the integrations by using the formula \eqref{formulaI} as
\begin{align}
\langle L_{\mathsf{p}^{(I)};\mathsf{q}}\rangle^{(l,1)_1}_C(z;\mathfrak{t})=\frac{1}{1-\mathfrak{t}^2}\delta_{\sum_{I=1}^{l+1}\mathsf{p}^{(I)},0}\mathfrak{t}^{|\mathsf{q}|}\sum_{m^{(1)}}\mathfrak{t}^{\sum_{I=2}^{l+1}|\sum_{J=I}^{l+1}\mathsf{p}^{(J)}-m^{(1)}|}z^{\sum_{I=2}^l(I-1)\mathsf{p}^{(I)}+l(-m^{(1)}+\mathsf{p}^{(l+1)})}.
\end{align}
After changing the dummy variable as $m^{(1)}=-m+\mathsf{p}^{(l+1)}$, we find the line defect index coincides with the line defect index of the ADHM theory with $l$ flavors \eqref{ADHMLCoulomb} under the following identification of the charges:
\begin{align}
\langle L_{\mathsf{p}^{(I)}=n^{(I)};\mathsf{q}=n}\rangle^{(l,1)_1}_C(z;\mathfrak{t})=\langle L_{\mathsf{p}=\sum_{I=1}^{l+1}n^{(I)};\mathsf{q}_0=\sum_{I=1}^{l+1}n^{(I)},\mathsf{q}_1=0,\mathsf{q}_{\alpha\ge 2}=\sum_{I=\alpha}^ln^{(I)}}\rangle^{\textrm{$U(1)$ ADHM-$[l]$}}_C(z;\mathfrak{t}).
\label{appendixADHMl11chargeidentification}
\end{align}

In the Higgs limit, the monopole charge $m^{(l+1)}$ is restricted to $m^{(l+1)}=m^{(1)}-\mathsf{q}$ and hence the line defect index of the $(l,1)_1$ model reduces to
\begin{align}
&\langle L_{\mathsf{p}^{(I)};\mathsf{q}}\rangle^{(l,1)_1}_H(x,y_\alpha;\mathfrak{t})=\sum_{m^{(1)},\cdots,m^{(l)}}\frac{1}{(1-\mathfrak{t}^2)^{l-1}}\prod_{I=1}^{l+1}\oint\frac{ds^{(I)}}{2\pi is^{(I)}}
(s^{(1)})^{m^{(1)}+\mathsf{p}^{(1)}}
\prod_{I=2}^l(s^{(I)})^{\mathsf{p}^{(I)}}\nonumber \\
&\quad\times (s^{(l+1)})^{-m^{(1)}+\mathsf{q}+\mathsf{p}^{(l+1)}}
\frac{1}{1-\mathfrak{t}(x\frac{s^{(l+1)}}{s^{(1)}})^{\pm 1}}
\mathfrak{t}^{\sum_{I=1}^{l-1}|m^{(I)}-m^{(I+1)}|+|m^{(l)}-m^{(1)}+\mathsf{q}}
y_2^{m^{(1)}}
\prod_{\alpha=2}^{l-1}\Bigl(\frac{y_{\alpha+1}}{y_\alpha}\Bigr)^{m^{(\alpha)}}
\nonumber \\
&\quad\times 
\Bigl(\frac{y_1}{y_l}\Bigr)^{m^{(l)}}
y_1^{-m^{(1)}+\mathsf{q}}.
\end{align}
By defining the new integration variable as $s^{(l+1)}=\frac{s^{(1)}\sigma}{x}$, we can perform the integrations by using the formula \eqref{formulaI} as
\begin{align}
&\langle L_{\mathsf{p}^{(I)};\mathsf{q}}\rangle^{(l,1)_1}_H(x,y_\alpha;\mathfrak{t})
=
\frac{1}{(1-\mathfrak{t}^2)^l}
\delta_{\mathsf{p}^{(1)}+\mathsf{p}^{(l+1)}+\mathsf{q},0}
\prod_{I=2}^l\delta_{\mathsf{p}^{(I)},0}
\sum_{m^{(1)},\cdots,m^{(l)}}
x^{m^{(1)}-\mathsf{q}-\mathsf{p}^{(l+1)}}\nonumber \\
&\quad\times \mathfrak{t}^{|m^{(1)}-\mathsf{q}-\mathsf{p}^{(l+1)}|+\sum_{I=1}^{l-1}|m^{(I)}-m^{(I+1)}|+|m^{(l)}-m^{(1)}+\mathsf{q}|}
y_2^{m^{(1)}}
\prod_{\alpha=2}^{l-1}\Bigl(\frac{y_{\alpha+1}}{y_\alpha}\Bigr)^{m^{(\alpha)}}
\Bigl(\frac{y_1}{y_l}\Bigr)^{m^{(l)}}
y_1^{-m^{(1)}+\mathsf{q}}.
\end{align}
After changing the dummy variables as $m^{(I)}=m_I'+m^{(1)}$ ($I=2,\cdots,l$), we can perform the summation over $m^{(1)}$ and obtain
\begin{align}
&\langle L_{\mathsf{p}^{(I)};\mathsf{q}}\rangle^{(l,1)_1}_H(x,y_\alpha;\mathfrak{t})
=
\frac{1}{(1-\mathfrak{t}^2)^{l-1}}
\prod_\pm\frac{1}{1-x^{\pm 1}\mathfrak{t}}
\delta_{\mathsf{p}^{(1)}+\mathsf{p}^{(l+1)}+\mathsf{q},0}
\prod_{I=2}^l\delta_{\mathsf{p}^{(I)},0}\nonumber \\
&\quad\times \sum_{m'_2,\cdots,m'_l}
\mathfrak{t}^{|m'_2|+\sum_{I=2}^{l-1}|m'_I-m'_{I+1}|+|m_l'+\mathsf{q}|}
\prod_{\alpha=2}^{l-1}\Bigl(\frac{y_{\alpha+1}}{y_\alpha}\Bigr)^{m_\alpha'}
\Bigl(\frac{y_1}{y_l}\Bigr)^{m_l'}
y_1^{\mathsf{q}}.
\end{align}
By comparing this expression with the line defect index of the ADHM theory \eqref{ADHMLHiggs} with the following identification of the dummy variables
\begin{align}
m'_2=-n_2,\quad m'_\alpha-m'_{\alpha-1}=-n_\alpha,\quad (\alpha=3,\cdots,l),
\end{align}
we find the two expressions agree with each other after taking into account of the constraint $\delta_{\mathsf{p}^{(1)}+\mathsf{p}^{(l+1)}+\mathsf{q},0}$ and $\delta_{\mathsf{p}^{(I)},0}$ ($I=2,\cdots,l$), under the identification of the charges \eqref{appendixADHMl11chargeidentification}.

\bibliographystyle{utphys}
\bibliography{ref}

\end{document}